\documentclass[final,3p,times]{elsarticle}

\usepackage{lineno,hyperref,amsmath}

\usepackage{subfigure}
\usepackage{tabularx}
\usepackage{multirow}

\journal{Journal of Computational Physics}

\begin{document}

\begin{frontmatter}

\title{A nonhydrostatic atmospheric dynamical core on cubed sphere using hybrid multi-moment finite-volume/finite difference methods: formulations and preliminary tests}

\author[XJTU]{Chungang Chen}

\author[CEMC]{Xingliang Li}

\author[Titech]{Feng Xiao}

\author[CEMC]{Xueshun Shen\corref{correspondingauthor}}
\cortext[correspondingauthor]{Corresponding author}
\ead{shenxs@cma.gov.cn}

\address[XJTU]{State Key Laboratory for Strength and Vibration
of Mechanical Structures and Department of Mechanics, School of Aerospace Engineering, Xi'an Jiaotong
University, Xi'an, China}
\address[CEMC]{Center for Earth System Modelling and Predication, China
Meteorological Administration, Beijing, China}
\address[Titech]{Department of Mechanical Engineering, Tokyo
Institute of Technology, Tokyo, Japan}

%%%%%%%%%%%%%%%%%% ABSTRACT

\begin{abstract}
A nonhydrostatic dynamical core has been developed by using the multi-moment finite volume method that ensures the rigorous numerical conservation. To represent the spherical geometry free of polar problems, the cubed-sphere grid is adopted. A fourth-order multi-moment discretization formulation is applied to solve the governing equations cast in the local curvilinear coordinates on each patch of cubed sphere through a gnomonic projection. In vertical direction, the height-based terrain-following grid is used to deal with the topography and a conservative finite difference scheme is adopted for the spatial discretization. The dynamical core adopts the nonhydrostatic governing equations. To get around the CFL stability restriction imposed by sound wave propagation and relatively small grid spacing in the vertical direction, the dimensional-splitting time integration algorithm using the HEVI (horizontally-explicit and vertically-implicit) strategy is implemented by applying the IMEX (implicit-explicit) Runge-Kutta method. The proposed model was checked by the widely-used benchmark tests in this study. The numerical results show that the multi-moment model has superior solution quality and great practical potential as a numerical platform for development of the atmospheric general circulation models.
\end{abstract}


\end{frontmatter}

%\linenumbers

%%%%%%%%%%%%%%%%%% INTRODUCTION SECTION

\section{Introduction}

The multi-moment methods were proposed by introducing two or more kinds of moments, which are quantities used to describe the spatial distributions of physical fields, such as the pointwise values, the volume (surface or line) integrated averages, the derivatives of different orders and so on. The different moments can be used as the model variables directly \cite{Xiao2004,Xiao2006} or the constraint conditions \citep{ii2009} to derive the updating formulations of the unknowns. With more local degrees of freedom (DOFs), the multi-moment schemes can accomplish the high-order spatial reconstructions within the compact stencils. As a result, they have better flexibility in dealing with the various grid topologies with the unified numerical framework and is promising to develop the highly scalable models running on the massive parallel clusters, like discontinuous Galerkin scheme, spectral element schemes among others. Furthermore, the moments defined in our schemes have clear physical meanings and the resulting discretization procedures are flexible in achieving the different numerical properties and simple to be implemented in various applications \citep{xiao2013}.

To develop the numerical models for atmospheric dynamics in spherical geometry, the computational meshes with quasi-uniform grid spacing, such as cubed-sphere grid \citep{Sadourny1972}, icosahedral geodesic grid \citep{Sadourny1968,Williamson1968} and Yin-Yang grid \cite{Kageyama2013}, gain more and more attentions in the past several decades \cite{Staniforth2012} due to the continuously increasing demands on refining the grid resolutions of global models. With the applications of multi-moment finite volume method, the unified high-order formulation for global shallow-water equations (SWEs) has been proposed on these three grids \citep{chen2014}. The numerical results of spherical shallow-water test cases verified that the compact spatial reconstructions realized by using multi-moment concept are helpful to suppress the extra numerical errors around the patch edges on cubed sphere and build the rigorous third-order model on icosahedral grid with the hexagonal and pentagonal elements. Following our previous studies, a three-dimensional nonhydrostatic model for atmospheric dynamics using multi-moment finite volume scheme is reported in this paper. The cubed-sphere grid is adopted in this study because the high-order schemes designed on Cartesian grid can be straightforwardly implemented on the structured-quadrilateral grid and the resulting model is more computationally efficient than those on the icosahedral-hexagonal grid. Additionally, in comparison with the overset Yin-Yang grid the numerical model on cubed sphere can naturally assure the numerical conservation without complex corrections if the adopted scheme is conservative and the flux-form governing equations are solved.

To extend the global shallow-water model to three-dimensional dynamical core, attentions should be paid in developing the proper vertical discretization scheme and the efficient time marching scheme to assure the available time step suited for the practical applications.

The multi-moment scheme can also be applied in vertical direction and a completely 4$^{th}$-order model are achieved as we have reported in \cite{chen2020global}. However, updating the DOF defined at the surface introduces some difficulties for practical models with the physical processes, e.g. the evaluation of derivatives of flux functions in vertical direction at surface requires finer grid resolution to assure the stability within the boundary layer and the physical parameterizations should to be modified to calculate the values of physical sources at cell center as well as its endpoints. Thus, a conservative three-point finite-difference scheme is designed in this study, where the DOFs are defined as pointwise values of unknowns only at cell centers in vertical direction. As no DOF is defined at surface, only the flux functions, excluding their derivatives, are evaluated there. Additionally, it is helpful to implement the dynamics-physics coupling in practice as what has been done in many existing models.

With very fine horizontal resolution, the hydrostatic approximation, widely used in many existing models, is no longer valid. Since the nonhydrostatic governing equations are adopted in this study, the propagation of sound wave in vertical direction is directly solved and should be carefully considered in designing the time marching scheme. The explicit model in three dimensions can only run with a very restrictive time step in comparison with the time scale of dominant phenomena due to the very large ratio between the grid spacings in horizontal and vertical directions. A dimensional-splitting scheme using horizontally-explicit and vertically-implicit (HEVI) strategy is adopted here to design a time marching scheme with an acceptable time integration step in the practical model. In this study, the implicit-explicit (IMEX) Runge-Kutta method \cite{ascher1997,CB15} is adopted. All terms related to the partial derivatives with respect to the vertical direction and the stiff source terms, e.g. the gravity force term, are treated in the implicit way. While all numerical operations requiring the data exchanging between the subdomains decomposed in horizontal directions for parallel computations are treated explicitly to preserve the high scalability. The resulting model can run with a time step determined by the CFL stability condition in horizontal directions. Since the high-order spatial and temporal discretization schemes are applied in vertical direction, it is expected that the proposed model is robust and accurate not only in simulating the quasi-hydrostatic large-scale atmospheric dynamics, but also in the non-hydrostatic multi-scale ones.

The rest of this paper is organized as follows. In section 2, the numerical formulations of the multi-moment nonhydrostatic dynamical core are described in details. Some widely-used benchmark tests are then checked to verify the performance of the proposed numerical model in section 3. And a short summary is finally given in section 4.

%%%%%%%%%%%%%%%%%% NUMERICAL FORMULATIONS

\section{Numerical formulations}

\subsection{Governing equations}

On each patch of cubed sphere, the nonhydrostatic governing equations for atmospheric dynamics with shallow-atmosphere assumption are written in the flux-form as \cite{mcore2012,clark1977}
\begin{equation}\label{eqs}
 \frac{\partial \boldsymbol{q}}                           {\partial t}
+\frac{\partial \boldsymbol{e}\left(\boldsymbol{q}\right)}{\partial \xi}
+\frac{\partial \boldsymbol{f}\left(\boldsymbol{q}\right)}{\partial \eta}
+\frac{\partial \boldsymbol{h}\left(\boldsymbol{q}\right)}{\partial \zeta}
=\boldsymbol{s}\left(\boldsymbol{q}\right),
\end{equation}
where $\left(\xi,\eta\right)$ are local horizontal coordinates on each patch of the cubed sphere, $\zeta$ is a height-based terrain-following coordinate in vertical direction, $\boldsymbol{q}$ are dependent variables (predicted variables), $\boldsymbol{e}\left(\boldsymbol{q}\right)$, $\boldsymbol{f}\left(\boldsymbol{q}\right)$ and $\boldsymbol{h}\left(\boldsymbol{q}\right)$ are flux functions in $\xi$, $\eta$ and $\zeta$ directions, respectively  and $\boldsymbol{s}\left(\boldsymbol{q}\right)$ denotes all source terms. The detailed expressions of governing equations is described with a brief introduction of transformation laws of curvilinear coordinates as follows.

In the horizontal directions, the coordinates are $\xi=R\alpha$ and $\eta=R\beta$, where $R$ is radius of the Earth and $\alpha$, $\beta$ are central angles for a gnomonic projection varying within $\left[-\frac{\pi}{4},\frac{\pi}{4}\right]$ for each patch (details can be referred to \cite{chen2008}).

In the vertical direction, $\zeta\in\left[0,z_t\right]$ is a uniform grid, where $z_t$ is altitude of model top. A non-uniform grid can be generated by a transformation $\hat{\zeta}=\mathcal{T}\left(\zeta\right)$, which has smaller grid spacing near the surface to better represent the effects of the topography and the atmospheric boundary layer. The formulations used to generated the non-uniform grid adopted in this study are described as follows. The smallest grid spacing of coordinate $\hat{\zeta}$ at surface is $\Delta \hat{\zeta}_{\min}$ and the largest one is $\Delta \hat{\zeta}_{\max}$ at model top. In the region close to the surface $\left[0,\zeta_1\right]$ or model top $\left[\zeta_2,z_t\right]$, several layers of uniform cells may be arranged. In the region $\zeta\in\left[\zeta_1,\zeta_2\right]$, the grid spacing is gradually increasing from $\Delta \hat{\zeta}_{\min}$ to $\Delta \hat{\zeta}_{\max}$. Additionally, we require the $2^{nd}$-order derivatives of transformation $\mathcal{T}$ are zero at $\zeta=\zeta_1$ and $\zeta=\zeta_2$ to make the transformation C2-continuous.Thus, the grid transformation can be derived as
\begin{equation}
\hat{\zeta}=\mathcal{T}\left(\zeta\right)=
\left\{
\begin{array}{ll}
\mathcal{T}_l=\frac{\Delta\hat{\zeta}_\mathrm{min}}{\Delta\zeta}\zeta&\mathrm{if\ }\zeta\le\zeta_1\\
\mathcal{T}_m=\displaystyle\sum_{r=0}^{5}c_r\zeta^{r}&\mathrm{if\ }\zeta_1<\zeta<\zeta_2\\
\mathcal{T}_h=z_t+\frac{\Delta\hat{\zeta}_\mathrm{max}}{\Delta\zeta}\left(\zeta-z_t\right)&\mathrm{otherwise}
\end{array}
\right.
\end{equation}
where the coefficients $c_r$ ($r=0$ to $5$) are determined with constraint conditions as
\begin{equation}
\left\{\begin{array}{l}
\mathcal{T}_m\left(\zeta_1\right)=\frac{\Delta\hat{\zeta}_\mathrm{min}}{\Delta \zeta}\zeta_1\\
\mathcal{T}_m\left(\zeta_2\right)=z_t+\frac{\Delta\hat{\zeta}_\mathrm{max}}{\Delta \zeta}\left(\zeta_2-z_t\right)\\
\mathcal{T}^\prime_m\left(\zeta_1\right)=\frac{\Delta\hat{\zeta}_\mathrm{min}}{\Delta \zeta}\\
\mathcal{T}^\prime_m\left(\zeta_2\right)=\frac{\Delta\hat{\zeta}_\mathrm{max}}{\Delta \zeta}\\
\mathcal{T}^{\prime\prime}_m\left(\zeta_1\right)=0\\
\mathcal{T}^{\prime\prime}_m\left(\zeta_2\right)=0
\end{array}\right..
\end{equation}

Considering the surface topography $z_s\left(\xi,\eta\right)$, the terrain-following coordinate is then built as  \cite{schar2002}
\begin{equation}
z=\hat{\zeta}+z_s\left(\xi,\eta\right)\frac{\sinh\left[\left(z_t-\hat{\zeta}\right)/S\right]}{\sinh\left(z_t/S\right)},\label{verticalmapping}
\end{equation}
where $z$ is altitude and the scale height $S=5000$ m is adopted in this study.

The horizontal transformation laws between the longitude-latitude ($\lambda-\phi$) grid and the local curvilinear coordinates on each patch of cubed sphere are defined as follows.

The contravariant base vectors $\boldsymbol{a}^\xi$ and $\boldsymbol{a}^\eta$ are
\begin{equation}
\left\{
\begin{array}{c}
\boldsymbol{a}^\xi=\boldsymbol{i}\frac{1}{R\cos\theta}\frac{\partial \xi}{\partial \lambda}+\boldsymbol{j}\frac{1}{R}\frac{\partial \xi}{\partial \phi}\\
\boldsymbol{a}^\eta=\boldsymbol{i}\frac{1}{R\cos\theta}\frac{\partial \eta}{\partial \lambda}+\boldsymbol{j}\frac{1}{R}\frac{\partial \eta}{\partial \phi}\\
\end{array}
\right..
\end{equation}

Above base vectors have different expressions on different patches and can be derived from the projection relations \cite{Nair2005a}.

The horizontal contravariant metric tensor is
\begin{equation}
\boldsymbol{G}^{ij}_H=\frac{\delta}{\left(1+X^2\right)\left(1+Y^2\right)}\left[\begin{array}{cc}1+Y^2&XY\\XY&1+X^2\end{array}\right],
\end{equation}
where $X=\tan{\alpha}$, $Y=\tan{\beta}$ and $\delta=\sqrt{1+X^2+Y^2}$.

The Jacobian of the horizontal transformation is
\begin{equation}
J_H=\left[\det\left({\boldsymbol{G}^{ij}_H}^{-1}\right)\right]^{\frac{1}{2}}=\left(1+X^2\right)\left(1+Y^2\right)\delta^{-3}.
\end{equation}

The contravariant velocity components are obtained by
\begin{equation}
\left\{
\begin{array}{c}
\tilde{u}=\boldsymbol{a}^\xi\cdot\boldsymbol{v}\\
\tilde{v}=\boldsymbol{a}^\eta\cdot\boldsymbol{v}\\
\end{array}
\right.,
\end{equation}
where $\boldsymbol{v}=\left(u_s,v_s\right)$ is the velocity vector on longitude-latitude grid.

The details of projection relations and transformation laws on cubed sphere can be referred to \cite{Nair2005a,Nair2005b,chen2008,mcore2012}.

In vertical direction, the governing equations in the height-based terrain-following coordinates can be derived through the chain rules \citep{clark1977}. The Jacobian of vertical transform is $J_V=\frac{\partial z}{\partial \zeta}$, which can be directly obtained through Eq. \eqref{verticalmapping}. The components of contravariant metric tensor related to vertical transformation are $G_V^{13}=\frac{\partial \zeta}{\partial \xi}\left.\right|_{z=constant}$ and $G_V^{23}=\frac{\partial \zeta}{\partial \eta}\left.\right|_{z=constant}$. For the idealized test cases in this study, they can be analytically evaluated.

The dependent variables adopted in this study are \cite{IFSFVM2019}
\begin{equation}
\boldsymbol{q}=\left[J\rho,J\rho \tilde{u}, J\rho \tilde{v}, J\rho w, J\rho\theta^\prime\right]^T,
\end{equation}
where $J$ is the Jacobian of the transformation $J=J_HJ_V$, $\rho$ is density, $\tilde{u}$ and $\tilde{v}$ are contravariant velocity components in horizontal directions, $w$ is vertical velocity, $\theta$ is potential temperature and the superscript prime denotes the deviation with respect to the hydrostatic reference state as
\begin{equation}
\theta^\prime\left(\xi,\eta,\zeta\right)=\theta\left(\xi,\eta,\zeta\right)-\overline{\theta}\left(\xi,\eta,\zeta\right)
\end{equation}
The reference state is derived through the hydrostatic balance in vertical direction as
\begin{equation}
\frac{\partial \overline{p}}{\partial z}=-g\overline{\rho}.
\end{equation}
It usually has an analytic expression from the initial condition in the benchmark tests.

The flux functions are written in three directions as
\begin{equation}
\boldsymbol{e}=J\left[\rho\tilde{u},\rho\tilde{u}^2+G_H^{11}p^\prime,\rho\tilde{u}\tilde{v}+G_H^{12}p^\prime,\rho\tilde{u}w,\rho\tilde{u}\theta^\prime \right]^T,
\end{equation}
\begin{equation}
\boldsymbol{f}=J\left[\rho\tilde{v},\rho\tilde{u}\tilde{v}+G_H^{21}p^\prime,\rho\tilde{v}^2+G_H^{22}p^\prime,\rho\tilde{v} w,\rho\tilde{v}\theta^\prime \right]^T,
\end{equation}
and
\begin{equation}
\boldsymbol{h}=J\left[\rho\tilde{w},\rho\tilde{u}\tilde{w}+M^1p^\prime,\rho\tilde{v}\tilde{w}+M^2p^\prime,\rho\tilde{w}^2+ J_V^{-1}p^\prime,\rho\tilde{w}\theta^\prime \right]^T,
\end{equation}
where $\tilde{w}=\frac{1}{J_V}w+G^{13}_V\tilde{u}+G^{23}_V\tilde{v}$, $M^{s}=\left(G_V^{13}G_H^{s1}+G_V^{23}G_H^{s2}\right)$ ($s=1\ \mathrm{to}\ 2$) and the deviation of pressure is
$p^\prime\left(\xi,\eta,\zeta\right)=p\left(\xi,\eta,\zeta\right)-\overline{p}\left(\xi,\eta,\zeta\right)$.

The source term is written as
\begin{equation}
\boldsymbol{s}=\boldsymbol{s}_{H1}+\boldsymbol{s}_{H2}+\boldsymbol{s}_P+\boldsymbol{s}_C+\boldsymbol{s}_G+\boldsymbol{s}_R.
\end{equation}

$\boldsymbol{s}_{H1}$ includes the derivatives of reference pressure $\overline{p}$ as
\begin{equation}
\boldsymbol{s}_{H1}=-J\left[
\begin{array}{c}
0\\
G_H^{11} \frac{\partial \overline{p}}{\partial \xi}+G_H^{12}\frac{\partial \overline{p}}{\partial \eta}+M^1\frac{\partial \overline{p}}{\partial \zeta}\\
G_H^{21} \frac{\partial \overline{p}}{\partial \xi}+G_H^{22}\frac{\partial \overline{p}}{\partial \eta}+M^2\frac{\partial \overline{p}}{\partial \zeta}\\
0\\
0
\end{array}
\right].
\end{equation}

$\boldsymbol{s}_{H2}$ includes the derivatives of reference potential temperature $\overline{\theta}$ as
\begin{equation}
\boldsymbol{s}_{H2}=-J\left[0,0,0,0,
\rho\tilde{u}\frac{\partial \overline{\theta}}{\partial \xi}+\rho\tilde{v}\frac{\partial \overline{\theta}}{\partial \eta}+\rho\tilde{w}\frac{\partial \overline{\theta}}{\partial \zeta}
\right]^T.
\end{equation}

$\boldsymbol{s}_P$ is the source term due to the horizontal grid transformation as \cite{mcore2012}
\begin{equation}
\boldsymbol{s}_P=\frac{2J}{R\delta^2}\left[0,AY\tilde{u},-BX\tilde{v},0,0\right]^T,
\end{equation}
and $\boldsymbol{s}_C$ is the source term representing the Coriolis force, having the form of \cite{mcore2012}
\begin{equation}
\boldsymbol{s}_C=\frac{2J\Omega}{\delta^2}\left[0,AY,BY,0,0\right]^T
\end{equation}
on patch one to four,
\begin{equation}
\boldsymbol{s}_C=\frac{2J\Omega}{\delta^2}\left[0,A,B,0,0\right]^T
\end{equation}
on patch five and
\begin{equation}
\boldsymbol{s}_C=-\frac{2J\Omega}{\delta^2}\left[0,A,B,0,0\right]^T,
\end{equation}
on patch six, where $\Omega$ is rotational speed of the Earth,
\begin{equation}
A=-XY\rho\tilde{u}+\left(1+Y^2\right)\rho\tilde{v}
\end{equation}
and
\begin{equation}
B=-\left(1+X^2\right)\rho\tilde{u}+XY\rho\tilde{v}.
\end{equation}

$\boldsymbol{s}_G$ is the source term representing the gravity force as
\begin{equation}
\boldsymbol{s}_G=\left[0,0,0,-Jg\rho^\prime,0\right]^T,
\end{equation}
where $g$ is gravitation constant.

$\boldsymbol{s}_R$ is the source term to introducing Rayleigh friction near model top, having the form of
\begin{equation}
\boldsymbol{s}_R=\tau_R\left(\zeta\right)\rho\left[0,\tilde{u}-\tilde{u}_f,\tilde{v}-\tilde{v}_f,w,0\right]^T.
\end{equation}
where coefficient $\tau_R$ determines the strength of Rayleigh friction and $\left(u_f,v_f,0\right)$ denotes a reference velocity field. Rayleigh friction is adopted to absorb the reflected waves from top boundary where a solid wall boundary condition is applied to assure the numerical conservation.

\subsection{Definition of degrees of freedom}

\begin{figure}[ht]
 \centering
 \includegraphics[width=0.6\textwidth]{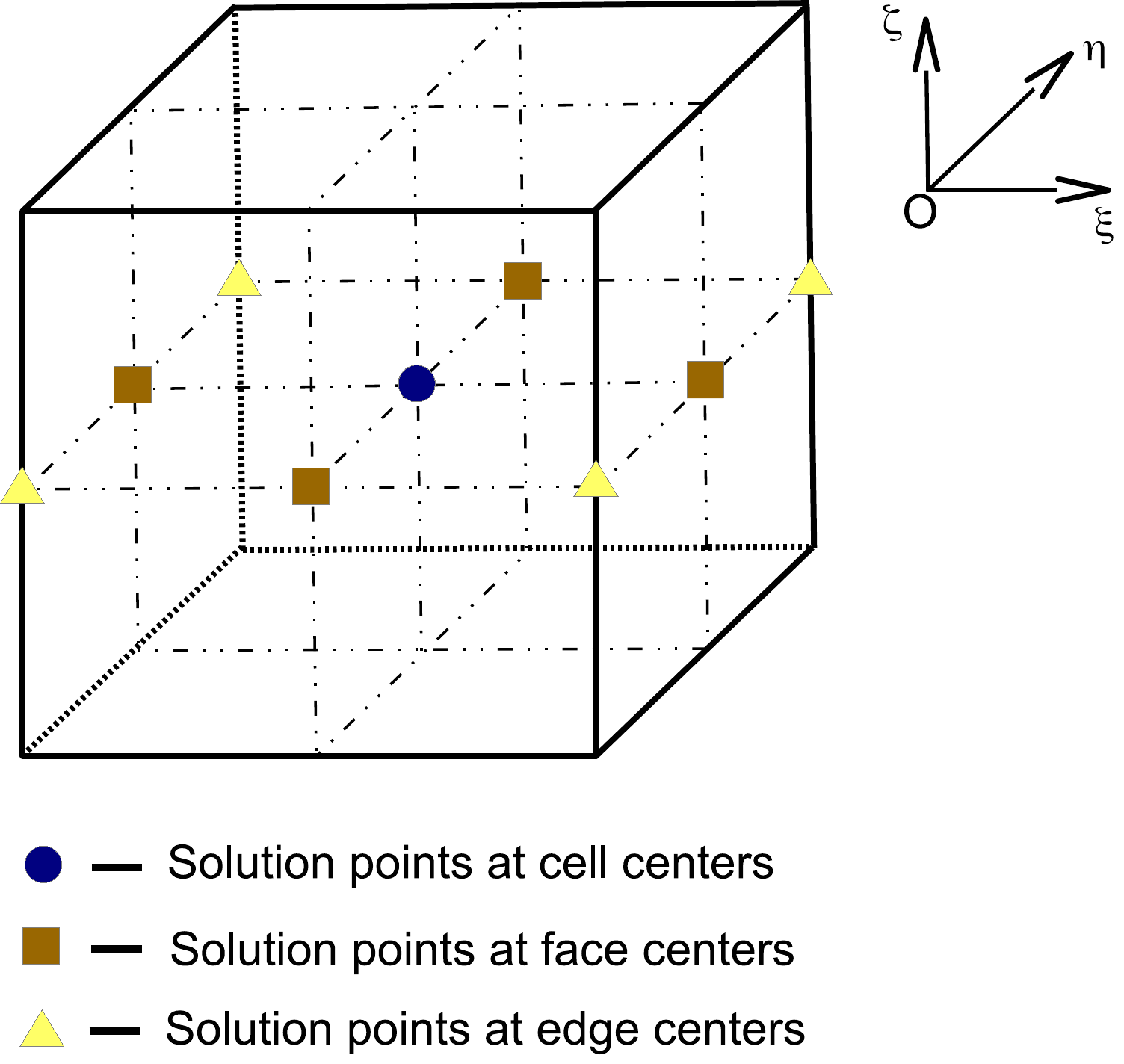}
  \caption{Definition of local DOFs within cell $\mathcal{C}^{ijkp}$.}\label{DOF}
\end{figure}

The 3-point multi-moment constrained finite volume (MCV) method \citep{ii2009} is adopted to implement the spatial discretization in horizontal directions. Nine pointwise values are defined as local DOFs within each cell to construct the 3-point MCV scheme in two dimensions, as shown in Fig. \ref{DOF} for cell $\mathcal{C}^{ijkp}$, where superscripts $i,j,k$ denote the indices in $\xi$, $\eta$ ($i,j=1\ \mathrm{to}\ N_h$) and $\zeta$ ($k=1\ \mathrm{to}\ N_v$) directions and $p=1\ \mathrm{to}\ 6$ the number of the patch. The solution points are equidistantly distributed within the cell and the DOFs defined at the cell surfaces are shared by adjacent cells. All local DOFs are defined at the centers of line segments in vertical direction, where the conservative finite difference scheme is adopted to accomplish the spatial discretizations.

The total number of computational cells adopted by the proposed model is $6{N_h}^2N_v$. The resolution in horizontal direction along the equator is $\frac{90^\circ}{N_h}$ in terms of number of computational cells and $\frac{45^\circ}{N_h}$ in terms of number of DOFs. In vertical directions, total number of layers is $N_v$. Hereafter, we denote the computational mesh by its resolution $N_h\times N_v$.

\subsection{Spatial discretizations}

At solution point $P^{ijkp}_{mn}$, where the superscripts denote the indices of cell, the subscripts $m,n=1\ \mathrm{to}\ 3$ are local indices of DOFs within the corresponding computational cell, the local DOF is updated through a differential-form formulation as
\begin{equation}
 \frac{\partial \boldsymbol{q}^{ijkp}_{mn}}                           {\partial t} =
-\widehat{\boldsymbol{e}}_\xi\left(\xi^{ip}_{m}\right)
-\widehat{\boldsymbol{f}}_\eta\left(\eta^{jp}_{n}\right)
-\widehat{\boldsymbol{h}}_\zeta\left(\zeta^{kp}\right)
+\boldsymbol{s}\left(\boldsymbol{q}^{ijkp}_{mn}\right),\label{updateDOF}
\end{equation}
where $\widehat{e_\xi}$, $\widehat{f_\eta}$ and $\widehat{h_\zeta}$ are numerical approximations of derivatives of flux functions in different directions at solution point.

\subsubsection{Spatial discretizations in horizontal directions}

The MCV scheme in multi-dimensional case can be implemented by applying the one-dimensional formulations in different directions one-by-one \citep{ii2009}. Thus, we describe the numerical procedure of spatial discretization in $\xi$-direction as follows. Similar formulations can be derived in $\eta$-direction and the details of multi-dimensional MCV discretization can be referred to \citep{ii2009}.

Considering the one dimensional governing equations in $\xi$-direction as
\begin{equation}
\left(\frac{\partial \boldsymbol{q}}{\partial t}\right)^\xi+\frac{\partial \boldsymbol{e\left(q\right)}}{\partial \xi}=0.
\end{equation}

\begin{figure}[ht]
\centering
\includegraphics[scale=0.7]{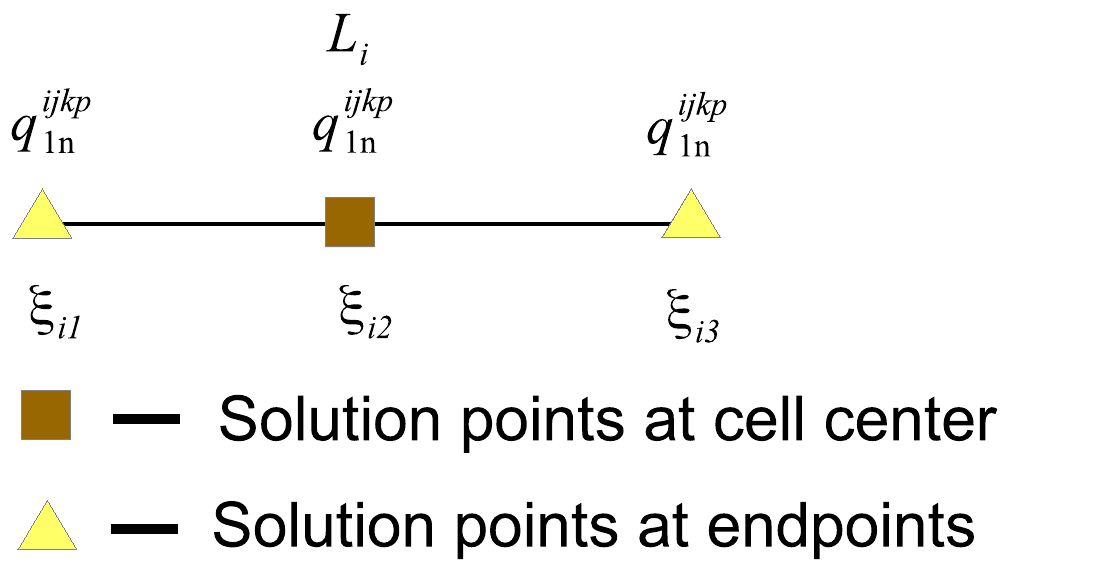}
\caption{Definition of DOFs in $xi$-direction (one-dimensional case).}\label{1dDOF}
\end{figure}

Three local DOFs are defined within line segment $\mathcal{L}^{ijkp}_{n}$ as shown in Fig. \ref{1dDOF} (one of 3 line segments along $\xi$-direction in Fig. \ref{DOF}), i.e., $\boldsymbol{q}^{ijkp}_{1n}$, $\boldsymbol{q}^{ijkp}_{3n}$ at cell interfaces (solid triangles) and $\boldsymbol{q}^{ijkp}_{2n}$ at cell center (solid square). Hereafter we use only the indices in $\xi$-direction for the sake of brevity. As shown in Eq. \eqref{updateDOF}, the semi-discrete formulation for each DOF is written as
\begin{equation}
 \left(\frac{\partial \boldsymbol{q}_{im}}{\partial t}\right)^\xi =-\widehat{\boldsymbol{e}}_\xi\left(\xi_{im}\right),\ \left(m=1,\ 3\right).\label{1DupdateDOF}
\end{equation}

\begin{figure}[ht]
\centering
\begin{subfigure}[Updating DOF defined at cell interfaces]
 {\includegraphics[scale=0.7]{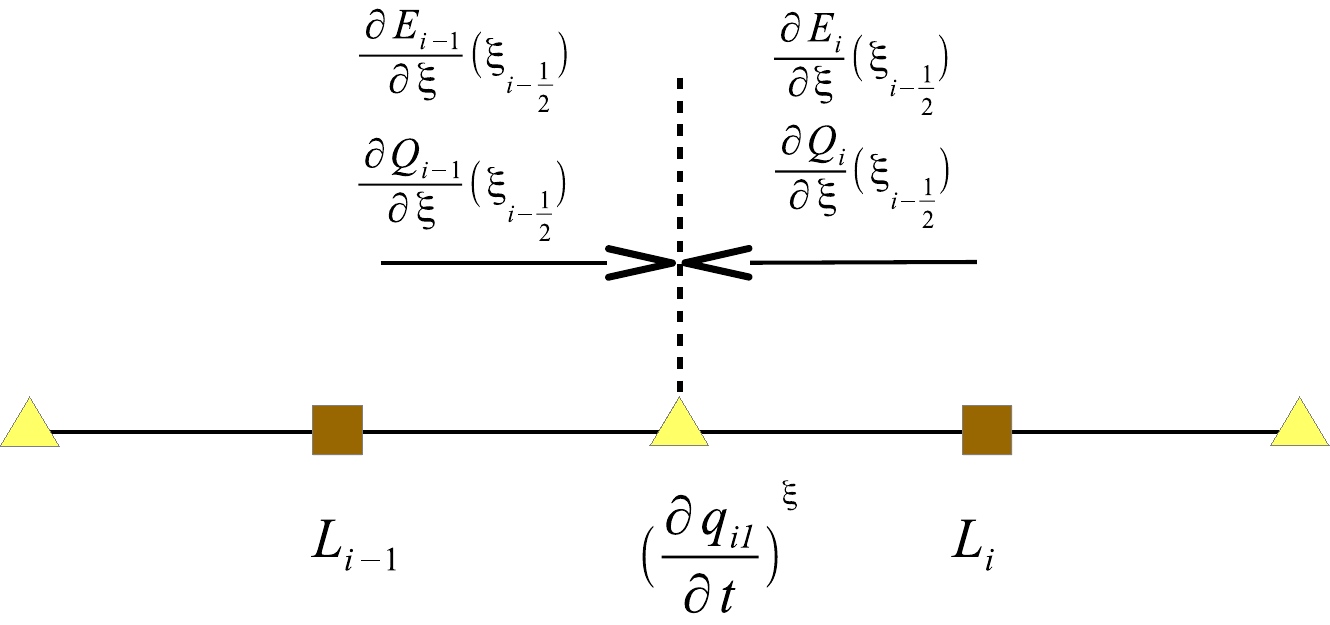}}
\end{subfigure}
\begin{subfigure}[Updating DOF defined at cell center]
 {\includegraphics[scale=0.7]{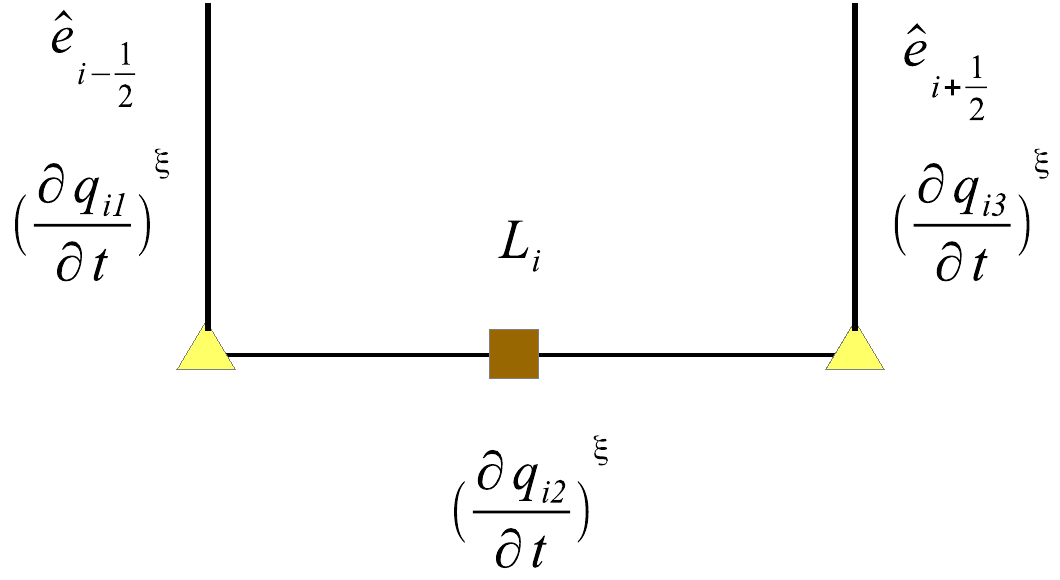}}
\end{subfigure}
\caption{Numerical scheme in one-dimensional case.}\label{1dFormulas}
\end{figure}

Different formulations are used to evaluated the derivatives of flux functions $\boldsymbol{e}$ at cell interfaces and center, as shown in Fig. \ref{1dFormulas} (a) and (b) respectively.

\begin{itemize}

\item Derivatives of flux functions $\boldsymbol{e}$ at cell interface ($\xi_{i1}$)

At cell interface, the derivatives of flux functions can be evaluated in two adjacent cells, i.e. $\mathcal{L}_{i-1}$ and $\mathcal{L}_{i}$ as shown in Fig. \ref{1dFormulas} (a). Generally, two different results are obtained. We then solve a derivative Riemann problem (DRP) to derive an upwind formulation as
\begin{equation}
\widehat{\boldsymbol{e}}_\xi\left(\xi_{i1}\right)=\frac{1}{2}\left[\frac{\partial  \boldsymbol{E}_{i-1}}{\partial \xi}\left(\xi_{i1}\right)+\frac{\partial  \boldsymbol{E}_{i}}{\partial \xi}\left(\xi_{i1}\right)\right]+\frac{1}{2}\boldsymbol{a}_\xi\left[\frac{\partial  \boldsymbol{Q}_{i-1}}{\partial \xi}\left(\xi_{i1}\right)-\frac{\partial  \boldsymbol{Q}_{i}}{\partial \xi}\left(\xi_{i1}\right)\right],
\end{equation}
where $\boldsymbol{Q}$ and $\boldsymbol{E}$ are piecewise spatial reconstruction of predicted variables $\boldsymbol{q}$ and flux functions $\boldsymbol{e}$, matrix $\boldsymbol{a}_\xi$ is determined by selected approximate Riemann solver in $\xi$-direction.

Using the multi-moment concept, several interpolation profiles \cite{chen2008,ii2009,chen2011,chen2015,BGS,WENO} for spatial reconstruction have been developed for the schemes with different numerical properties. Considering the trade-off between the accuracy and the efficiency, the fourth-order profile developed in \citep{chen2008} is adopted in this study. The spatial reconstruction for line segment $L_{i-1}$ is a Lagrangian interpolation polynomial using four pointwise values of predicted variables or flux functions at $\xi_{i-1,1}$, $\xi_{i-1,2}$, $\xi_{i-1,3}$ and $\xi_{i2}$ as constraint conditions. And the pointwise values at $\xi_{i-1,2}$, $\xi_{i1}$, $\xi_{i2}$ and $\xi_{i3}$ are adopted to build the spatial reconstruction within line segment $L_i$. The resulting multi-moment scheme is of fourth-order accuracy \cite{chen2008}.

Three approximate Riemann solvers are investigated in \cite{Paul2010} in solving atmospheric dynamics. Considering the significance influence from the effects of the Coriolis force and the gravity force in atmospheric dynamics, specially for those large-scale atmospheric flows,  the waves propagate in a different way in comparison with the Euler equations for gas dynamics. The adopted Riemann solver should be carefully considered to accurately reproduce the wave propagation in atmosphere. A modified local Lax-Friedrichs (LLF) approximate Riemann solver is used in this study for its simplicity. With the LLF solver, matrix $\boldsymbol{a}_\xi$ is simplified to be the maximal absolute value of eigenvalues of Jacobian matrix, i.e. $\frac{\partial \boldsymbol{e\left(\boldsymbol{q}\right)}}{\partial \boldsymbol{q}}$, which represents the maximal propagation speed related to the sound wave. In $\xi$-direction, it is written as
\begin{equation}
a_\xi=\left|\tilde{u}\right|+c_\xi,
\end{equation}
where the sound speed in the transformed coordinates is
\begin{equation}
c_\xi=\sqrt{G_H^{11}}c
\end{equation}
and the sound speed in physical space $c=\sqrt{\gamma\frac{p}{\rho}}$.

In this study, the LLF solver is then modified by adopted a much smaller parameter $a_\xi$, which is specified as
\begin{equation}
a_\xi=\left|\tilde{u}\right|+K_h c_\xi,
\end{equation}
where $K_h$ is a parameter to adjust the effective of numerical viscosity.

Since the physically-significant waves for large-scale atmospheric dynamics propagate much slower than the sound wave, this modification is expected to improve the accuracy of the proposed global model.

Analogously in $\eta$-direction, modified LLF solver is applied with $a_\eta=\left|\tilde{v}\right|+K_hc_\eta$, where the sound speed in transformed coordinates
are
\begin{equation}
c_\eta=\sqrt{G_H^{22}}c.
\end{equation}

In this study, parameter $K_h=0.2$ is adopted.

\item Derivatives of flux functions $\boldsymbol{e}$ at cell center ($\xi_{i2}$)

To guarantee the numerical conservation, the updating formulation of DOF at cell center is derived through the constraint condition based on the line-integrated average of the predicated variables, defined as
\begin{equation}
\overline{^{L_{\xi}}\boldsymbol{q}}_i=\frac{1}{\Delta \xi}\displaystyle\int_{\xi_{i1}}^{\xi_{i3}}\boldsymbol{q}\left(\xi\right)\mathrm{d}\xi,
\end{equation}

which can be approximated as
\begin{equation}
\overline{^{L_{\xi}}\boldsymbol{q}}_i=\frac{1}{6}\boldsymbol{q}_{i1}+\frac{2}{3}\boldsymbol{q}_{i2}+\frac{1}{6}\boldsymbol{q}_{i3}\label{1DVIA}
\end{equation}
using above spatial reconstruction polynomial.

Thus, the updating formulation for DOF $\boldsymbol{q}_{i2}$ at cell center can be written as
\begin{equation}
\left(\frac{\partial \boldsymbol{q}_{i2}}{\partial t}\right)^\xi=\frac{3}{2}\left(\frac{\partial\overline{^{L_\xi}\boldsymbol{q}}_i}{\partial t}\right)^\xi-\frac{1}{4}\left[\left(\frac{\partial \boldsymbol{q}_{i1}}{\partial t}\right)^\xi+\left(\frac{\partial \boldsymbol{q}_{i3}}{\partial t}\right)^\xi\right],
\end{equation}
where the updating formulations of DOFs at cell interfaces have been obtained above and the line-integrated average is updated using a flux-form formulation as
\begin{equation}
\left(\frac{\partial\overline{^{L_\xi}\boldsymbol{q}}_i}{\partial t}\right)^\xi=-\frac{1}{\Delta \xi}\left(\hat{\boldsymbol{e}}_{i+\frac{1}{2}}-\hat{\boldsymbol{e}}_{i-\frac{1}{2}}\right)
\end{equation}
with the flux functions at cell interfaces estimated by known DOFs defined there directly.

The resulting scheme is conservative in terms of line-integrated average calculated through Eq. \eqref{1DVIA}.

\end{itemize}

\subsubsection{Spatial discretizations in vertical direction}

In vertical direction, a conservative finite difference scheme is developed to solve the equations
\begin{equation}
\left(\frac{\partial \boldsymbol{q}}{\partial t}\right)^\zeta+\frac{\partial \boldsymbol{h\left(q\right)}}{\partial \zeta}=0.
\end{equation}

\begin{figure}[ht]
 \centering
 \includegraphics[scale=0.55]{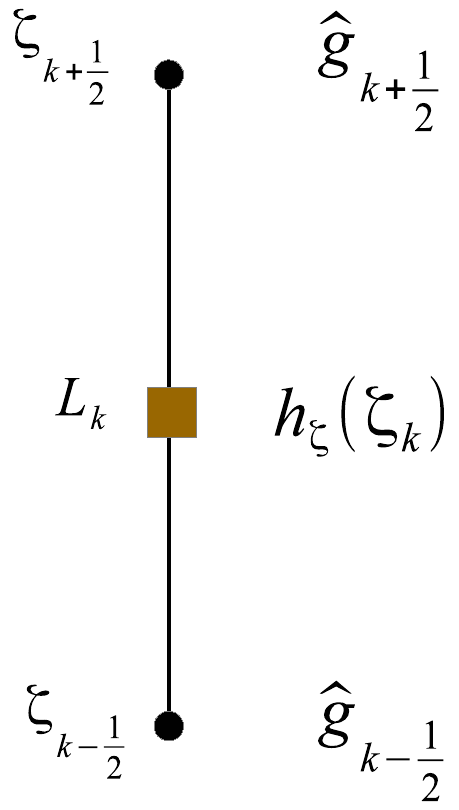}
 \caption{Definition of DOF in vertical direction.}\label{1dverFormulas}
\end{figure}

The key task here is again to  evaluate the derivatives of flux functions $\boldsymbol{h}$ at the center of line segment $L_k$ (shown in Fig. \ref{1dverFormulas}) as
\begin{equation}
\left(\frac{\partial \boldsymbol{q}_k}{\partial t}\right)^\zeta=-\widehat{\boldsymbol{h}}_\zeta\left(\zeta_k\right).\label{verticaldis}
\end{equation}
Here, we still omit the indices in horizontal directions for the sake of brevity.

To design a conservative scheme, we define auxiliary variables $\boldsymbol{g}\left(\zeta\right)$, which satisfy the relation
\begin{equation}
\boldsymbol{h}\left(\zeta\right)=\frac{1}{\Delta \zeta}\int_{\zeta-\frac{1}{2}\Delta \zeta}^{\zeta+\frac{1}{2}\Delta \zeta}\boldsymbol{g}\left(\zeta^\prime\right)\mathrm{d}\zeta^\prime.
\end{equation}

Then the derivatives of flux functions can be calculated by a flux-form formulation as
\begin{equation}
\boldsymbol{h}_\zeta\left(\zeta\right)=\frac{1}{\Delta \zeta}\left[\boldsymbol{g}\left(\zeta+\frac{1}{2}\Delta \zeta\right)-\boldsymbol{g}\left(\zeta-\frac{1}{2}\Delta \zeta\right)\right].\label{auxiliary}
\end{equation}

Considering the relation Eq. \eqref{auxiliary}, updating formulation Eq. \eqref{verticaldis} is recast as
\begin{equation}
\left(\frac{\partial \boldsymbol{q}_k}{\partial t}\right)^\zeta=-\frac{1}{\Delta \zeta}\left(\widehat{\boldsymbol{g}}_{k+\frac{1}{2}}-\widehat{\boldsymbol{g}}_{k-\frac{1}{2}}\right).\label{verticaldis2}
\end{equation}

The updating formulation Eq. \eqref{verticaldis2} is of flux-form and the resulting model is numerically conservative.

At endpoint $\zeta_{k+\frac{1}{2}}$, the values of auxiliary variables $\boldsymbol{g}$ are determined by solving Riemann problem as
\begin{equation}
\widehat{\boldsymbol{g}}_{k+\frac{1}{2}}=\frac{1}{2}\left[\boldsymbol{G}_k\left(\zeta_{k+\frac{1}{2}}\right)+\boldsymbol{G}_{k+1}\left(\zeta_{k+\frac{1}{2}}\right)\right]+\frac{1}{2}\boldsymbol{a}_\zeta\left[\boldsymbol{Q}_{k}\left(\zeta_{k+\frac{1}{2}}\right)-\boldsymbol{Q}_{k+1}\left(\zeta_{k+\frac{1}{2}}\right)\right],
\end{equation}
where $\boldsymbol{G}$ and $\boldsymbol{Q}$ are one-dimensional piecewise polynomial for auxiliary variables $\boldsymbol{g}$ and predicted variables $\boldsymbol{q}$, the similar modified LLF Remann solver is applied with the parameter
\begin{equation}
\boldsymbol{a}_\zeta=\left|\tilde{w}\right|+\boldsymbol{K}_vc_\zeta\ \mathrm{with}\ c_\zeta=\sqrt{\left(J_V^{-2}+M^1+M^2\right)}c
\end{equation}
and
$\boldsymbol{K}_v=\mathrm{diag}\left[k^S_{v},k^S_{v},k^S_{v},k^B_{v},k^S_{v}\right]^T$ in this study.

A three-point stencil is used for spatial reconstruction in vertical direction. Two polynomials for spatial reconstruction can be obtained, including
\begin{itemize}

\item a quadratic polynomial $\left(2\leq k\leq N_{v-1}\right)$ as
\begin{equation}
\boldsymbol{G}_k\left(\zeta\right)=c_0+c_1\left(\zeta-\zeta_k\right)+c_2\left(\zeta-\zeta_k\right)^2,
\end{equation}
where the coefficients are determined by following constraint conditions
\begin{equation}
\left\{\begin{array}{l}
\int_{k-\frac{3}{2}}^{k-\frac{1}{2}}\boldsymbol{G}_k\left(\zeta\right)d\zeta=\boldsymbol{h}_{k-1}\\
\int_{k-\frac{1}{2}}^{k+\frac{1}{2}}\boldsymbol{G}_k\left(\zeta\right)d\zeta=\boldsymbol{h}_{k}\\
\int_{k+\frac{1}{2}}^{k+\frac{3}{2}}\boldsymbol{G}_k\left(\zeta\right)d\zeta=\boldsymbol{h}_{k+1}
\end{array}\right.,
\end{equation}

\item and a linear polynomial as
\begin{equation}
\boldsymbol{G}_k\left(\zeta\right)=\boldsymbol{h}_k+\boldsymbol{d}_k\left(\zeta-\zeta_k\right),\label{linearrecons}
\end{equation}
where
\begin{equation}
d_k=\left\{\begin{array}{ll}
\frac{1}{2\Delta\zeta}\left(\boldsymbol{h}_{k+1}-\boldsymbol{h}_{k-1}\right),&\mathrm{if\ } k\neq 1\mathrm{\ and\ }k\neq N_v\\
\frac{1}{\Delta\zeta}\left(\boldsymbol{h}_{k+1}-\boldsymbol{h}_{k}\right),&\mathrm{if\ } k= 1\\
\frac{1}{\Delta\zeta}\left(\boldsymbol{h}_{k}-\boldsymbol{h}_{k-1}\right),&\mathrm{if\ } k= N_v
\end{array}\right..\label{linearrecons1}
\end{equation}

\end{itemize}

In this study, the linear polynomial is adopted to calculate the benchmark tests since no notable improvement on the computational accuracy was observed by using the quadratic polynomial, meanwhile the higher order polynomial sometimes introduces the non-physical numerical oscillations. The parameters used to modify the LLF Riemann solver in vertical direction are selected as $K^S_v=2 \times 10^{-5}$, $K^B_v=0.1$ in the nonhydrostatic case, $K^B_v=4$ in the Held-Saurez long-term integration test and $K^B_v=1$ in other hydrostatic cases.

As the leading term of numerical diffusion term of above upwind finite difference scheme using linear reconstruction is proportional to the $4^{th}$-order derivative of dependent variable, enlarging the coefficient $K^B_v$ is equivalent to adding the $4^{th}$-order vertical diffusion in vertical momentum equation (w-equation). Thus, small value of $K_v^B$ is used in nonhydrostatic case to improve the accuracy, while the relatively large values are chosen for hydrostatic cases for robustness of the proposed model. In numerical experiments, we found the additional numerical diffusion is helpful in the proposed model to suppress the numerical oscillations and stabilize the proposed model in some cases, including the baroclinic wave test and Held-Saurez test.

\subsubsection{Boundary condition}

In horizontal direction, one layer of ghost cells are supplemented for each patch. With enough ghost cells, the updating procedure is applied on each patch independently. The DOFs within ghost cells are evaluated by a single-cell based polynomial over the cell in adjacent patch. Furthermore, some DOFs, which are defined along the patch boundaries, are updated in two or three patches and the different results may obtained during the simulation. A correction operation is applied by averaging the results from different patches. The construction of ghost cells in horizontal direction and the implementation of result correction along the patch boundaries can be accomplished for a three-dimensional model by applying the numerical manipulation we have developed for the global shallow water model \cite{chen2008} at each model layer.

In vertical direction, the one-sided formulations are applied at surface and model top for spatial reconstruction in $\zeta$-direction (Eqs. \eqref{linearrecons} and \eqref{linearrecons1}). Additionally, the slip-wall condition are applied in vertical direction, i.e. $\tilde{w}=0$ at surface and model top. Rayleigh friction is adopted in momentum equations near model top to assure the non-reflective boundary at model top and the strength of Rayleigh friction is given as \cite{Durran1983}
\begin{equation}
\tau_R=
\left\{\begin{array}{ll}
0&\mathrm{if\ }z<z_D\\
\frac{\tau_0}{2}\left[1-\cos\left(\frac{z-z_D}{z_t-z_D}\pi\right)\right] & \mathrm{if\ }z_D\leq z\leq \frac{z_D+z_t}{2}\\
\frac{\tau_0}{2}\left[1+\sin\left(\frac{z-z_D}{z_t-z_D}\pi-\frac{\pi}{2}\right)\right] & \mathrm{otherwise}
\end{array}\right..
\end{equation}

\subsection{Time marching scheme}

Due to the very large ratio between the horizontal and the vertical grid spacings, the very small time step of an explicit scheme will be determined by the sound speed and the smallest grid spacing in vertical direction, e.g. it has a magnitude less than one second in the practical applications with the vertical grid spacing of dozens of meters near the surface. In this study, we use the dimensional-splitting scheme based on horizontally explicit and vertically implicit (HEVI) strategy to implement an efficient time marching scheme. The terms related to the spatial discretization in vertical direction and the stiff source terms including gravity force and Rayleigh friction are implicitly integrated. To preserve the high-order accuracy, the implicit-explicit (IMEX) Runge-Kutta scheme is adopted to couple the explicit and implicit time marching. The time step of resulting scheme is decided by the stability condition in horizontal direction. With adopted 3-point $4^{th}$-order MCV scheme and $3^{rd}$-order Runge-Kutta scheme, the maximal CFL number is about 0.45 in two dimensions, which is calculated by
\begin{equation}
\mathrm{CFL}_{\max}=\frac{2N_h \Delta t}{\pi R}\max\left(\tilde{u}_{\max},\tilde{v}_{\max}\right),
\end{equation}
where $\tilde{u}_{\max}$ and $\tilde{v}_{\max}$ are maximal contravariant velocity components within the computational domain.

The time marching in the proposed model is accomplished from time step $n_t$ ($t=n_t\Delta t$) to $n_t+1$ as
\begin{equation}
\boldsymbol{q}^{n_t+1}=\boldsymbol{q}^{n_t}+\Delta t\sum_{s=0}^{S}\left[b_s\mathcal{H}\left(\boldsymbol{q}^{\left(s\right)}\right)+\tilde{b}_s\mathcal{V}\left(\boldsymbol{q}^{\left(s\right)}\right)\right],\label{imex}
\end{equation}
where
\begin{equation}
\boldsymbol{q}^{\left(s\right)}=\boldsymbol{q}^{n_t}+\Delta t\sum_{r=0}^{s-1}\left[a_{sr}\mathcal{H}\left(\boldsymbol{q}^{\left(r\right)}\right)\right]+\Delta t\sum_{r=0}^{s}\left[\tilde{a}_{sr}\mathcal{V}\left(\boldsymbol{q}^{\left(r\right)}\right)\right],
\end{equation}
and $\mathcal{H}$ and $\mathcal{V}$ denote the explicit and implicit parts of MCV discretization.

At the $s^{th}$ substep, a nonlinear equation set, having the form of
\begin{equation}\label{nonlineareqset}
\boldsymbol{y}\left(\boldsymbol{x}\right)=-\frac{1}{\Delta t}\boldsymbol{x}+\boldsymbol{B}+\tilde{a}_{rr}\mathcal{V}\left(\boldsymbol{x}\right)=0
\end{equation}
is solved to determine $\boldsymbol{q}^{\left(s\right)}$ by Newton's method, where $\boldsymbol{B}$ includes the known quantities at $s^{th}$ substep
\begin{equation}
\boldsymbol{B}=\frac{1}{\Delta t}\boldsymbol{q}^{n_t}+\sum_{r=0}^{s-1}\left[a_{sr}\mathcal{H}\left(\boldsymbol{q}^{\left(s\right)}\right)+\tilde{a}_{sr}\mathcal{V}\left(\boldsymbol{q}^{\left(s\right)}\right)\right].
\end{equation}

The solution is approximately determined through the iteration as
\begin{equation}\label{lineareqset}
\left(\frac{1}{\Delta t}\boldsymbol{I}-\tilde{a}_{ss}\frac{\partial\mathcal{V}}{\partial \boldsymbol{x}}\left(\boldsymbol{x}_{iter}\right)\right)\left(\boldsymbol{x}_{iter+1}-\boldsymbol{x}_{iter}\right)=\boldsymbol{y}\left(\boldsymbol{x}_{iter}\right).
\end{equation}

The initial guess is chosen as $\boldsymbol{x}_0=\boldsymbol{q}^{n_t }$ and the linear system Eq. \eqref{lineareqset} is solved using a Gaussian elimination algorithm designed for a sparse system corresponding to the finite difference vertical discretization. Jacobian matrix of the linear system is determined by analytically calculating the derivatives of spatial discretization formulations of the implicit-part with respect to the dependent variables.

The application of various IMEX Runge-Kutta scheme in the global atmospheric modelling to accomplish HEVI time marching was recently investigated in \cite{weller2013,Gardner2018}. In this study, a 3-stage, $3^{rd}$-order, L-stable DIRK scheme ($S=3$ in Eq. \eqref{imex}) introduced in \cite{ascher1997} is adopted. In this study, the Newton iteration is only conducted for one time in every Runge-Kutta substep, i.e. the non-linear system derived from implicit time marching is linearized. The numerical experiments show this simplification hardly alter the result and obviously save the computational overheads.

%%%%%%%%%%%%%%%%%% TESTS AND RESULTS

\section{Tests and results}

In this section, the widely used benchmark test cases were carried out to verify the proposed dynamical core. These test were described in detail in \cite{jablonowski2008,DCMIP2012,HS94}, including both hydrostatic and non-hydrostatic ones. All tests, except the Held-Saurez's long-term integration test, were conducted with horizontal resolution of $1^\circ$ (along the Equator) in terms of DOF ($N_h=45$), while Held-Saurez test adopted a little coarser grid with horizontal resolution of $1.5^\circ$ ($N_h=30$)). In vertical direction, we constructed the computational grids using the parameters shown in Table \ref{verticalgridTable}. In 3D Rossby-Haurwitz wave, gravity wave and nonhydrostatic mountain wave cases, the uniform grids were adopted. In other cases, we used the non-uniform vertical grids. The quality of numerical results of dynamical core is related to selected vertical grid to some extend. For the practical applications with physical processes, it is worth further investigations on designing the proper grid transformation in vertical direction to represent the behaviors of real atmosphere. The proposed dynamical core is applicable for various grid transformation formulations. In numerical experiments, the results are often displayed on the isobaric surfaces using the longitude-latitude grid in horizontal directions. The linear polynomial is used to calculate the geopotential height of the isobaric surface and interpolates other predicted variables from the height-based vertical coordinate to the prescribed isobaric surfaces. Similarly, bilinear interpolation is applied in horizontal directions to evaluate the predicted variables on longitude-latitude grid. Though the spatial interpolation based on the linear polynomial may degrade the numerical accuracy in the post-processing calculations, it is adopted in this study for it does not generate the new extrema in comparison with other high-order interpolations. The time step is $\Delta t=200$ s on grid $N_h=45$ and scaled on other grids to maintain the same value of $N_h\Delta t$.

\subsection{3D Rossby-Haurwitz wave}

This test case is an three-dimensional extension of Rossby-Haurwitz wave test proposed for global SWE model in \cite{Williamson1992}. The horizontal velocity components are identical on each layer in vertical direction, which have the same form as those defined in \cite{Williamson1992} and the details are described in \cite{jablonowski2008}. The vertical velocity component is zero.

The initial condition preserves hydrostatic relation and the thermodynamic variables are derived from temperature profile as
\begin{equation}
T=T_0-\Gamma \tilde{z},
\end{equation}
where $\tilde{z}$ is equivalent height, $T_0=288$ K, $\Gamma=0.0065$ K/m.

By integrating the hydrostatic relation, we have
\begin{equation}
p=p_{\mathrm{ref}}\left(1-\frac{\Gamma \tilde{z}}{T_0}\right)^{\frac{g}{\Gamma R_d}},
\end{equation}
where $p_{\mathrm{ref}}=955$ hPa is the pressure at $\tilde{z}=0$.

Equivalent height are related with altitude by
\begin{equation}
\tilde{z}=z-\frac{\Phi^\prime\left(\lambda,\phi\right)}{g},
\end{equation}
where the perturbation of geopotential can be referred to \cite{jablonowski2008}.

\begin{figure}[h]
\centering
\begin{subfigure}[850hPa zonal wind]
  { \includegraphics[width=0.48\textwidth]{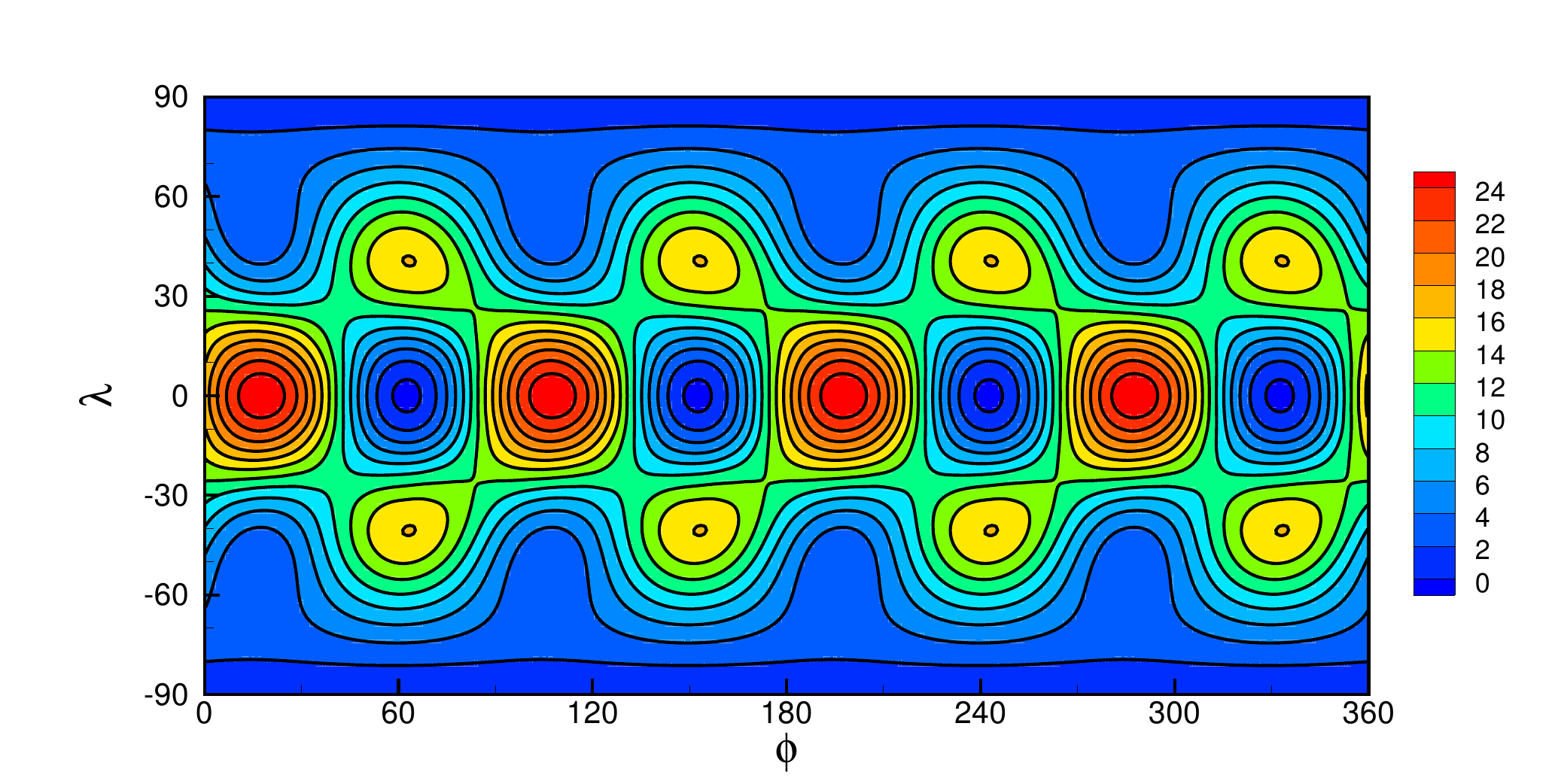}}
\end{subfigure}
\begin{subfigure}[850hPa meridional wind]
   { \includegraphics[width=0.48\textwidth]{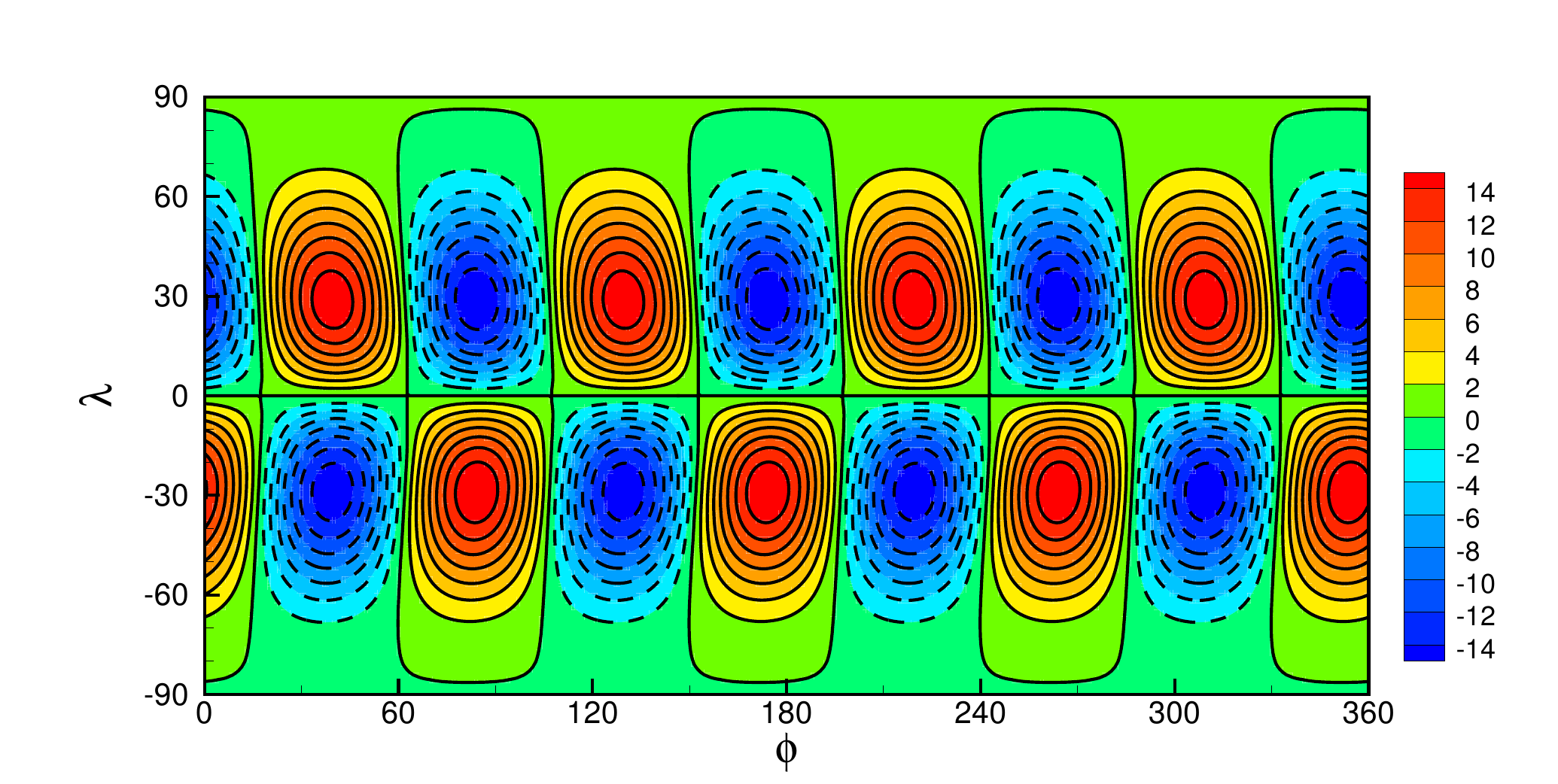}}
\end{subfigure}
\begin{subfigure}[Surface pressure]
  { \includegraphics[width=0.48\textwidth]{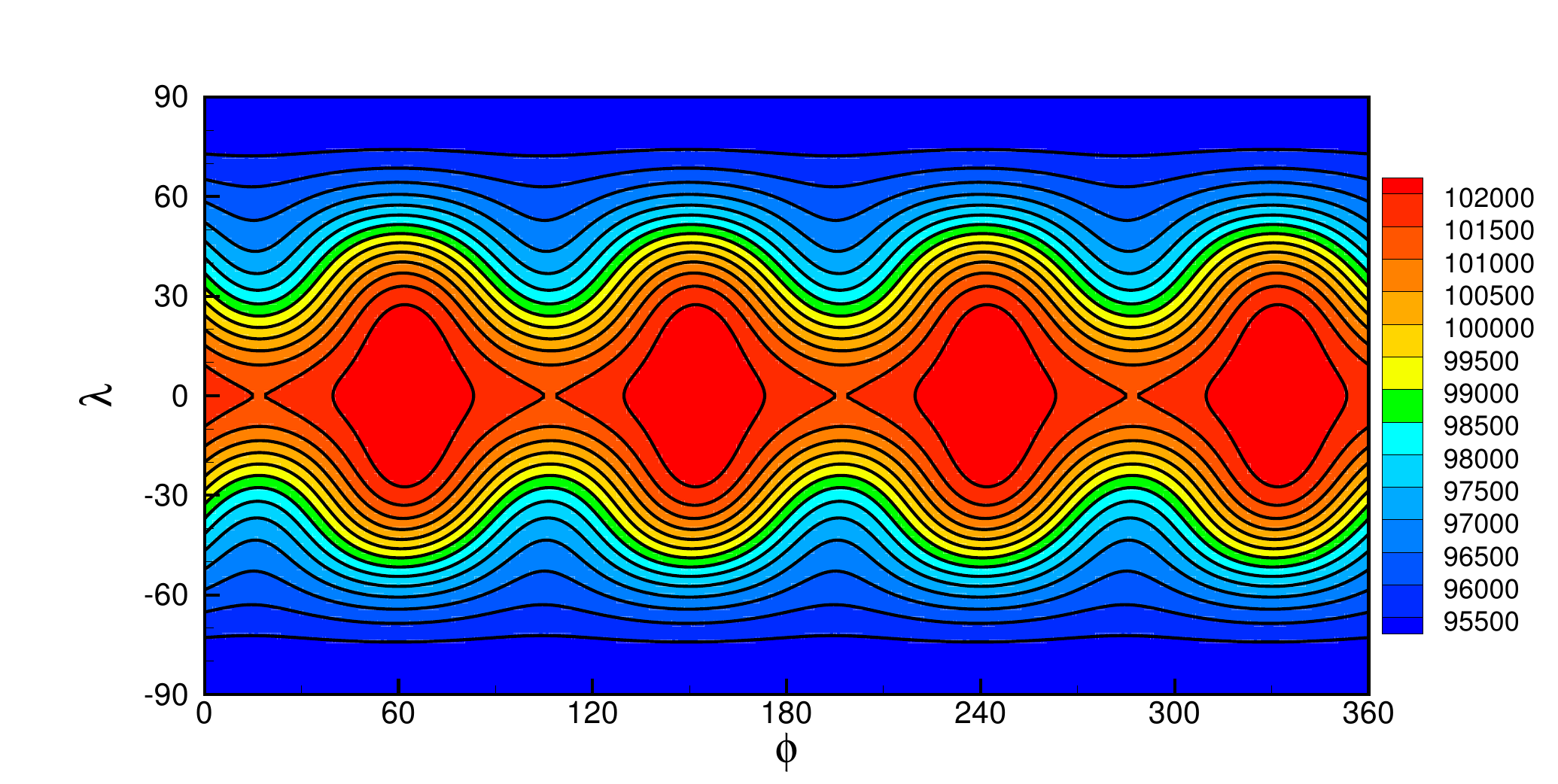}}
\end{subfigure}
\begin{subfigure}[500hPa geopotential height]
  { \includegraphics[width=0.48\textwidth]{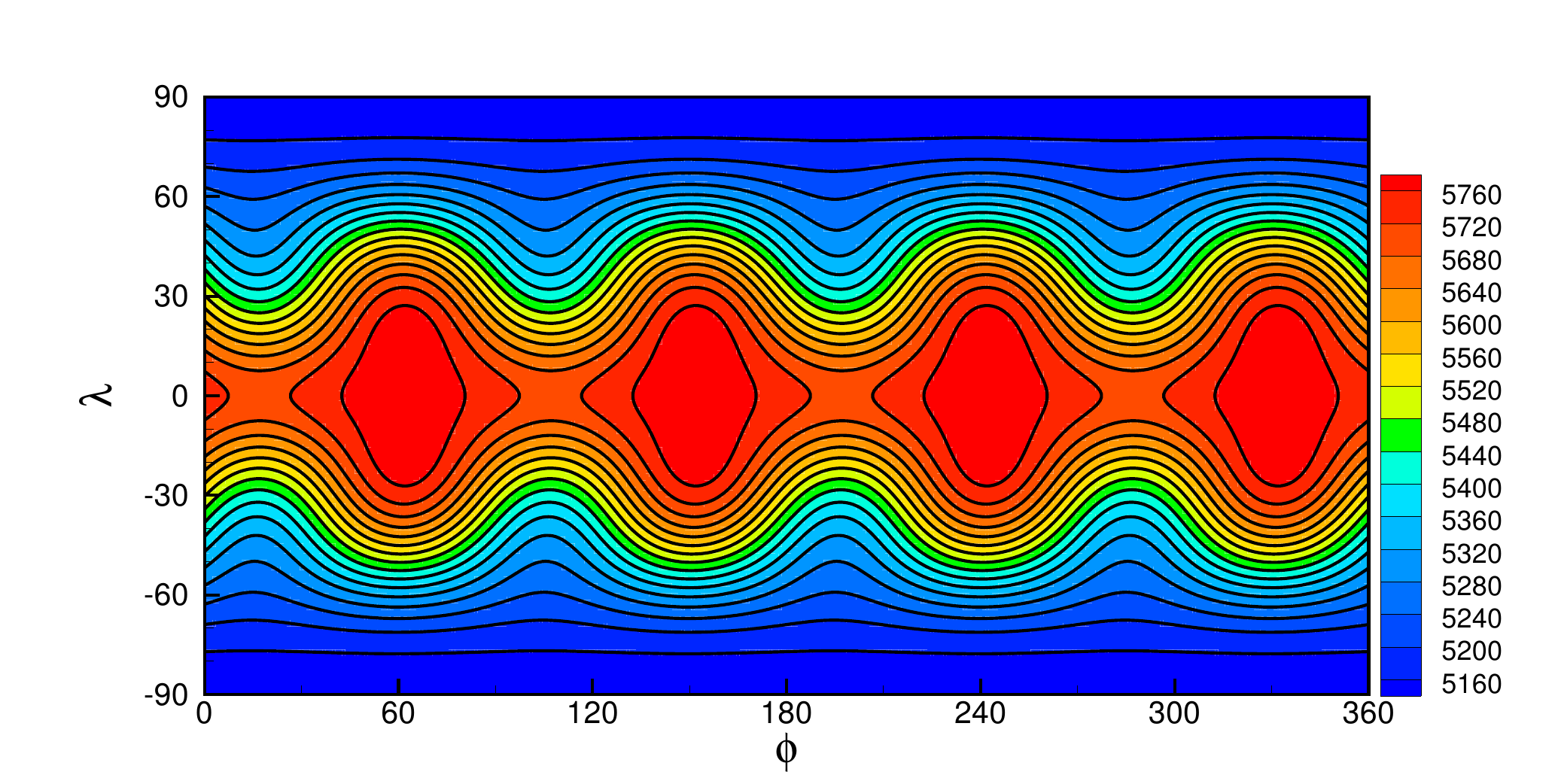}}
\end{subfigure}
\caption{Contour plots of numerical results of 3D Rossby-Haurwitz wave at day 15. Shown are 850 hPa zonal wind (panel (a)), meridional wind (panel (b)), Surface pressure (panel (c)) and 500hPa geopotential height (panel (d)). The dashed lines denote the negative values.}\label{RossbyWave}
\end{figure}

\begin{figure}[h]
\centering
\includegraphics[width=0.7\textwidth]{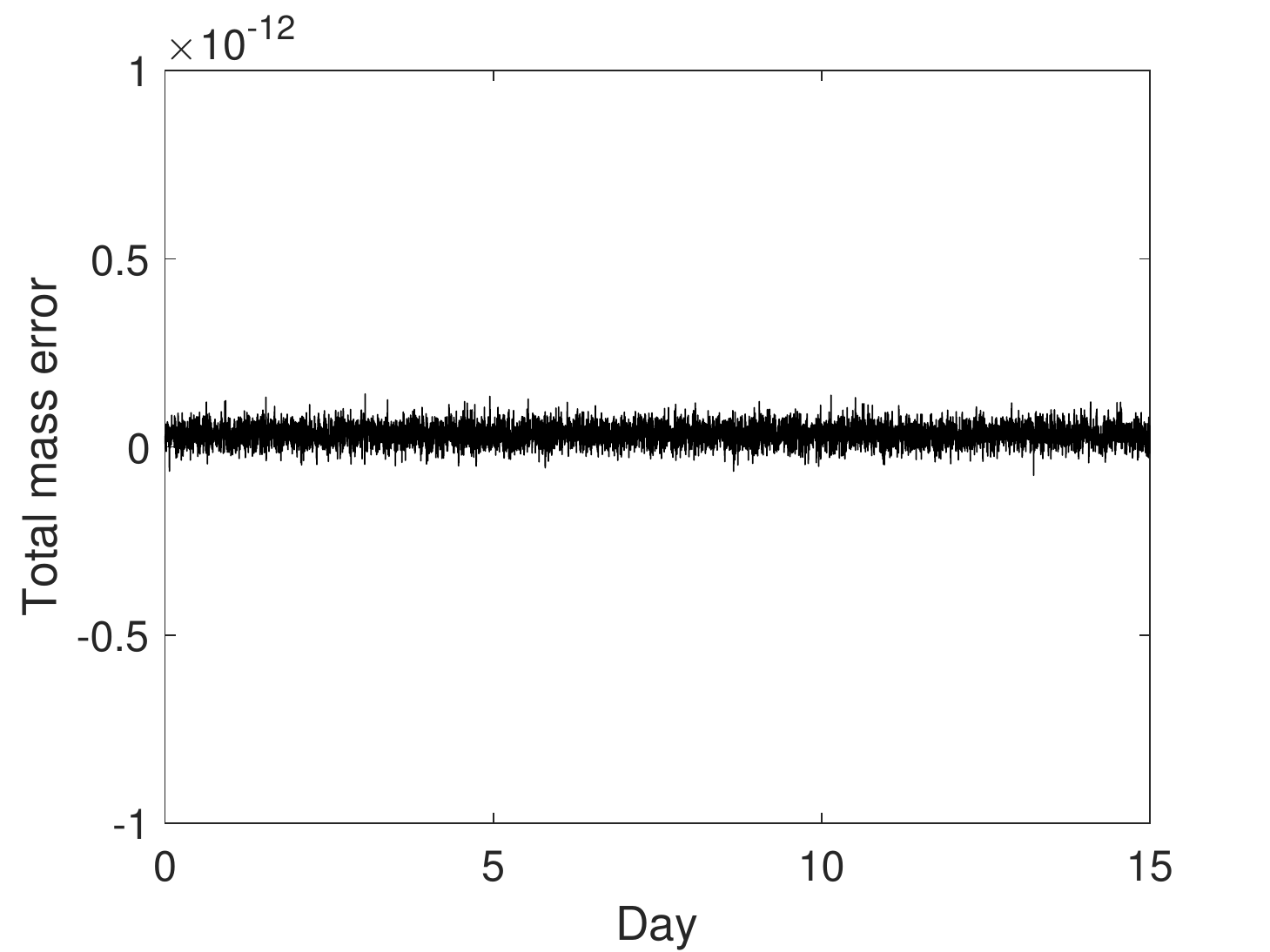}
\caption{Relative total mass error of 3D Rossby-Haurwitz wave during 15 days. }\label{RossbyWave2}
\end{figure}

\begin{figure}[h]
\centering
\begin{subfigure}[850hPa zonal wind]
  { \includegraphics[width=0.48\textwidth]{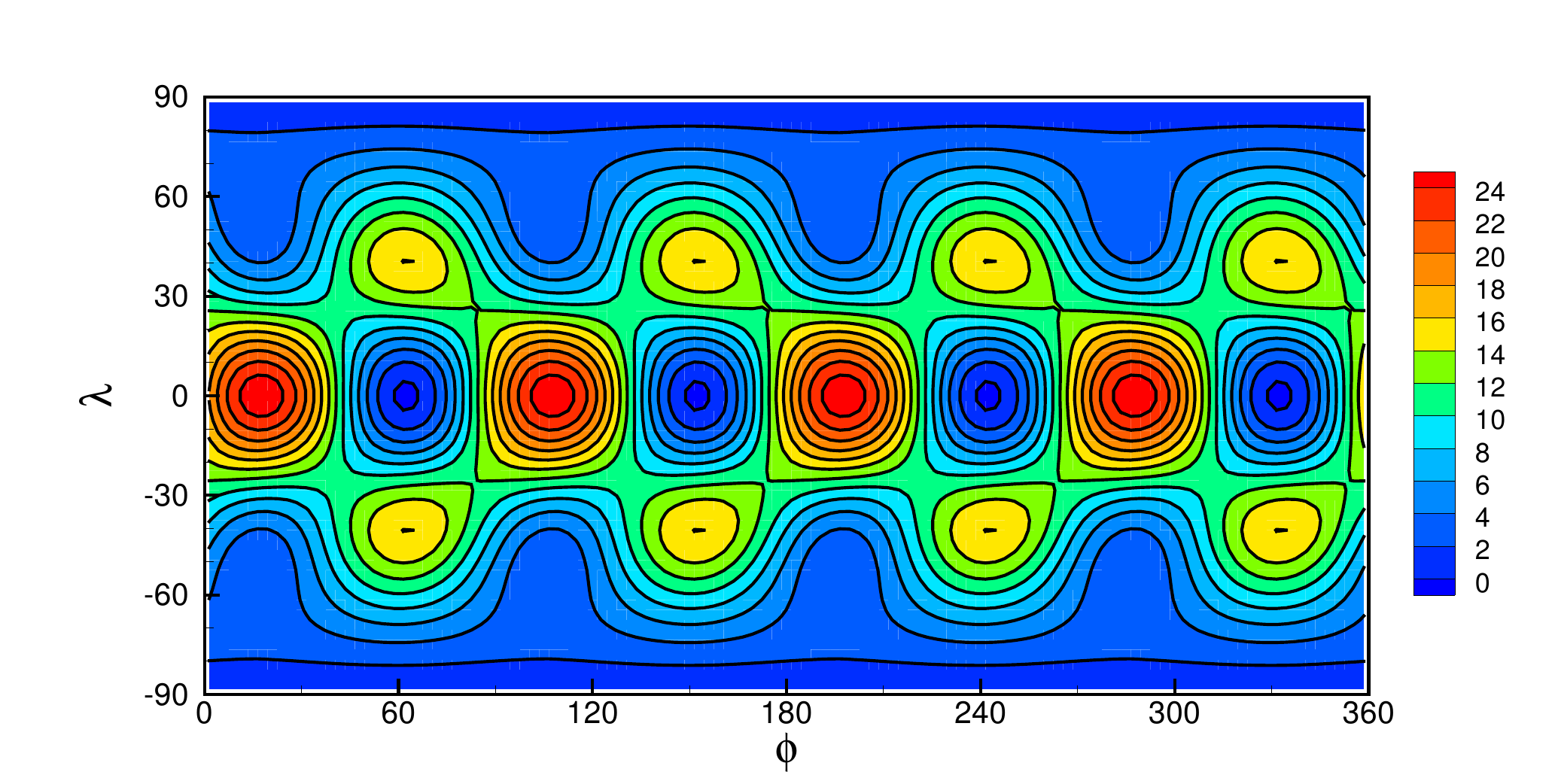}}
\end{subfigure}
\begin{subfigure}[850hPa meridional wind]
   { \includegraphics[width=0.48\textwidth]{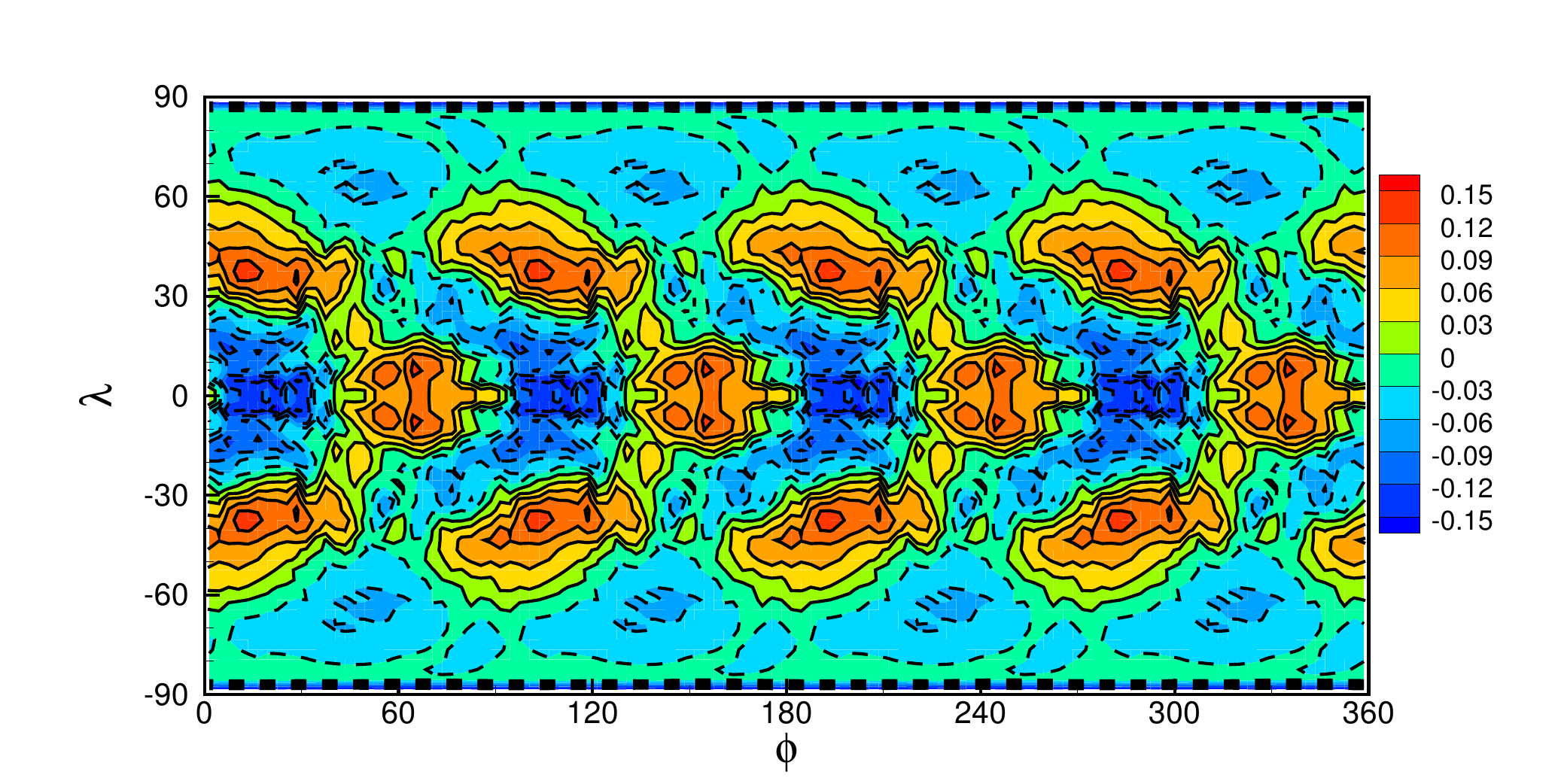}}
\end{subfigure}
\begin{subfigure}[500hPa geopotential height]
  { \includegraphics[width=0.48\textwidth]{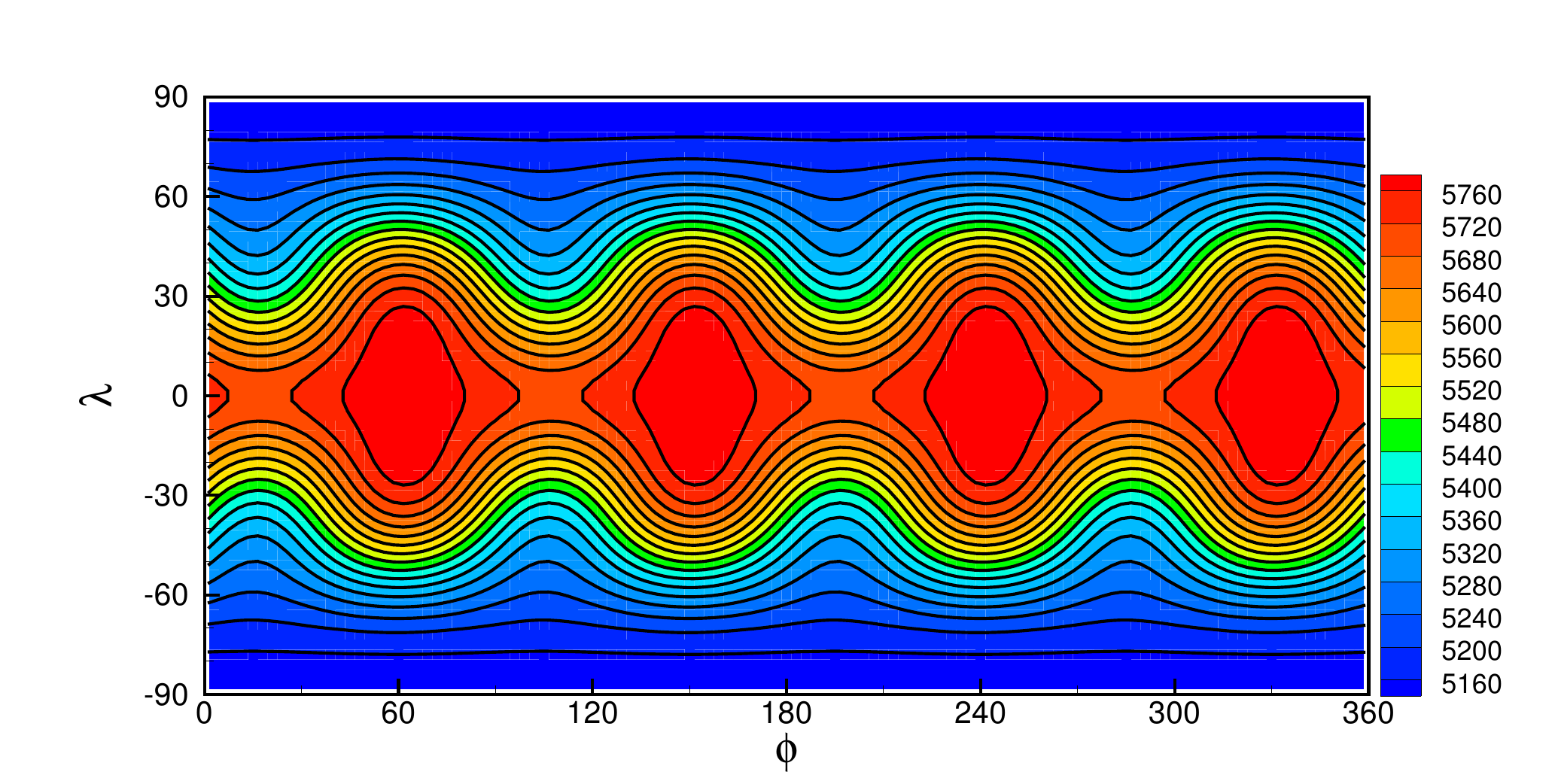}}
\end{subfigure}
\begin{subfigure}[500hPa geopotential height]
  { \includegraphics[width=0.48\textwidth]{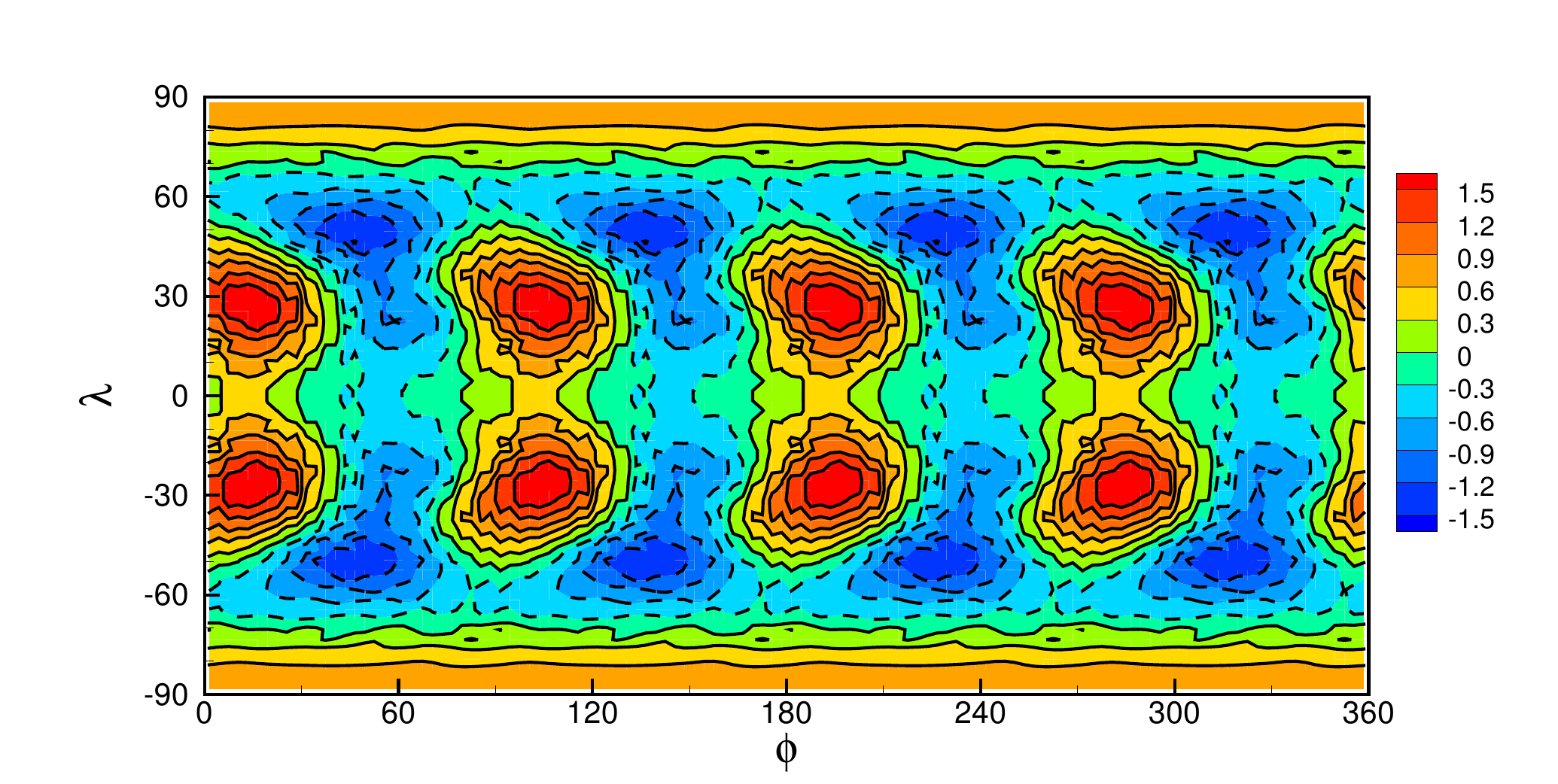}}
\end{subfigure}
\caption{Contour plots of numerical results of 3D Rossby-Haurwitz wave on a coarse grid ($N_h=15$) at day 15. Shown are 850 hPa zonal wind (panel (a)), 500hPa geopotential height (panel (c)) and their absolute differences in comparison with the solutions on grid $N_h=45$ (panels (b) and (d)). The dashed lines denote the negative values.}\label{RossbyWave3}
\end{figure}

The numerical results of Rossby-Haurwitz wave are given in Fig. \ref{RossbyWave} for the test on grid $N_h=45$. Shown are horizontal velocity components at 850 hPa level, geopotential height at 500 hPa level and surface pressure at day 15. The surface pressure is not a predicted quantity and extrapolated using a linear polynomial based on the pressure of the first and second model layers. Both shape and phase shift of the shown quantities agree well with the results in existing literatures. The proposed model is conservative and the relative total mass error has a value of machine precision as shown in Fig. \ref{RossbyWave2}. This test is also checked on a coarser grid with $N_h=15$ and the numerical results are depicted in Fig. \ref{RossbyWave3}. As the high-order MCV scheme is adopted in horizontal directions in this model, the considerably large-scale wave propagation in this test is accurately reproduced on this very coarse grid and the differences in comparison with those on grid $N_h=45$ are less than 1\%.

\subsection{Gravity wave without Earth's rotation}

The static atmosphere is given by specifying a horizontally uniform pressure field as
\begin{equation}
p\left(z\right)=p_0\left[\left(1-\frac{S}{T_0}\right)+\frac{S}{T_0}\exp\left(-\frac{N^2z}{g}\right)\right]^{\frac{1}{\kappa}},
\end{equation}
where Brunt-V$\ddot{\mathrm{a}}$is$\ddot{\mathrm{a}}$l$\ddot{\mathrm{a}}$ frequency $N=0.01$ s$^{-1}$, $p_0=1000$ hPa, $T_0=300$ K and $S=\frac{g^2}{c_pN^2}$.

The background potential temperature is obtained from hydrostatic relation as
\begin{equation}
\overline{\theta}\left(z\right)=T_0\exp\left(\frac{N^2 z}{g}\right).
\end{equation}

A perturbation of potential temperature is then added in the steady background field to trigger the hydrostatic gravity wave as
\begin{equation}
\theta^\prime\left(\lambda,\phi,z\right)=\Delta \theta s\left(\lambda,\phi\right)\sin\left(\frac{2\pi z}{L_z}\right),
\end{equation}
where function $s\left(\lambda,\phi\right)$ defines a cosine bell as
\begin{equation}
s\left(\lambda,\phi\right)=
\left\{
\begin{array}{ll}
\frac{1}{2}\left[1+\cos\left(\frac{\pi r}{R}\right)\right]&\mathrm{if\ } r < r_0\\
0&\mathrm{otherwise}
\end{array}
\right.,
\end{equation}
$r$ is great-circle distance to bell center $\left(\pi,0\right)$, $r_0=\frac{R}{3}$ and vertical wave length $L_z$=20 km.

\begin{figure}[h]
 \centering
 \begin{subfigure}[Hour 6]
  { \includegraphics[width=0.48\textwidth]{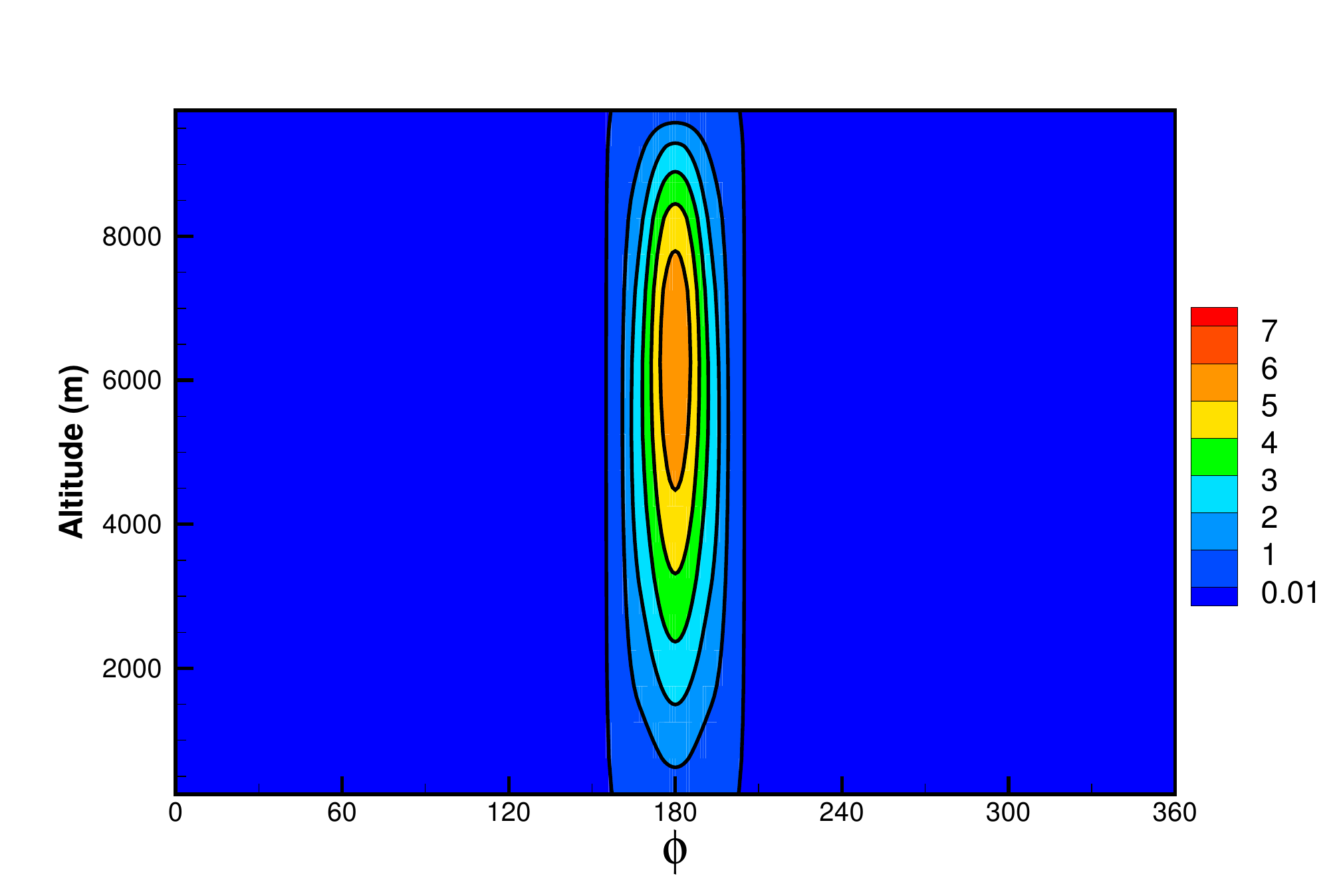}}
  \end{subfigure}
   \begin{subfigure}[Hour 12]
   { \includegraphics[width=0.48\textwidth]{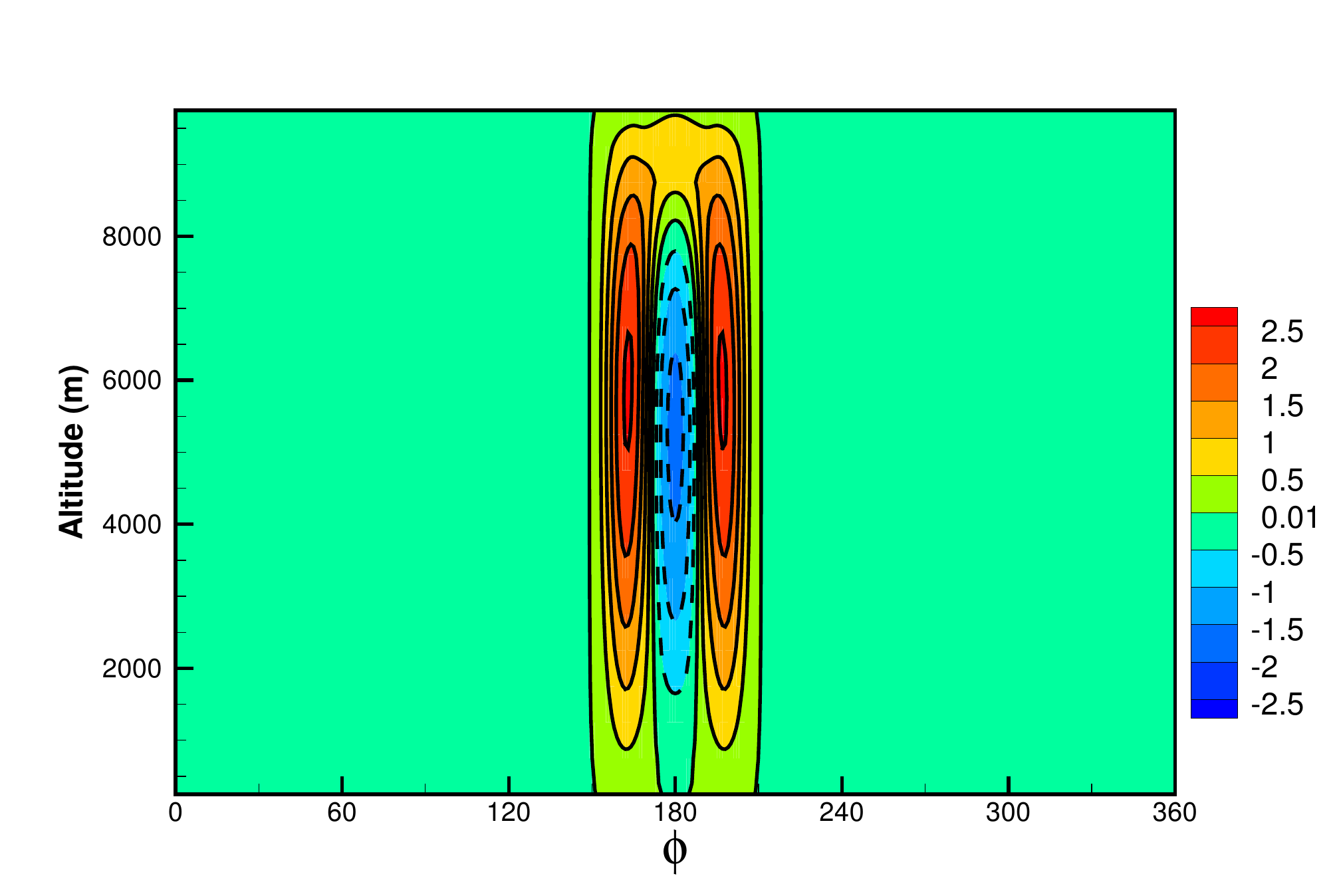}}
   \end{subfigure}
\begin{subfigure}[Hour 24]
  { \includegraphics[width=0.48\textwidth]{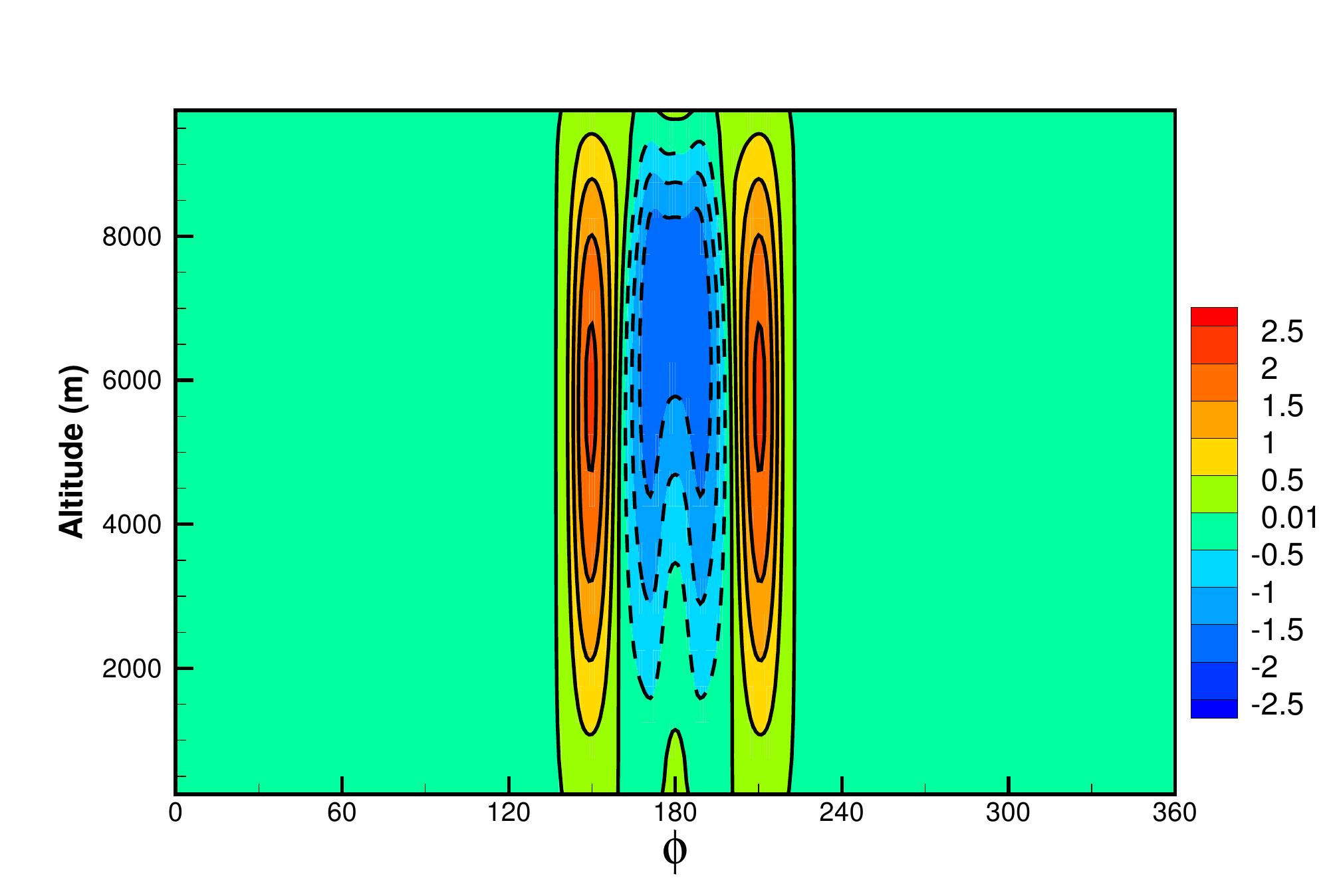}}
  \end{subfigure}
   \begin{subfigure}[Hour 48]
   { \includegraphics[width=0.48\textwidth]{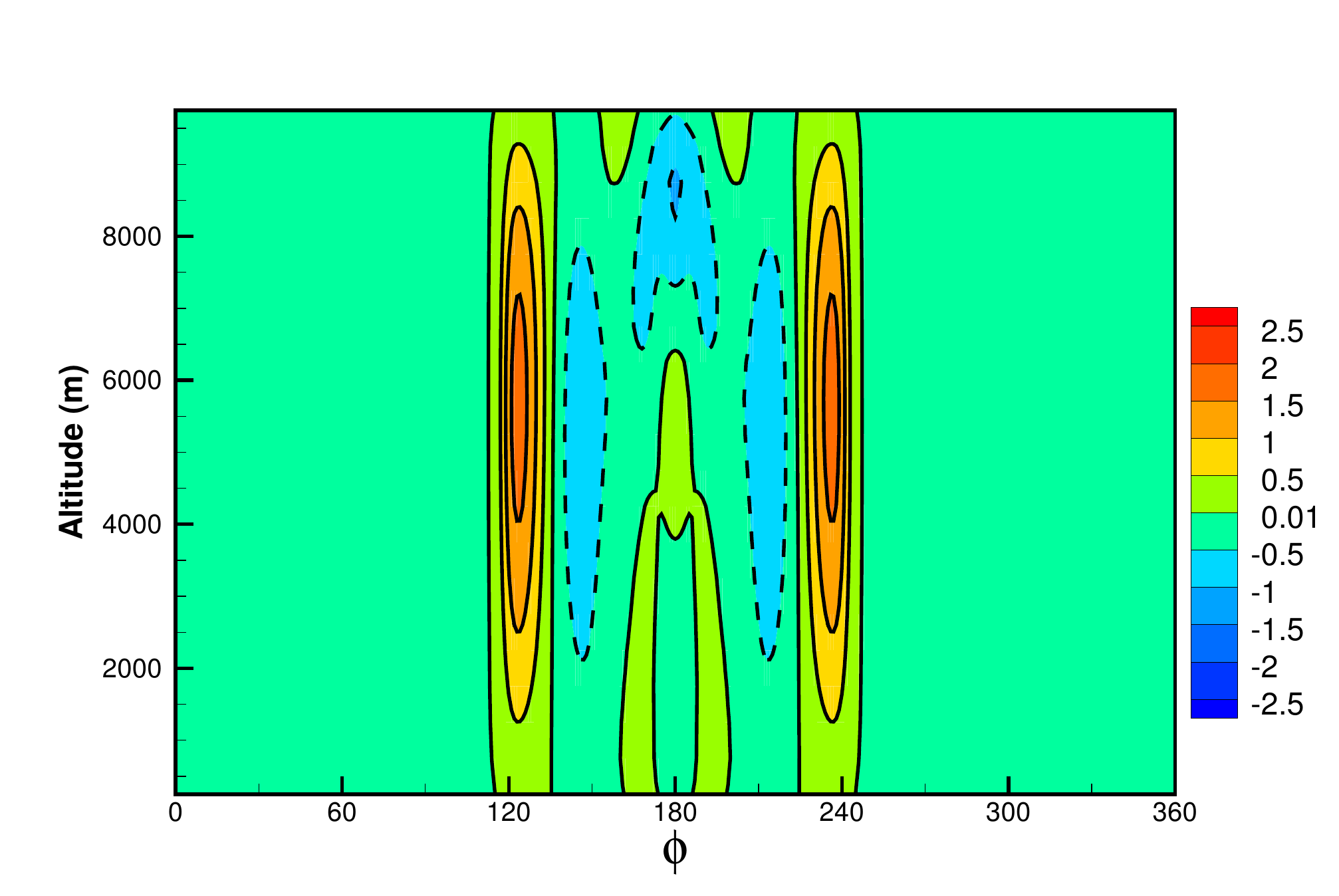}}
   \end{subfigure}
   \begin{subfigure}[Hour 72]
  { \includegraphics[width=0.48\textwidth]{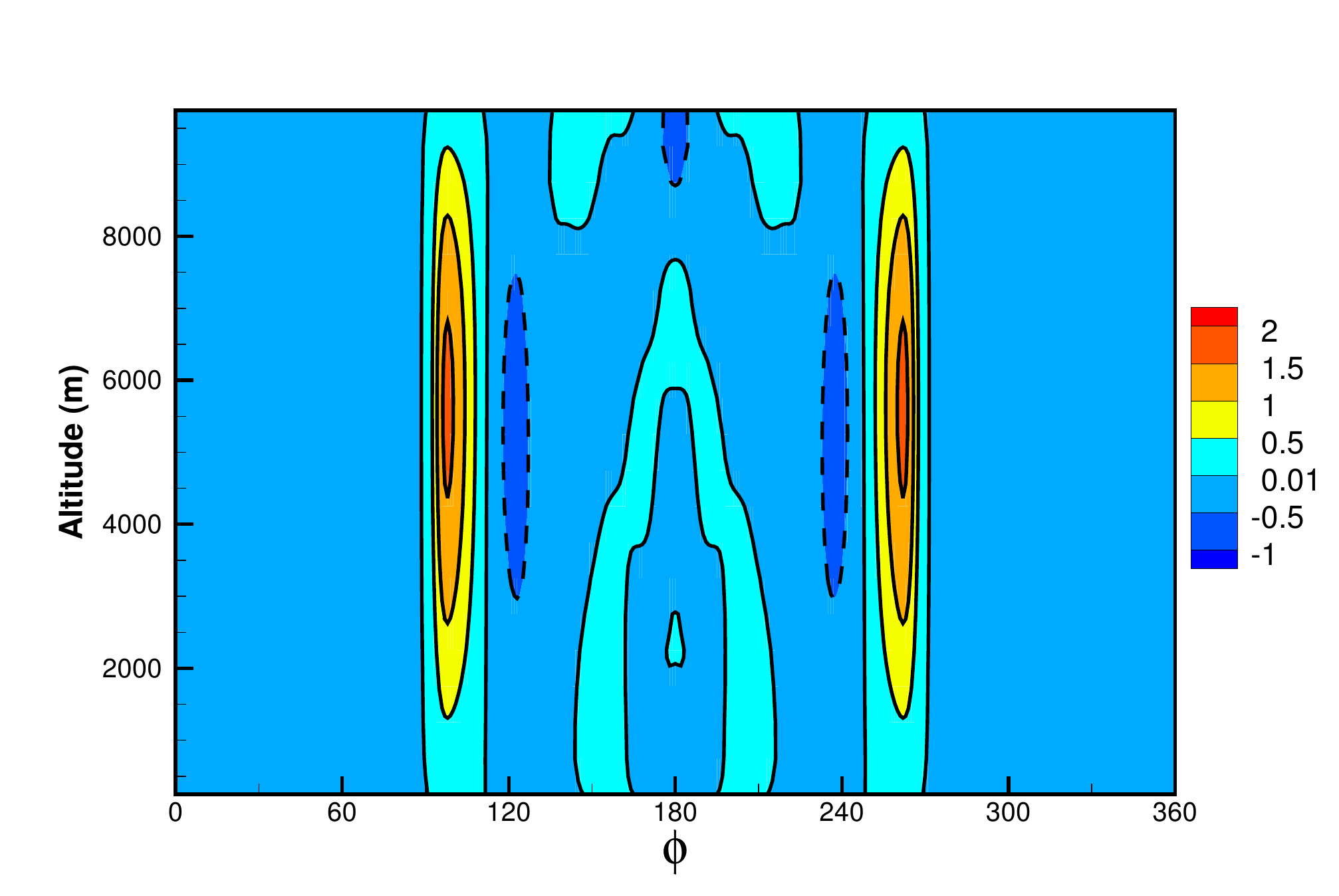}}
  \end{subfigure}
   \begin{subfigure}[Hour 96]
   { \includegraphics[width=0.48\textwidth]{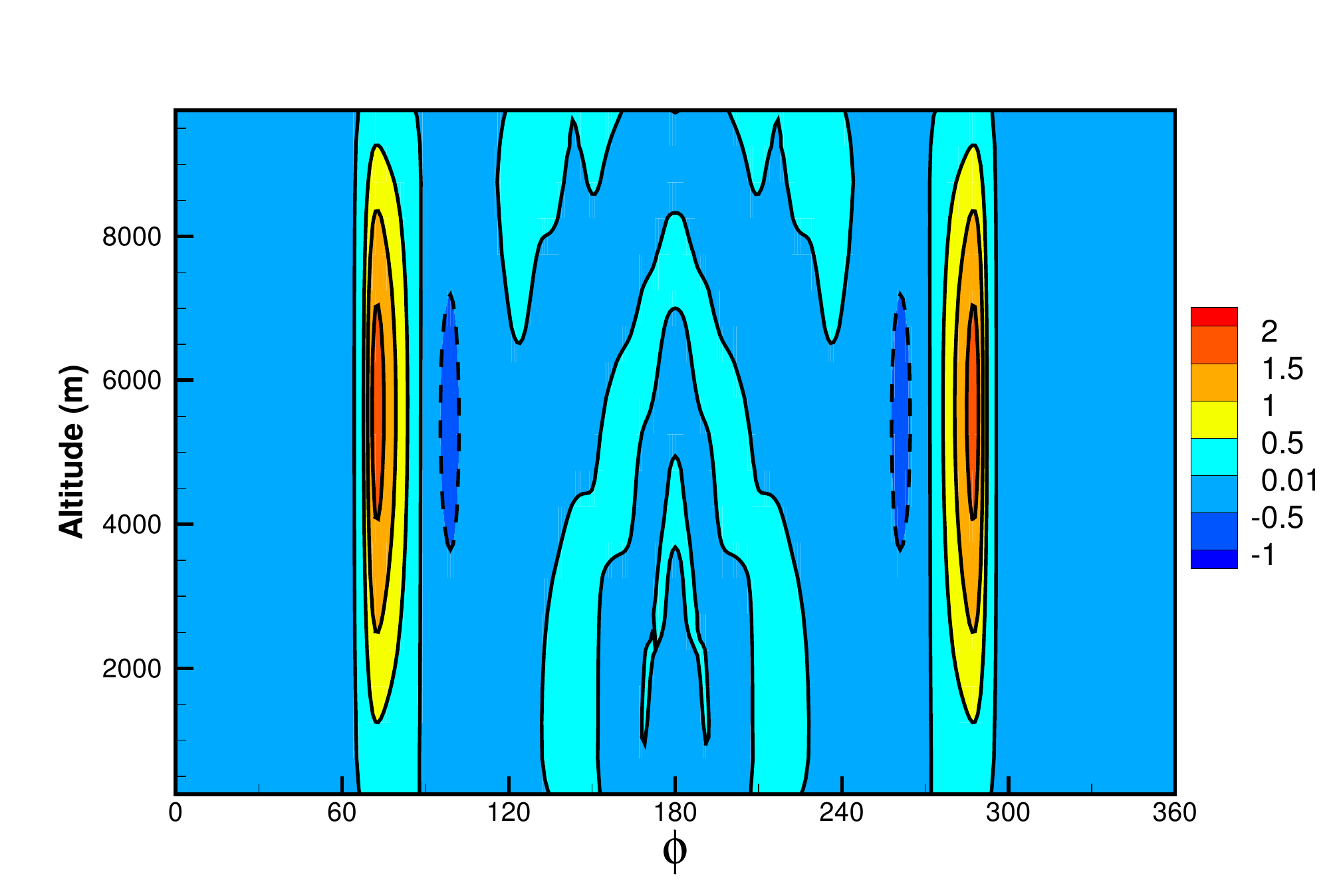}}
   \end{subfigure}
  \caption{Contour plots of numerical results of gravity wave test. Shown are potential temperature perturbation along the Equator and the dashed lines denote the negative values.}\label{GravityWave}
\end{figure}

The numerical results of perturbations of potential temperature along the Equator at different hours are shown in Fig. \ref{GravityWave}. No interpolation operation is applied for post-processing in this test. The wave horizontally propagates in two opposite directions and the keeps symmetrical shape. As the initial perturbation is specified having a shape of cosine bell in horizontal directions, non-physical numerical oscillations exist in the results of any unlimited high-order model. Thus, the 0 contour line is replaced by 0.01 in Fig. \ref{GravityWave}. Current results are competitive to those given in \cite{mcore2012} by $4^{th}$-order finite volume scheme on the same cubed-sphere grid. The results reproduce more details of wave structures in comparison with those of CAM-EUL and CAM FV using artificial diffusion or divergence damping (given in Fig. 10 in \cite{mcore2012}).

\subsection{Mountain-induced Rossby wave-train}

Without bottom mountain, the balanced initial condition is first specified as a steady geostrophic flow. The horizontal velocity components in longitude-latitude grid are
\begin{equation}
\left\{\begin{array}{l}
u_\lambda=u_0\cos\phi\\
u_\phi=0
\end{array}\right.,
\end{equation}
where $u_0=20$ m/s.

The thermodynamic variables are derived from hydrostatic relation considering a isothermal atmosphere with $T_0=288$ K as
\begin{equation}
p=p_s\exp\left(-\frac{g}{R_dT_0}z\right),
\end{equation}
where $p_s$ is the pressure at surface, specified to preserve geostrophic balance as
\begin{equation}
p_s=p_p \exp \left[ -\frac{1}{R_dT_0}\left(\frac{{u_0}^2}{2} + a\omega u_0\right) \left(\sin^2 \phi-1\right)-\frac{g}{R_dT_0}z\right],
\end{equation}
and $p_p=939$ hPa.

A bottom topography is then involved as
\begin{equation}
z_s=h_0 \exp\left[\left(-\frac{r}{d}\right)^2\right]
\end{equation}
where $h_0=2000$ m, $r$ is great-circle distant to the mountain center $\left(\frac{\pi}{2},\frac{\pi}{6}\right)$ and $d=1500$ km.

The Brunt-V$\ddot{\mathrm{a}}$is$\ddot{\mathrm{a}}$l$\ddot{\mathrm{a}}$ frequency $N=\sqrt{\frac{g^2}{c_pT_0}} \approx0.0182\ \mathrm{s}^{-1}$ and the flow is hydrostatic due to the nondimensional quantity $\frac{Nd}{u_0} >> 1$.

\begin{figure}[h]
 \centering
 \begin{subfigure}[$l_2$ error]
  {\includegraphics[width=0.7\textwidth]{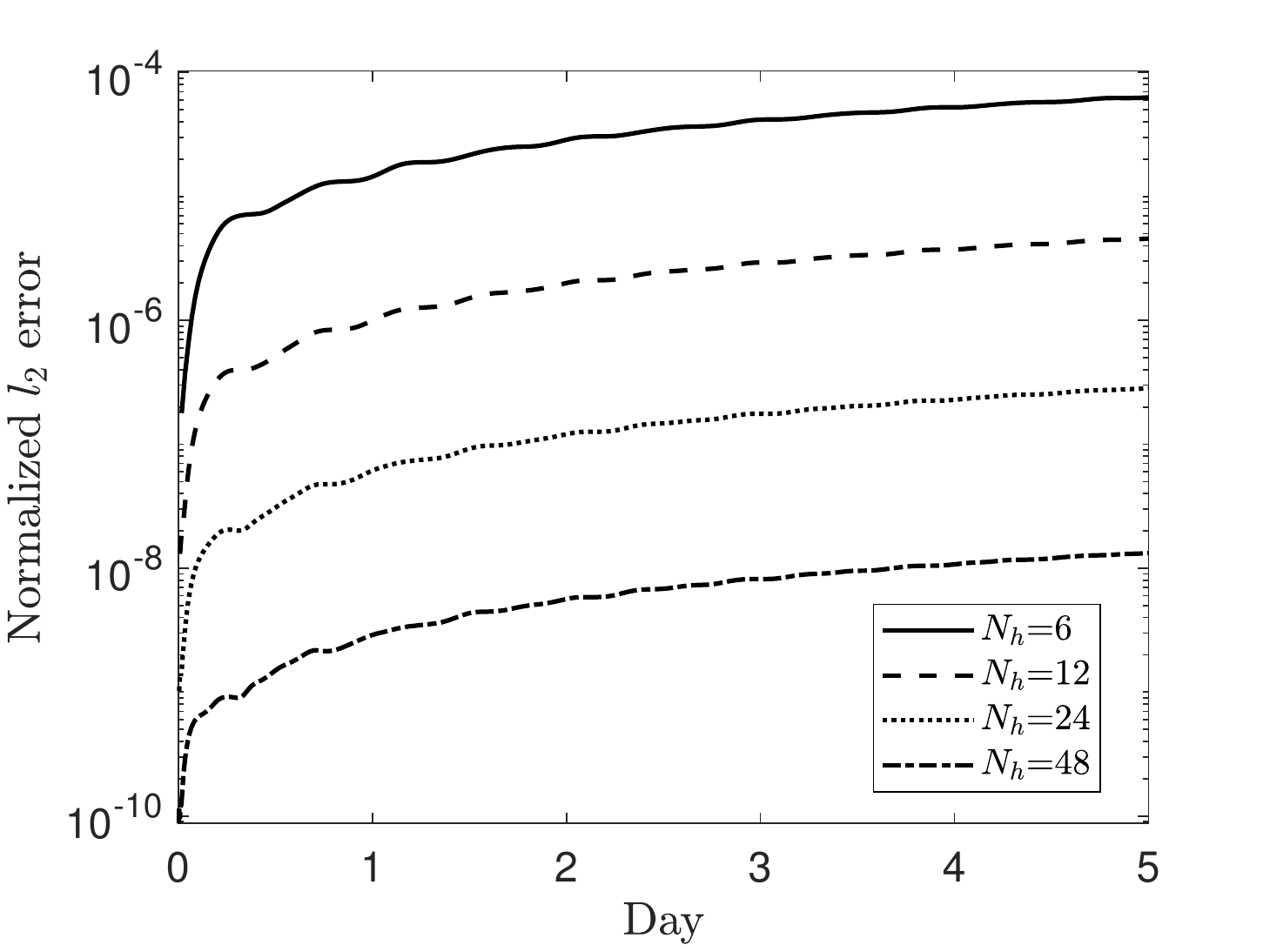}}
 \end{subfigure}
 \begin{subfigure}[Convergence rate]
  { \includegraphics[width=0.7\textwidth]{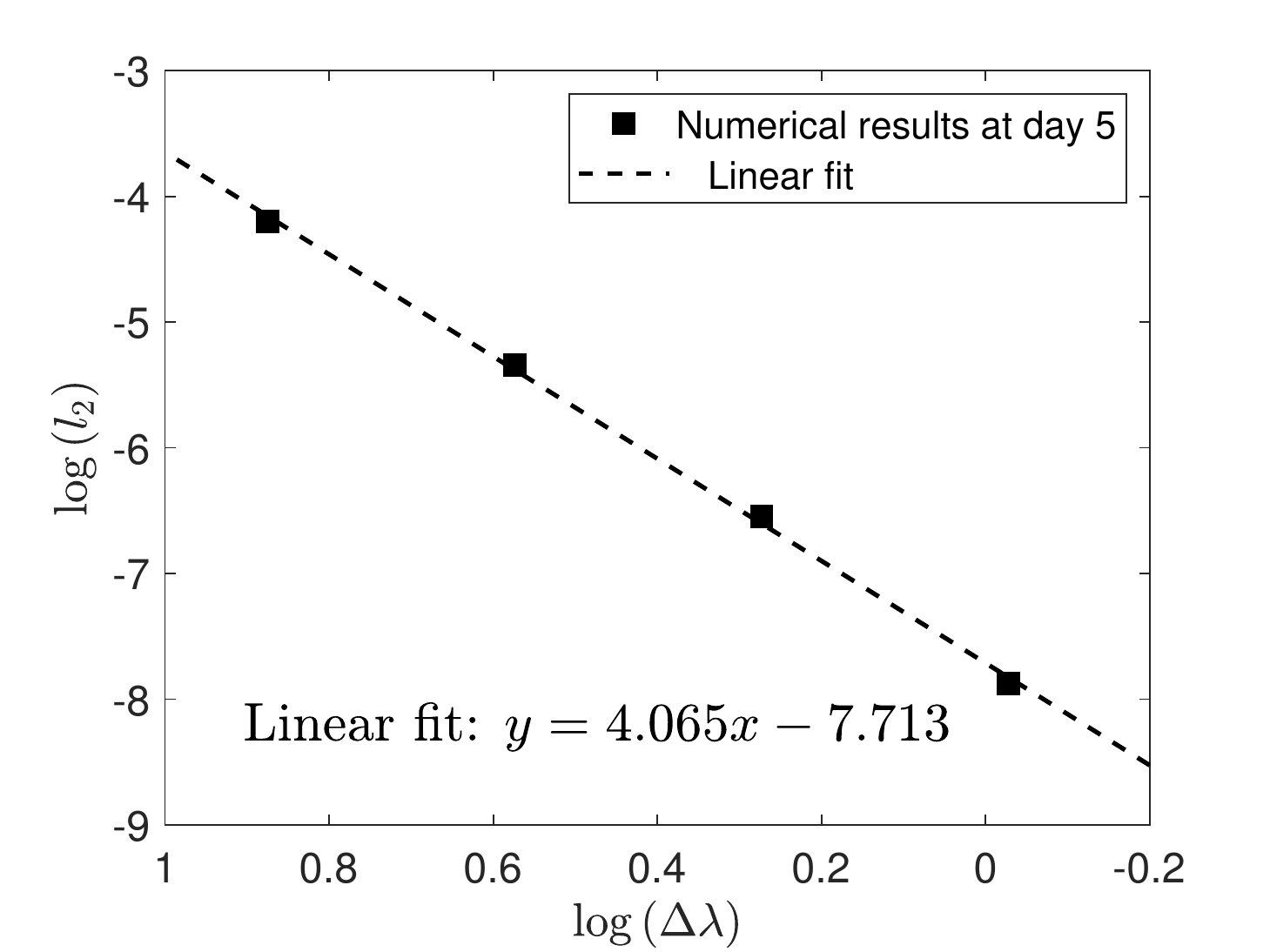}}
  \end{subfigure}
\caption{Normalized $l_2$ errors and the convergence rate of density in the balanced test case on a series of refining grids.}\label{Balanced}
\end{figure}

This test is first checked excluding the bottom mountain. The exact solution of this balanced flow is same as the initial condition. As the initial distribution is considerably smooth, the convergence test is conducted on a series of refining grids. The normalized $l_2$ errors (following the definition in \cite{Williamson1992}) of density and the convergence rates are given in Fig. \ref{Balanced}. In this balanced test, only 10 layers of computational cells are equidistantly arranged in vertical direction. With the current test setting, the errors are dominated by the spatial discretization in horizontal directions and the $4^{th}$-order convergence rate is achieved in spherical geometry with losing the theoretical accuracy of adopted MCV scheme.

\begin{figure}[h]
 \centering
 \begin{subfigure}[Geopotential height]
  {\includegraphics[width=0.48\textwidth]{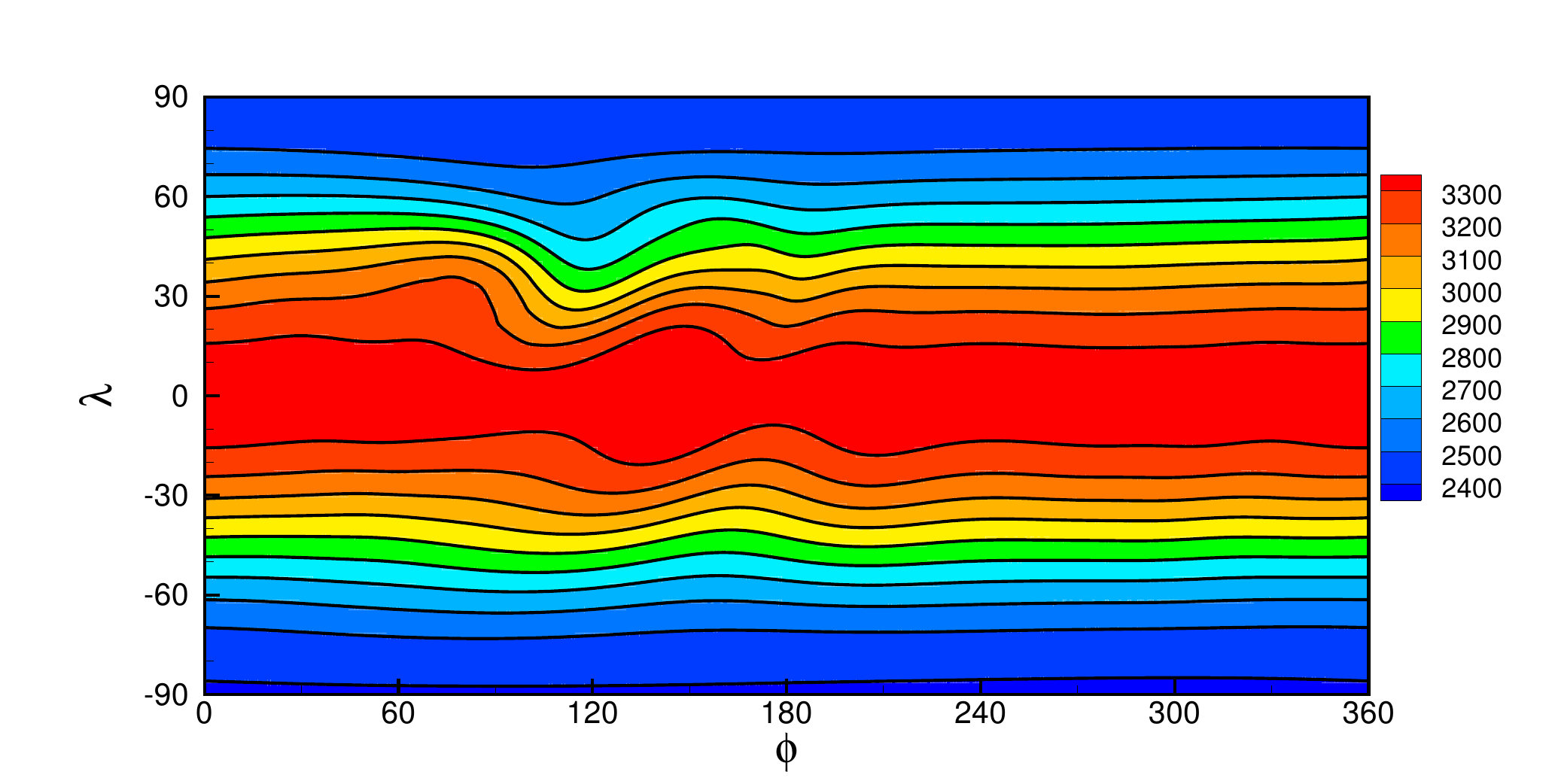}}
 \end{subfigure}
 \begin{subfigure}[Temperature]
  { \includegraphics[width=0.48\textwidth]{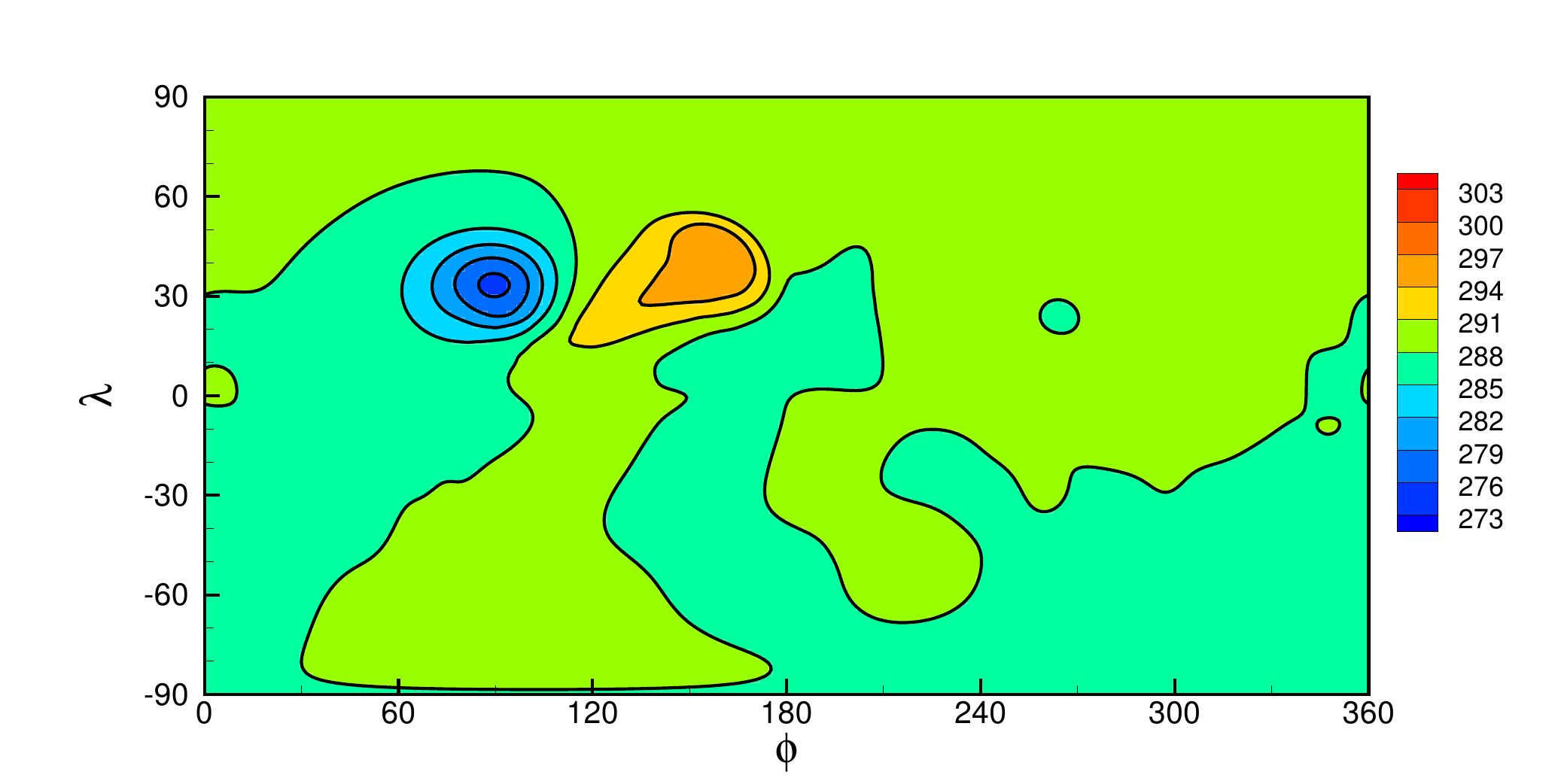}}
  \end{subfigure}
 \begin{subfigure}[Zonal wind]
  {\includegraphics[width=0.48\textwidth]{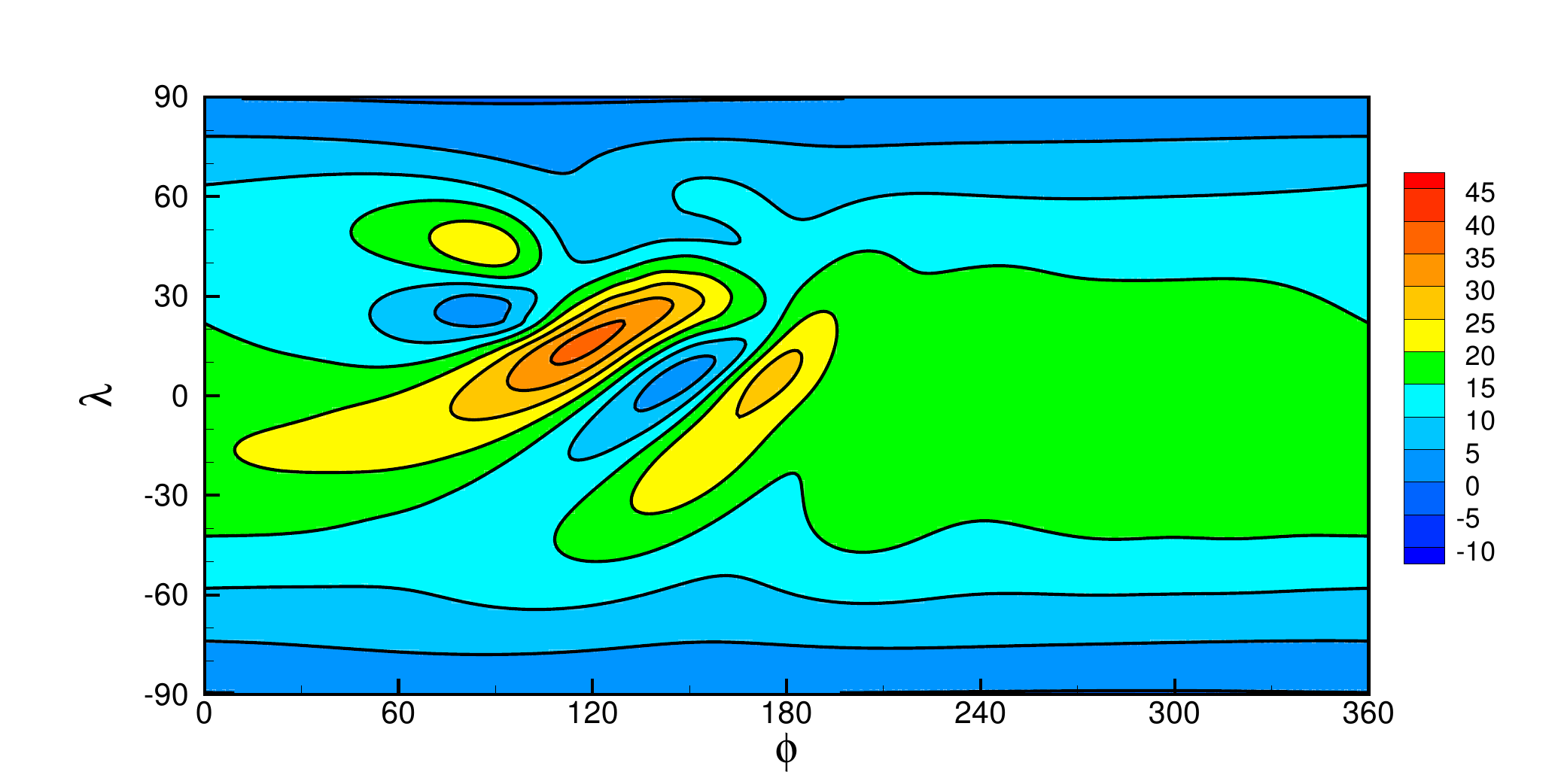}}
 \end{subfigure}
 \begin{subfigure}[Meridional wind]
  { \includegraphics[width=0.48\textwidth]{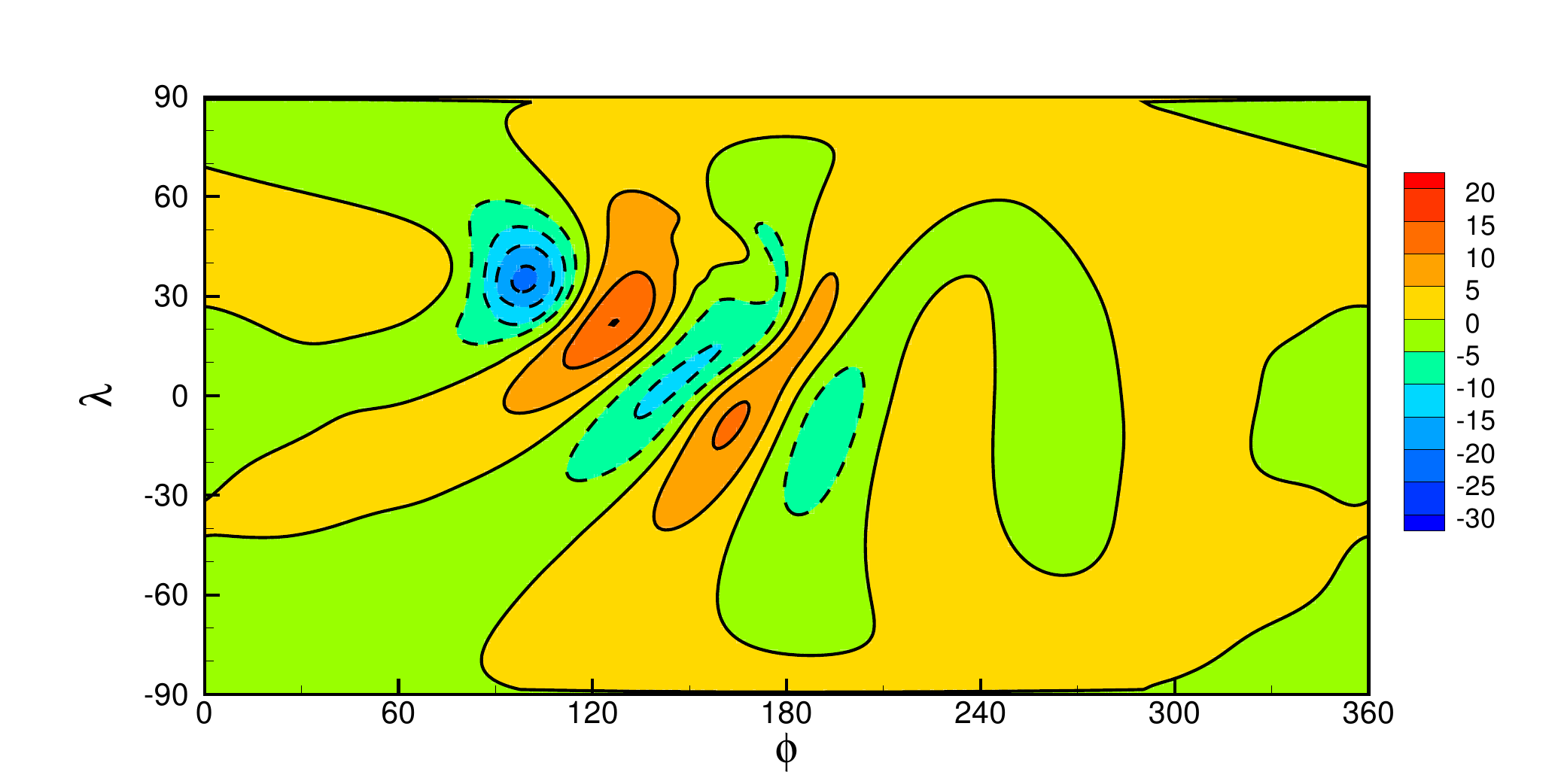}}
  \end{subfigure}
\caption{Contour plots of numerical results of mountain-induced Rossby wave-train at day 5. Shown are 700hPa height (panel (a)), temperature (panel (b)) and horizontal wind field (panels (c) and (d)) and the dashed lines denote the negative values.}\label{Mountain5}
\end{figure}

\begin{figure}[h]
 \centering
 \begin{subfigure}[Geopotential]
  {\includegraphics[width=0.48\textwidth]{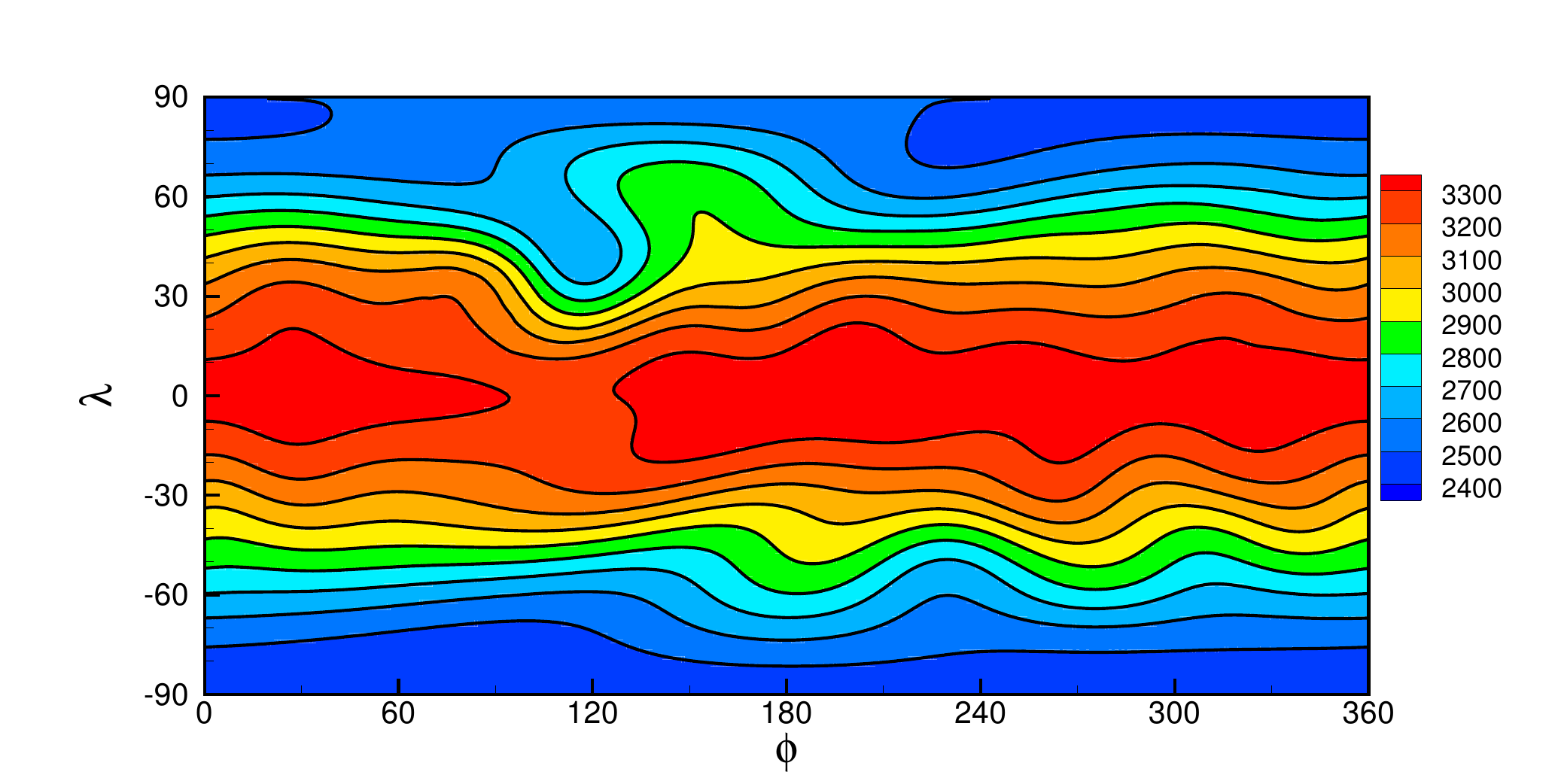}}
 \end{subfigure}
 \begin{subfigure}[Temperature]
  { \includegraphics[width=0.48\textwidth]{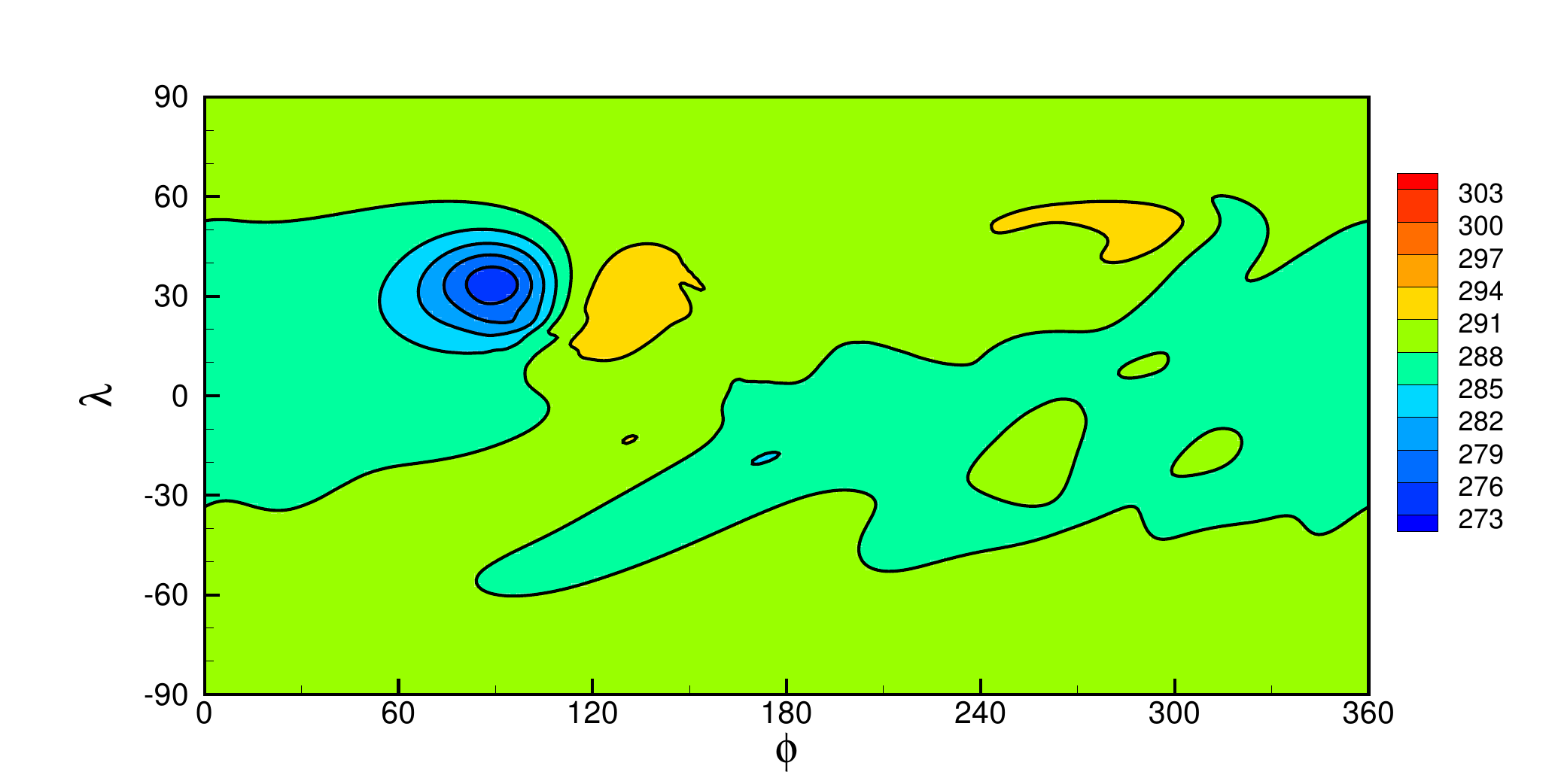}}
  \end{subfigure}
 \begin{subfigure}[Zonal wind]
  {\includegraphics[width=0.48\textwidth]{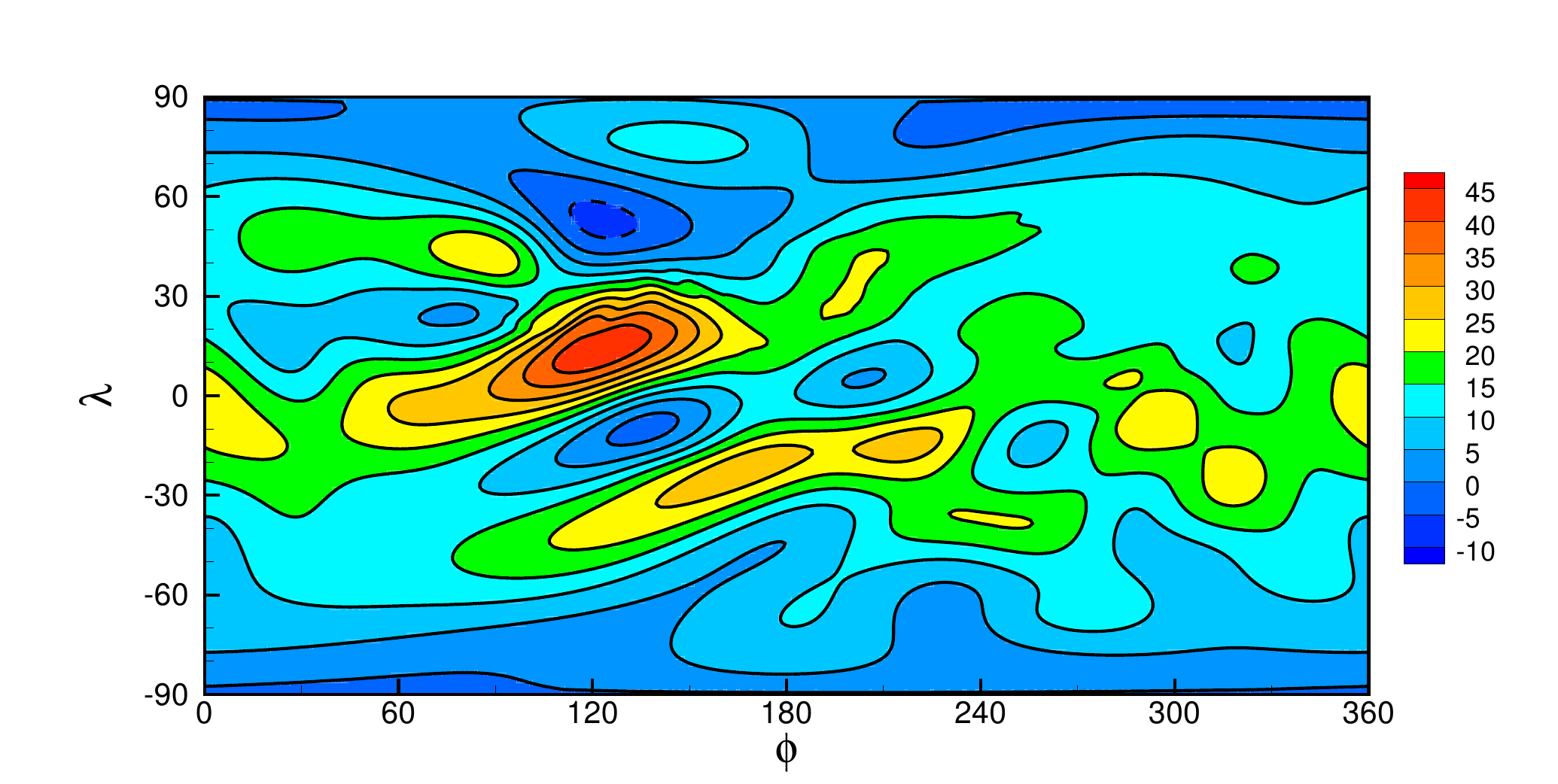}}
 \end{subfigure}
 \begin{subfigure}[Meridional wind]
  { \includegraphics[width=0.48\textwidth]{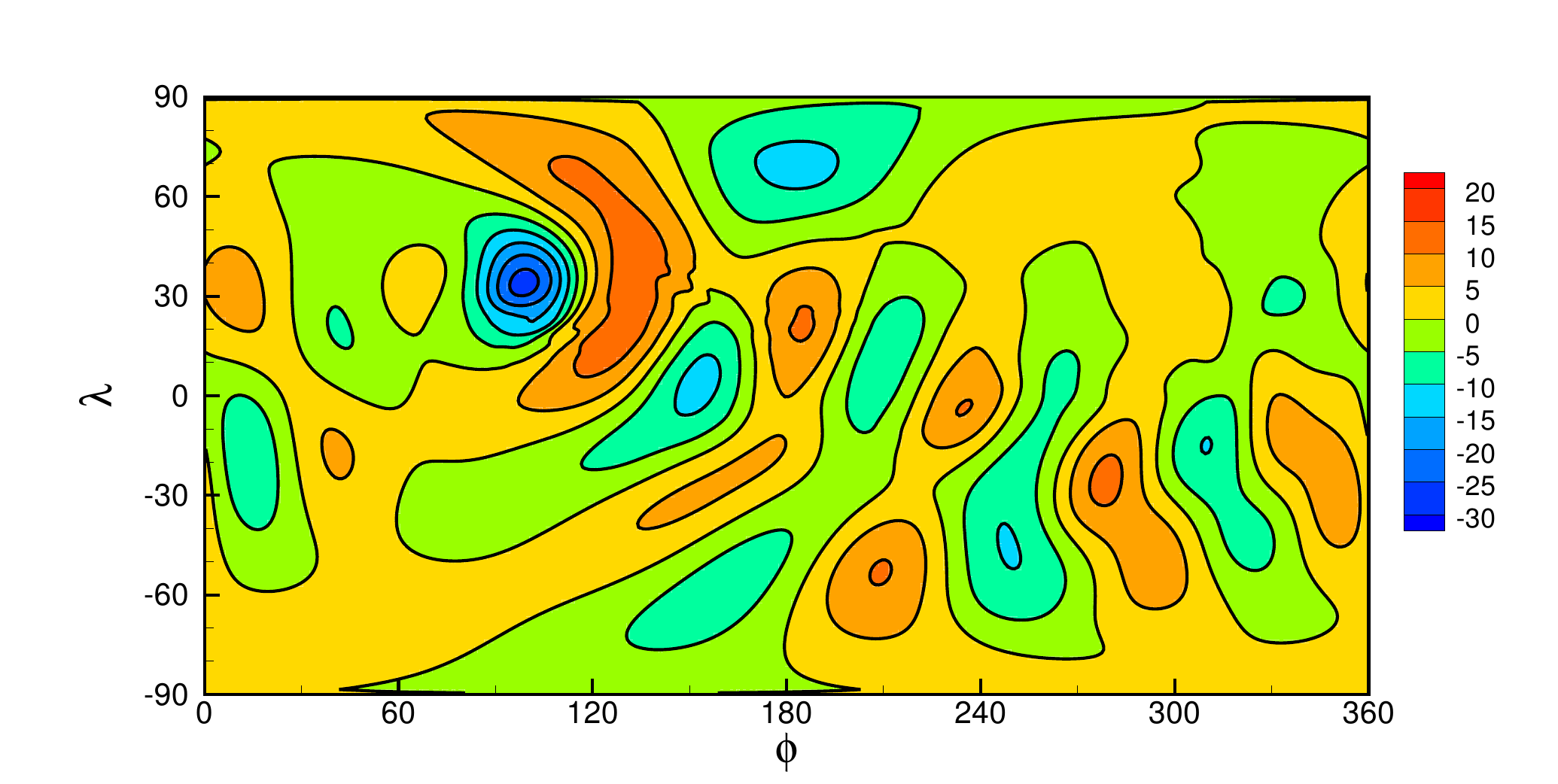}}
  \end{subfigure}
\caption{Same as Fig. \ref{Mountain5}, but for numerical results at day 15.}\label{Mountain15}
\end{figure}

The numerical results including the effect of the bottom mountain are shown in Figs. \ref{Mountain5} and \ref{Mountain15} for predicted 700 hPa geopotential height, temperature and horizontal wind fields at day 5 and day 15. The balanced state is destroyed by the topography effect, which triggers a propagation of Rossby wave-train. This test is a challenging case to verify the robustness of dynamical cores since relatively large deviations from the initial conditions are generated, specially in the horizontal wind field. The results by the proposed model are visibly identical to those given in \cite{jablonowski2008}, except the broken 3300 m contour line of geopotential height and some numerical oscillations found in horizontal wind field at day 15. Similar differences are also found in nonhydrostatic finite volume dynamical core \cite{mcore2012}.

\subsection{Baroclinic wave}

A balanced initial condition is first specified in pressure-based grid $\left(\lambda,\phi,\tilde{\eta}\right)$ ($\tilde{\eta}=\frac{p}{p_0}$) with bottom mountain in this test \cite{JW2006}. The horizontally averaged temperature profile is given as
\begin{equation}
\overline{T}\left(\tilde{\eta}\right)=\left\{\begin{array}{ll}
T_0\tilde{\eta}^{\frac{R_d\Gamma}{g}}&\mathrm{if\ }\tilde{\eta}_t\leq\tilde{\eta}\leq 1\\
T_0\tilde{\eta}^{\frac{R_d\Gamma}{g}}+\Delta T\left(\tilde{\eta}_t-\tilde{\eta}\right)^5&\mathrm{otherwise}
\end{array}\right.,
\end{equation}
where $T_0=288$ K, $\Gamma=0.005$ K/m, $\tilde{\eta}_t=0.2$ and $\Delta T=4.8\times 10^{5}$ K. This distribution is close to the vertical profile of real atmosphere.

The zonal velocity component is given as
\begin{equation}
u_\lambda=u_0\cos^{\frac{3}{2}}\tilde{\eta}_v\sin^2\left(2\phi\right),
\end{equation}
where $u_0=35$ m/s and $\tilde{\eta}_v=\frac{\pi}{2}\left(\tilde{\eta}-\tilde{\eta}_0\right)$ with $\tilde{\eta}_0=0.252$.

The details of three-dimensional temperature distribution $T\left(\lambda,\phi,\tilde{\eta}\right)$ and corresponding geopotential $\Phi\left(\lambda,\phi,\tilde{\eta}\right)$, which gives a balanced steady state, are found in \cite{jablonowski2008}.

To set up the initial condition in height-based grid, coordinate $\tilde{\eta}$ is first determined at any solution point $\left(\lambda,\phi,z\right)$ by solving the equation
\begin{equation}
y=\Phi\left(\lambda,\phi,\tilde{\eta}\right)-gz=0.
\end{equation}

This equation is solved by Newton iteration. With known $\tilde{\eta}^k$ at $k^{th}$ iteration, the next guess is
\begin{equation}
\tilde{\eta}^{k+1}=\tilde{\eta}^{k+1}-\frac{y\left(\tilde{\eta}^k\right)}{y_{\tilde{\eta}}\left(\tilde{\eta}^k\right)},
\end{equation}
where
\begin{equation}
y_{\tilde{\eta}}=-\frac{R_d T\left(\lambda,\phi,\tilde{\eta}\right)}{\eta}
\end{equation}
according to the hydrostatic relation.

The detailed procedure can be referred to Appendix D in \cite{jablonowski2008}.

\begin{figure}[h]
 \centering
 \includegraphics[width=0.7\textwidth]{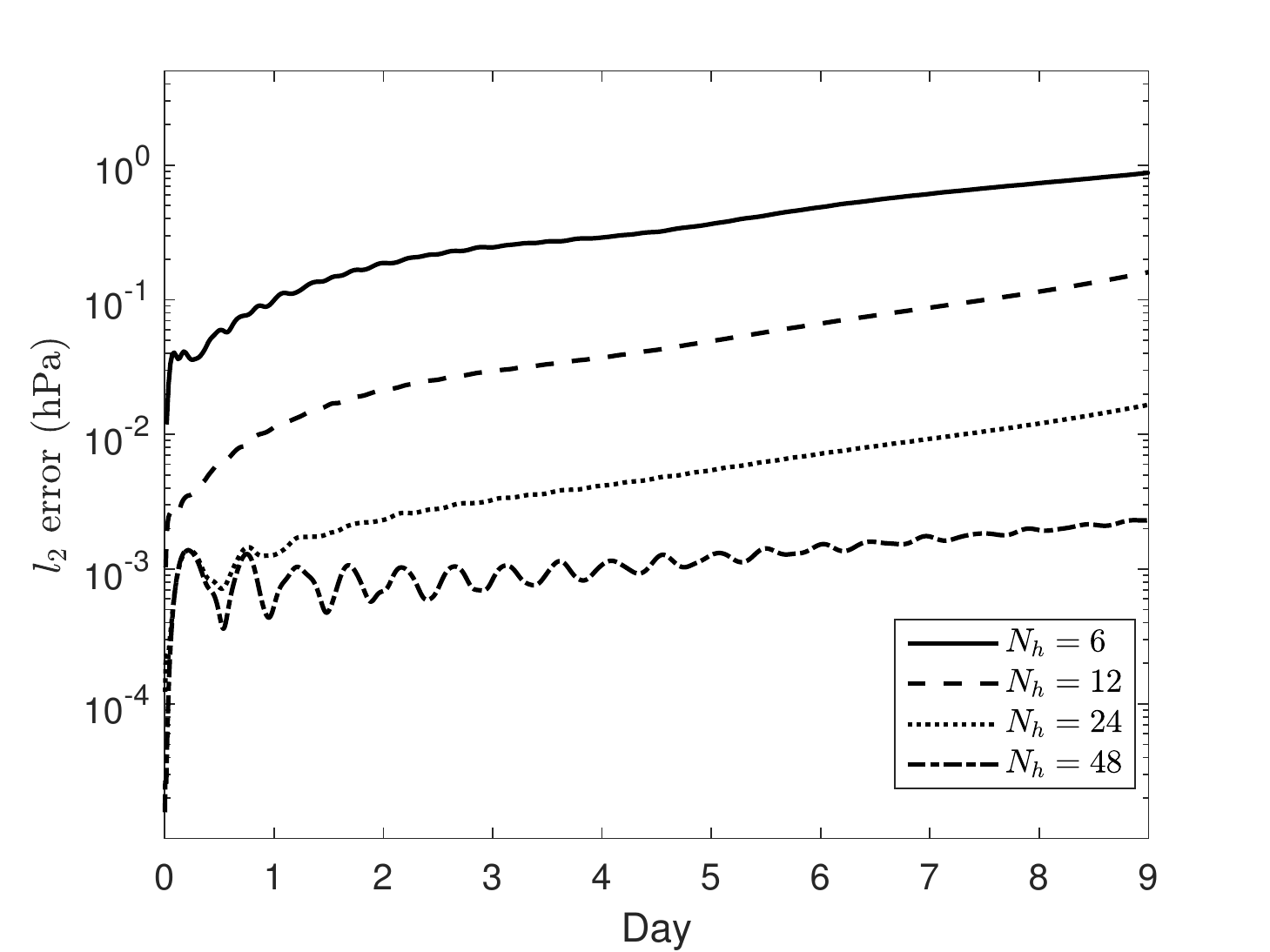}
 \caption{$l_2$ errors of pressure at first model layer in the balanced case on a series refining grid.}\label{BaroclinicWave3}
 \end{figure}

\begin{figure}[h]
 \centering
  \begin{subfigure}[$N_H=12$]
  { \includegraphics[width=0.48\textwidth]{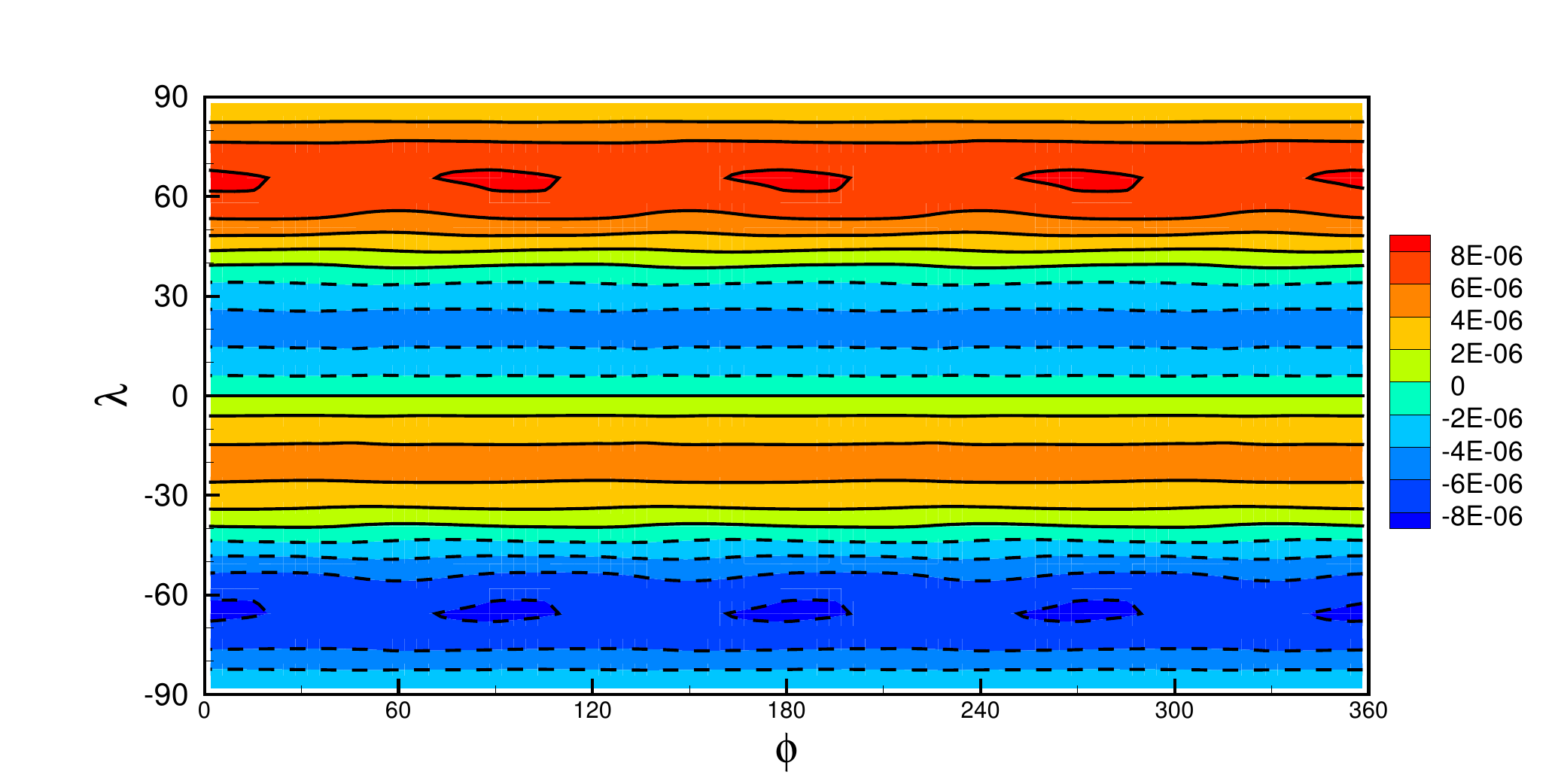}}
  \end{subfigure}
  \begin{subfigure}[$N_H=48$]
  { \includegraphics[width=0.48\textwidth]{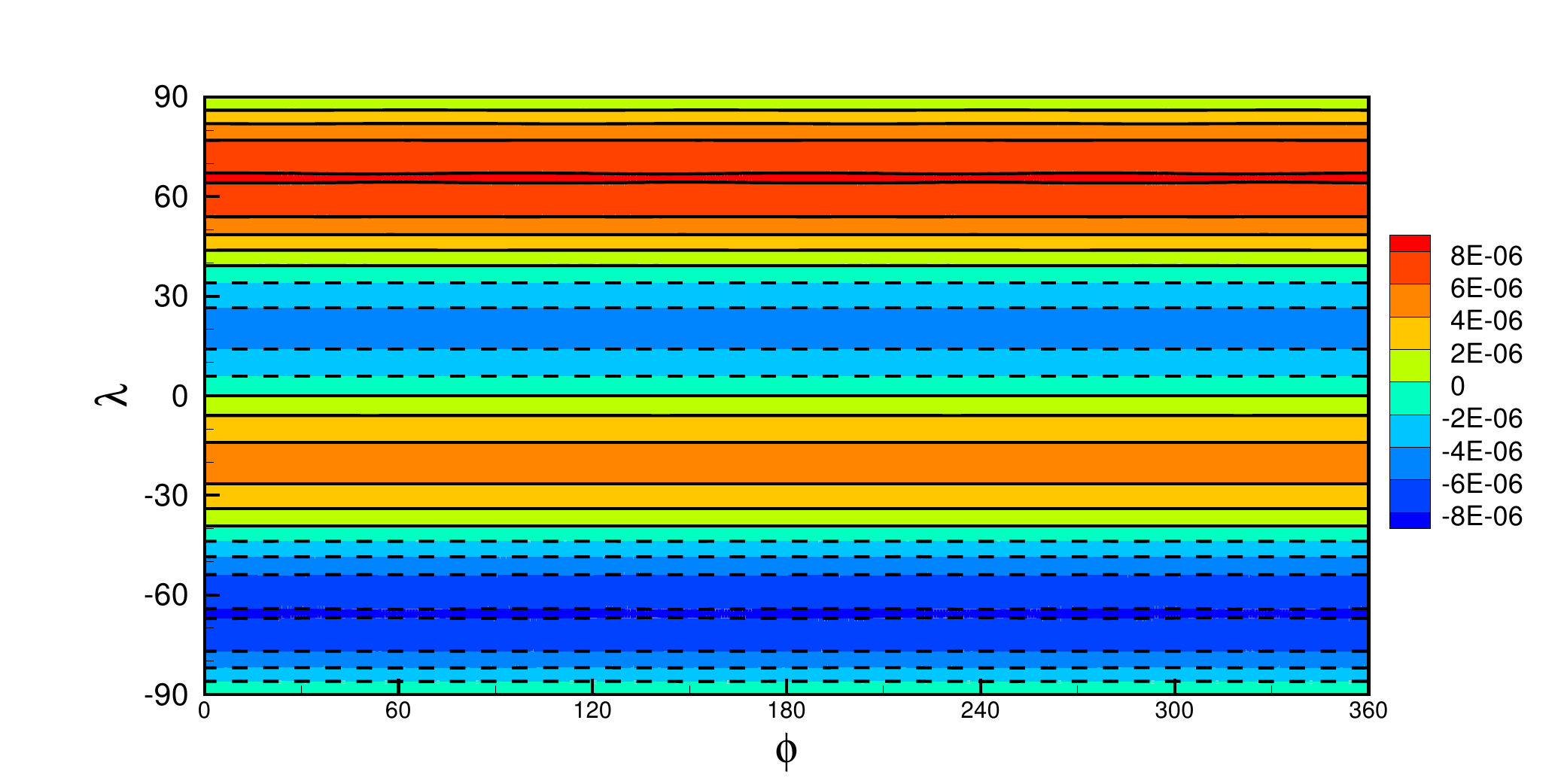}}
  \end{subfigure}
  \caption{Contour plots of relative vorticity of the balanced case on grids $N_h=12$ (panel (a)) and $N_h=48$ (panel (b)).}\label{BaroclinicWave2}
 \end{figure}

Same as above test, the balanced initial condition is first checked. Due to the grid lines are not coincide with the wind direction, specially on two polar pathes, 4-wave errors may be observed on coarse grid \cite{JW2006}. Thus, we run the model on a series of refining grid to verify the grid-imprinting errors can be effectively suppressed by increasing the grid resolution. The $l_2$ error of predicted pressure of first layer (shown in Fig. \ref{BaroclinicWave3}) is evaluated following the definition in \cite{JW2006} to evaluate the quality of numerical results. At beginning, a large jump is observed on all grids due to the initial condition is not balanced in the discrete form. Then $l_2$ errors are gradually increasing with the time. The relative vorticity fields at day 9 on grids $N_h=12$ and $N_h=48$ are shown in Fig. \ref{BaroclinicWave2}. As expected, the 4-wave structure (error) is observed on coarse grid and visibly disappeared on fine one.

Then a perturbation is added in zonal wind to trigger the baroclinic wave, specified as
\begin{equation}
u^\prime=u_p\exp\left(-\frac{r^2}{{r_0}^2}\right),
\end{equation}
where $u_p=1$ m/s, $r_0=\frac{R}{10}$ and $r$ is great-circle distance to $\left(\frac{\pi}{9},\frac{2\pi}{9}\right)$.

\begin{figure}[h]
 \centering
  \begin{subfigure}[Surface pressure at day 7]
  { \includegraphics[width=0.48\textwidth]{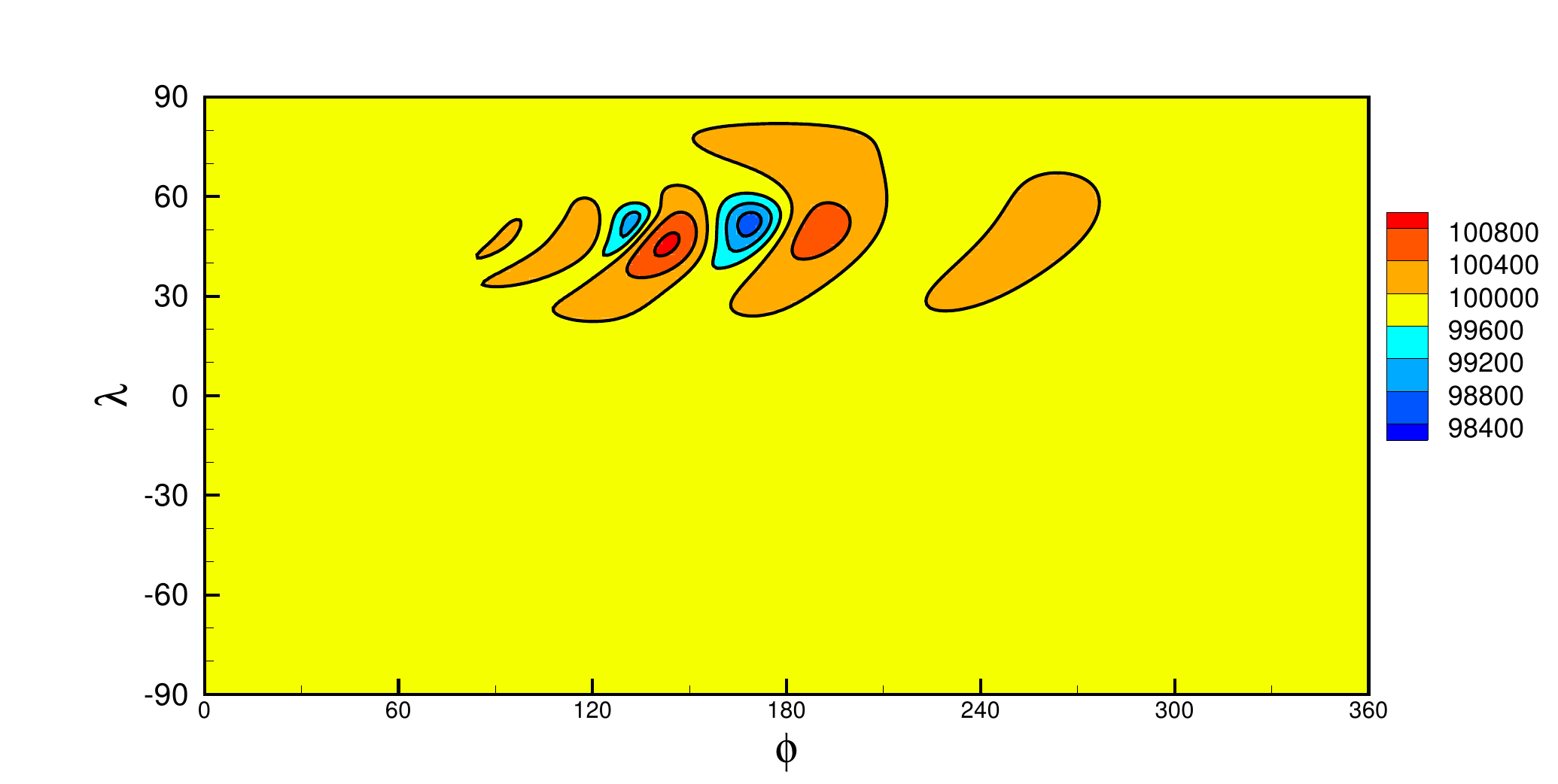}}
  \end{subfigure}
  \begin{subfigure}[Surface pressure at day 9]
  { \includegraphics[width=0.48\textwidth]{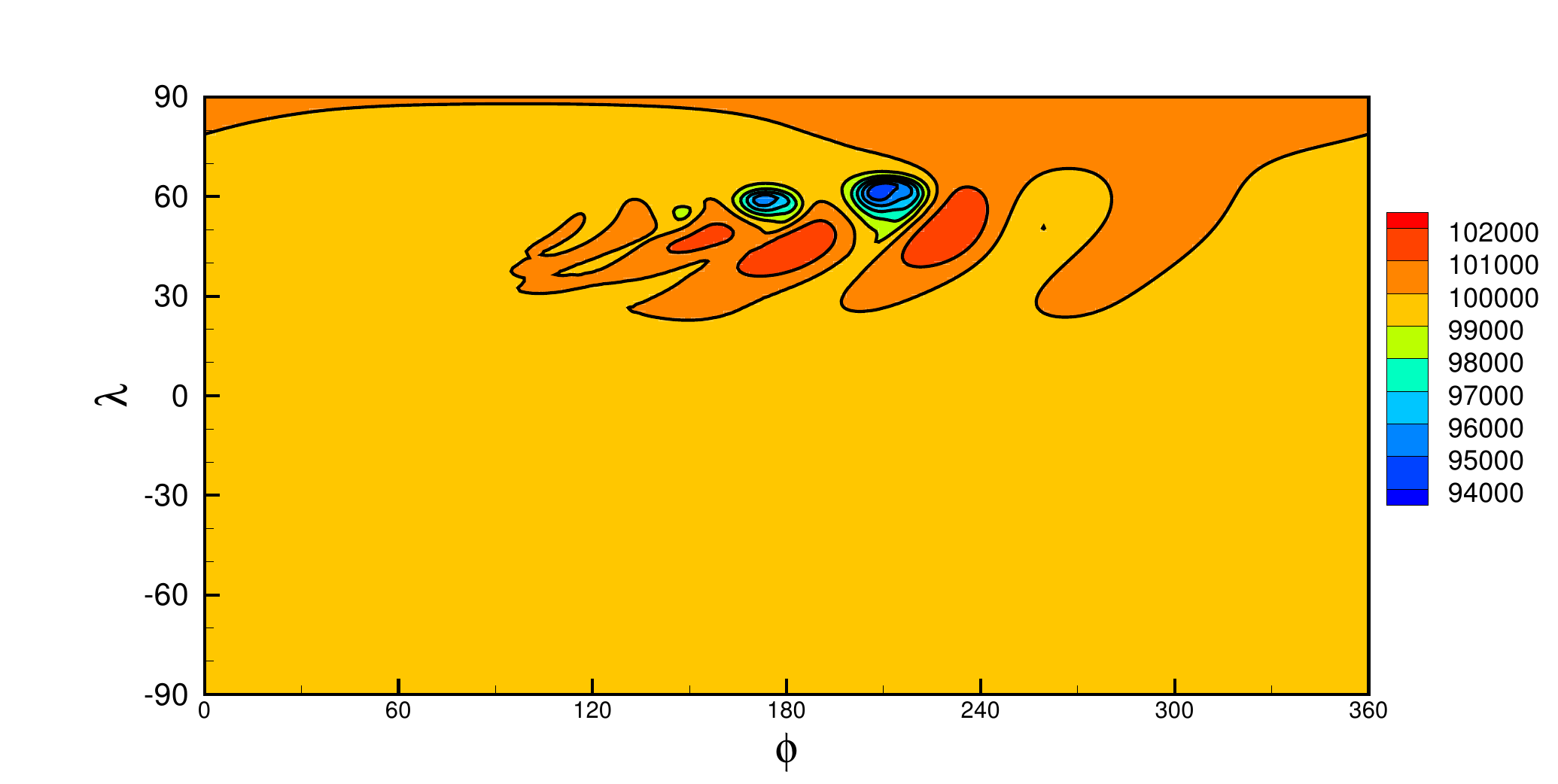}}\\
  \end{subfigure}
  \begin{subfigure}[850hPa temperature at day 7]
   { \includegraphics[width=0.48\textwidth]{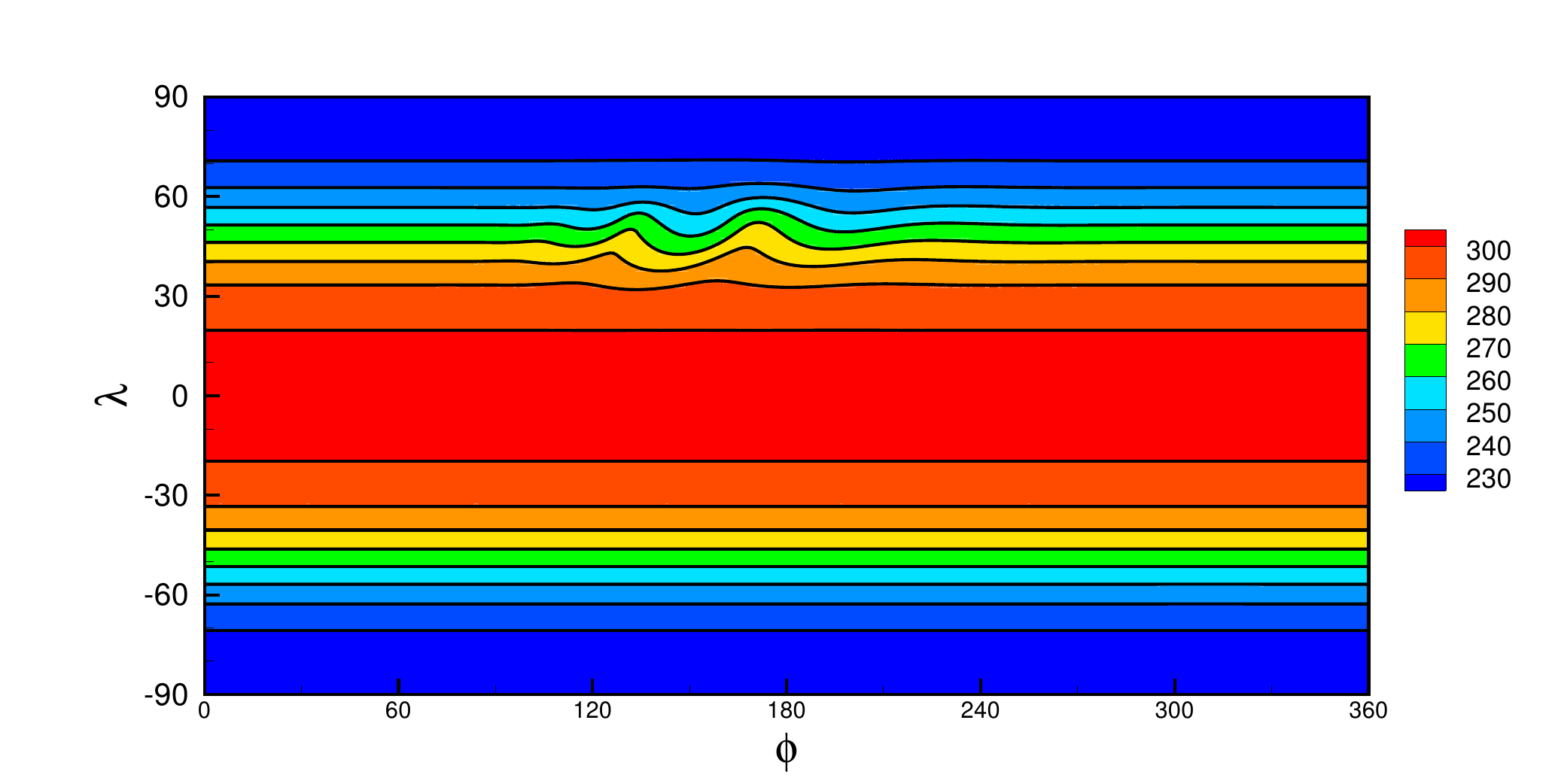}}
  \end{subfigure}
  \begin{subfigure}[850hPa temperature at day 9]
   { \includegraphics[width=0.48\textwidth]{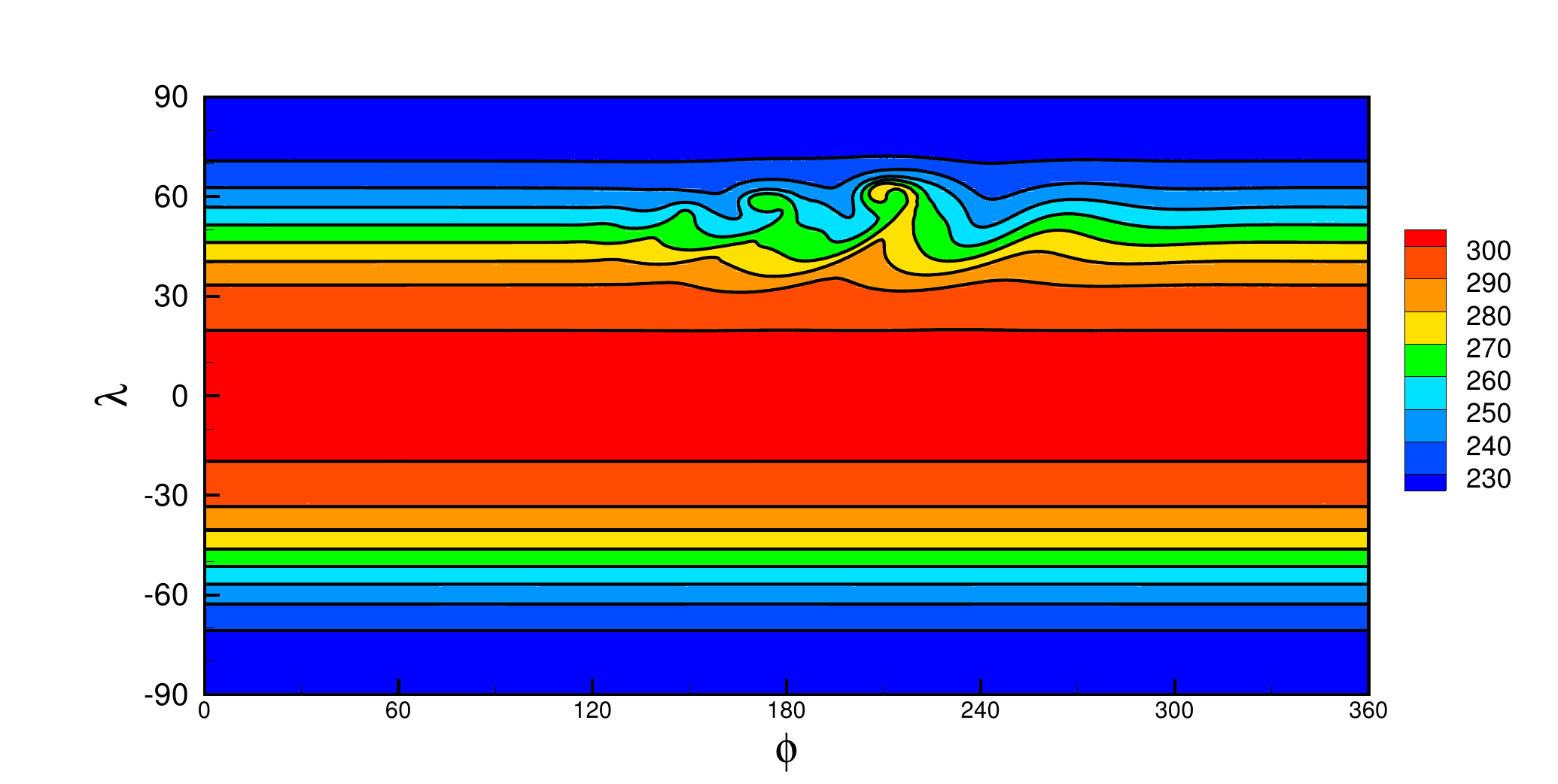}}\\
  \end{subfigure}
  \begin{subfigure}[850hPa relative vorticity at day 7]
  { \includegraphics[width=0.48\textwidth]{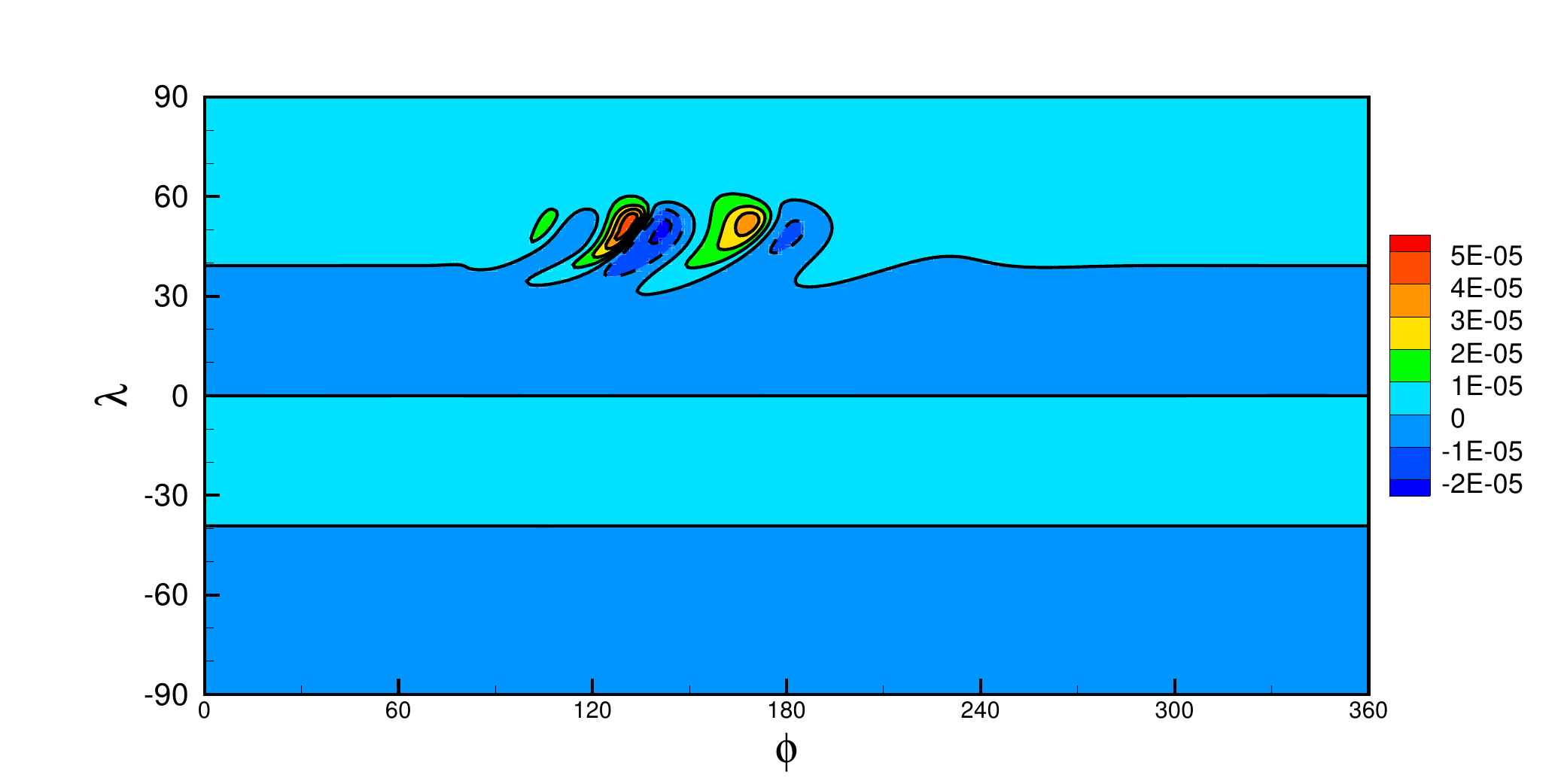}}
  \end{subfigure}
  \begin{subfigure}[850hPa relative vorticity at day 9]
  { \includegraphics[width=0.48\textwidth]{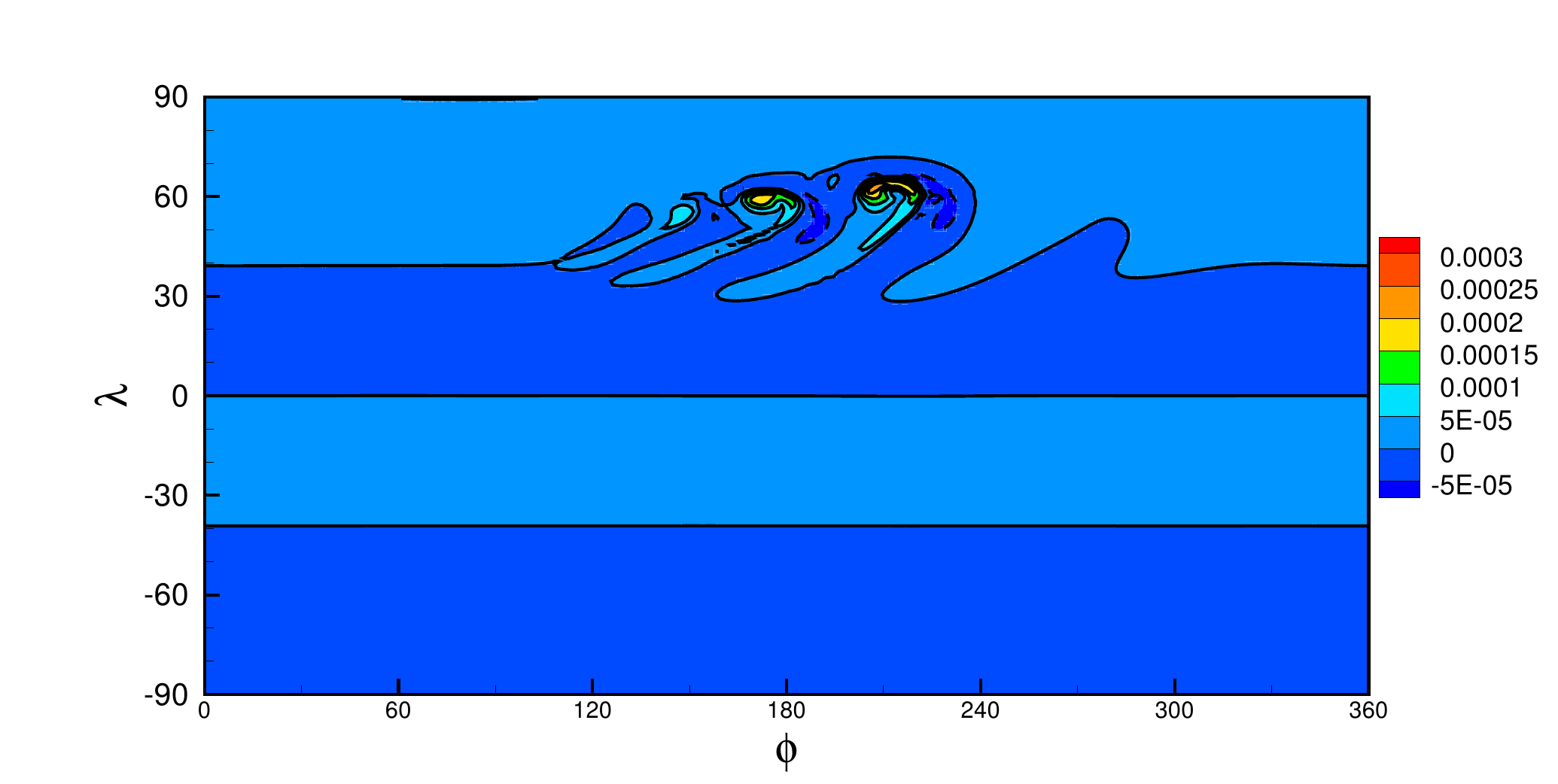}}
  \end{subfigure}
  \caption{Contour plots of numerical results of baroclinic wave test at day 7 and day 9. Shown are surface pressure (panels (a) and (b)), 850 hPa temperature (panels (c) and (d)) and 850 hPa relative vorticity (panels (e) and (f)). The dashed lines denote the negative values.}\label{BaroclinicWave}
\end{figure}

The numerical results at day 7 and day 9 are given in Fig. \ref{BaroclinicWave}. No analytic solution is available for this test. In comparison with numerical results of some representative models given in \cite{JW2006,Lauritzen2010}, our results accurately reproduce the propagation of baroclinic wave. As suggested in \cite{JW2006}, we also calculated $l_2$ error of pressure field of first layer. The numerical result on high resolution grid ($N_h=180$) is used as the reference solution. At day 9, $l_2$ errors are 0.37 hPa, 0.20 hPa, 0.14 hPa and 0.07 hPa for the results on grids $N_h=30$, $45$, $60$ and $90$.

\subsection{Non-hydrostatic mountain waves over a  Sch$\ddot{\mathrm{a}}$r-type Mountain}

In this test, the radius of the Earth is scaled to simulate the nonhydrostatic flow over the bottom mountain (tests 2-1 and 2-2 in \cite{DCMIP2012}). Numerical model is carried out on a non-rotating reduced-size Earth with radius $R^\prime=\frac{R}{500}$.

The topography is specified as a Sch$\ddot{\mathrm{a}}$r-type mountain, having the form of
\begin{equation}
z_s\left(\lambda,\phi\right)=h_0\exp\left(-\frac{r^2}{d^2}\right)\cos^2\left(\frac{\pi r}{L}\right),
\end{equation}
where $r$ is great-circle distance to mountain center $\left(\frac{\pi}{4},0\right)$, $d=5$ km is  Sch$\ddot{\mathrm{a}}$r-type mountain half-width and $L=4$ km is  Sch$\ddot{\mathrm{a}}$r-type mountain wavelength.

The strength of Rayleigh friction is also specified in this test case as
\begin{equation}
\tau_R=
\left\{\begin{array}{ll}
0&\mathrm{if\ }z<z_D\\
\tau_0\sin^2\left[\frac{\pi}{2}\left(\frac{z-z_D}{z_t-z_D}\right) \right]& \mathrm{otherwise}
\end{array}\right.,
\end{equation}
where $\tau_0=0.04 \mathrm{s}^{-1}$, $z_D=20$ km and $z_t=30$ km.

The hydrostatic pressure distribution is written as
\begin{equation}
p\left(\lambda,\phi,z\right)=p_0\exp\left(-\frac{u_0^2}{2R_dT_0}\sin^2\phi-\frac{gz}{R_dT\left(\phi\right)}\right),
\end{equation}
where $u_0=20$ m/s and $T_0$=300 K.

The temperature field depends on latitude and is uniform in vertical direction as
\begin{equation}
T\left(\phi\right)=T_0\left(1-\frac{cu_0}{g}\sin^2\phi\right),
\end{equation}
and the initially balanced zonal velocity is
\begin{equation}
u\left(\lambda,\phi,z\right)=u_0\cos\phi\sqrt{\frac{2T_0}{T\left(\phi\right)}cz+\frac{T\left(\phi\right)}{T_0}},
\end{equation}
where parameter $c$ denotes a prescribed vertical wind shear of the zonal velocity field at the surface.

Two velocity fields are used corresponding different values of $c$, including a non-sheared background flow ($c=0$) and a sheared one with $c=2.5\times 10^{-4}$ m/s.

Numerical results are shown in Figs. \ref{NHMountrain1} and \ref{NHMountrain2} for vertical wind and temperature perturbation at different simulation time for non-sheared and sheared cases, respectively. For this nonhydrostatic test case, we choose $k_v^B=0.1$ to improve the computational accuracy of vertical wind by reducing the numerical diffusion added in vertical momentum equation. Several models provided their numerical results of this tests in the Dynamical Core Model Intercomparison Project (DCMIP 2012 http://earthsystemcog.org/projects/dcmip-2012/). The numerical results from different models look a little divergent in this case, while our results agree well with those of ENDGame model \cite{wood2013}.

\subsection{Held-Saurez Climate test}

In this test, the idealized physical source terms are added, which are specified as \cite{HS94}
\begin{equation}
\left\{\begin{array}{l}
\frac{\partial \boldsymbol{v}}{\partial t}=\cdots-k_v\left(p\right)\boldsymbol{v}\\
\frac{\partial \theta}{\partial t}=\cdots-k_{\theta}\left(\phi,p\right)\left(\theta-\theta_{eq}\right)
\end{array}\right.,
\end{equation}
where the heating/cooling source in energy equation forces the model to a radiative equilibrium temperature field and the Rayleigh friction term in momentum equations represents the effect of boundary-layer friction near the surface.

Two coefficients $k_v$ and $k_\theta$ are determined by latitude and pressure of solution point as
\begin{equation}
\left\{\begin{array}{l}
k_v=k_f\max\left(0,\frac{\sigma-\sigma_b}{1-\sigma_b}\right)\\
k_f=k_a+\left(k_s-k_a\right)\max\left(0,\frac{\sigma-\sigma_b}{1-\sigma_b}\right)\cos\left(\phi\right)^4
\end{array}\right.,
\end{equation}
the radiative equilibrium potential temperature is
\begin{equation}
\theta_{eq}=\max\left[200\left(\frac{p_0}{p}\right)^{\kappa},315-\Delta T_y\sin\left(\phi\right)^2-\Delta \theta_z\log\left(\frac{p}{p_0}\right)\cos\left(\phi\right)^2\right],
\end{equation}
where $\sigma=\frac{p}{p_s}$, $p_s$ is surface pressure and all other parameters are identical to those adopted in \cite{HS94}.

The source terms specified in this test are treated implicitly. The model runs for 1200 days in this test. During the first 200 days, the model spins up and reaches to a state of statistical equilibrium. The numerical results are then averaged over the followed 1000 days to check the performance of the proposed model on reproducing the long-term statistical characteristics of atmospheric dynamics. The multi-moment dynamical core are integrated on $30\times 30$ grid with the time step of288 s. The details of non-uniform vertical grid are given in Table \ref{verticalgridTable}. The reference state is derived from hydrostatic relation using the specified radiative equilibrium temperature and the integration starts with the static atmosphere having the density and potential temperature distributions identical to the reference state.

The numerical results output once a day. The predicted quantities are then interpolated to a post-processing coordinate system $\left(\lambda,\phi,p\right)$, i.e. the press-based vertical coordinate and the longitude-latitude horizontal coordinates, to calculate the time-averaged zonal mean quantities. The 1000-day averages of zonal mean temperature, zonal velocity, eddy momentum flux, eddy kinetic energy, eddy heat flux and temperature variance at different isobaric surfaces are shown in Fig. \ref{HS94}. This test is widely checked by many dynamical cores. The results of the proposed model show good agrement with spectral transform solution given in \cite{wan2008}.

%%%%%%%%%%%%%%%%%% SUMMARY

\section{Summary}

A fourth-order non-hydrostatic dynamical core for global atmospheric model is proposed in this study by using multi-moment finite volume method. Through introducing two kinds of moments as model variables, the high-order numerical scheme is constructed over a more compact spatial stencil in comparison with the traditional finite volume method. The resulting model is very flexible in dealing with the computational meshes with complex topologies and can effectively suppress the extra grid-imprinting errors due to the discontinuous coordinates along the inner patch boundaries. Considering the practical dynamics-physics coupling, the finite difference scheme is adopted in vertical direction. The benchmark tests proposed in \citep{jablonowski2008,DCMIP2012,HS94} were carefully checked, including both hydrostatic and nonhydrostatic ones. The numerical results are promising and achieve the expected accuracy in global simulations in comparison with reference solutions of existing advanced models. The proposed model is proven to be capable of accurately reproducing the atmospheric dynamics. Currently, a new high-resolution numerical weather prediction model is under development using the proposed dynamical core.
\section*{Acknowledgments}

This work is supported by National Key Research and Development Program of China (grant nos. 2017YFC1501901 and 2017YFA0603901), National Natural Science Foundation of China (grant no. 41522504).

\bibliography{ref}

\clearpage

\listoftables

\clearpage

\begin{table}[ht]
{
\caption{Parameters for constructing non-uniform vertical grid in different  test cases.}\label{verticalgridTable}
\begin{tabular}{ccccccccc}
\hline\hline
&  Case                              & Top (km) & $N_v$ & $\Delta \zeta (km)$ & $\frac{\Delta\hat{\zeta}_{\min}}{\Delta\zeta}$ & $\frac{\Delta\hat{\zeta}_{\max}}{\Delta\zeta}$ & $\zeta_1$ & $\zeta_2$ (km)  \\ \hline
& 3D Rossby-Haurez wave              & 30     & 26    & 1.154      & 1              & 1               & 0         & 30 \\
& Gravity wave                       & 10     & 20    & 0.5        & 1              & 1               & 0         & 10 \\
& Mountain induced Rossby wave-train & 30     & 26    & 1.154      & 0.1            & 2               & 0         & 30 \\
& Baroclinic wave                    & 30     & 26    & 1.154      & 0.1            & 2               & 0         & 30 \\
& Nonhydrostatic mountain wave       & 30     & 60    & 0.5        & 1              & 1               & 0         & 30 \\
& Held-Saurez test                   & 30     & 30    & 1          & 0.1            & 2               & 0         & 30 \\ \hline
\end{tabular}
}
\end{table}

\clearpage

\listoffigures

\clearpage

\clearpage

\clearpage

\clearpage

\clearpage

\clearpage

\clearpage

\clearpage

\clearpage

\clearpage

\clearpage

\clearpage

\clearpage

\clearpage

\clearpage

\begin{figure}[h]
 \centering
 \includegraphics[width=0.7\textwidth]{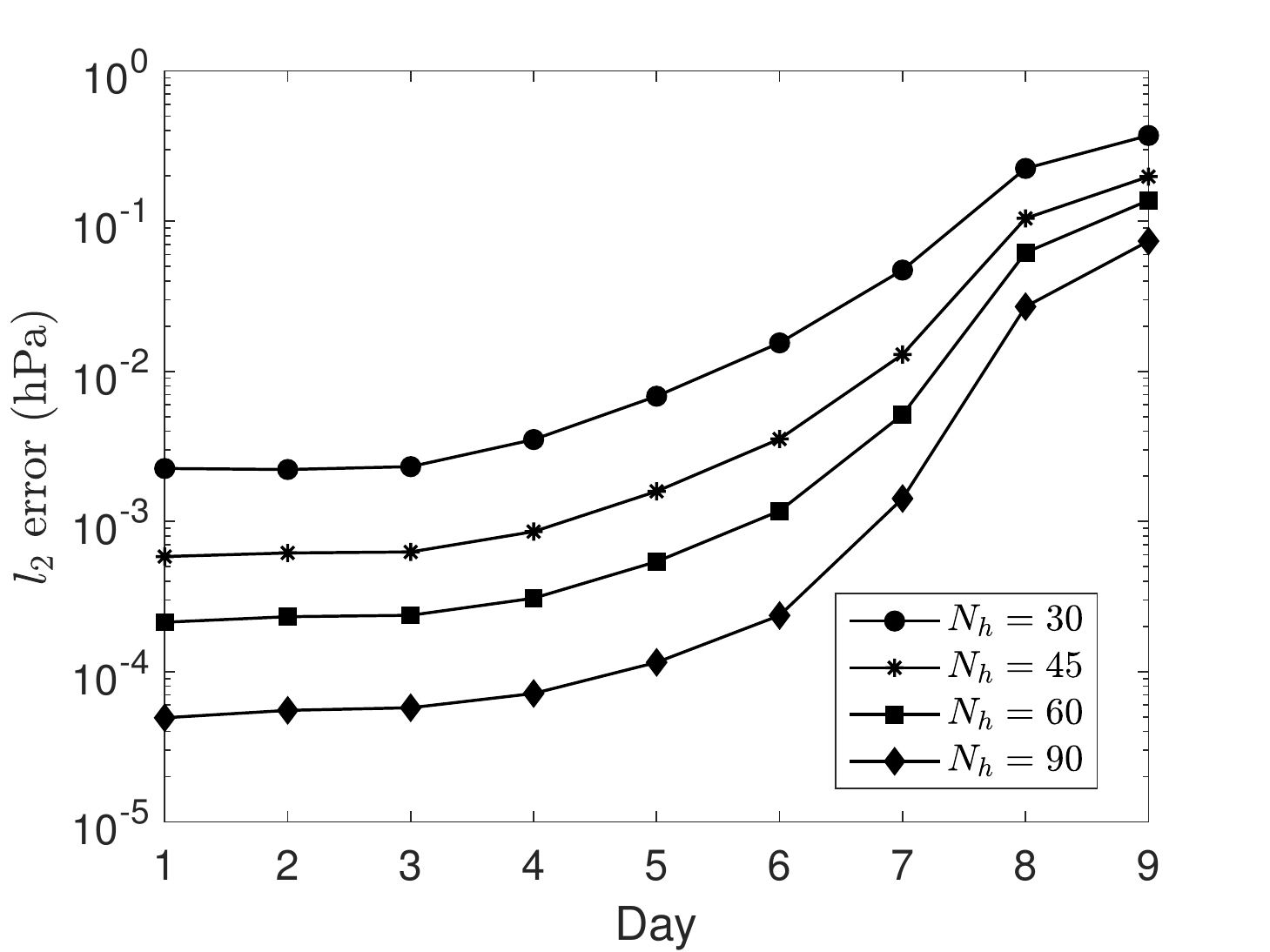}
 \caption{$l_2$ errors of pressure at first model layer on a series of refining grid in comparison with the reference solution calculated on grid $N_h=180$.}\label{BaroclinicWave4}
 \end{figure}

\clearpage

\begin{figure}[h]
 \centering
  \begin{subfigure}[Perturbation of temperature at $t=2400$ s]
  { \includegraphics[width=0.48\textwidth]{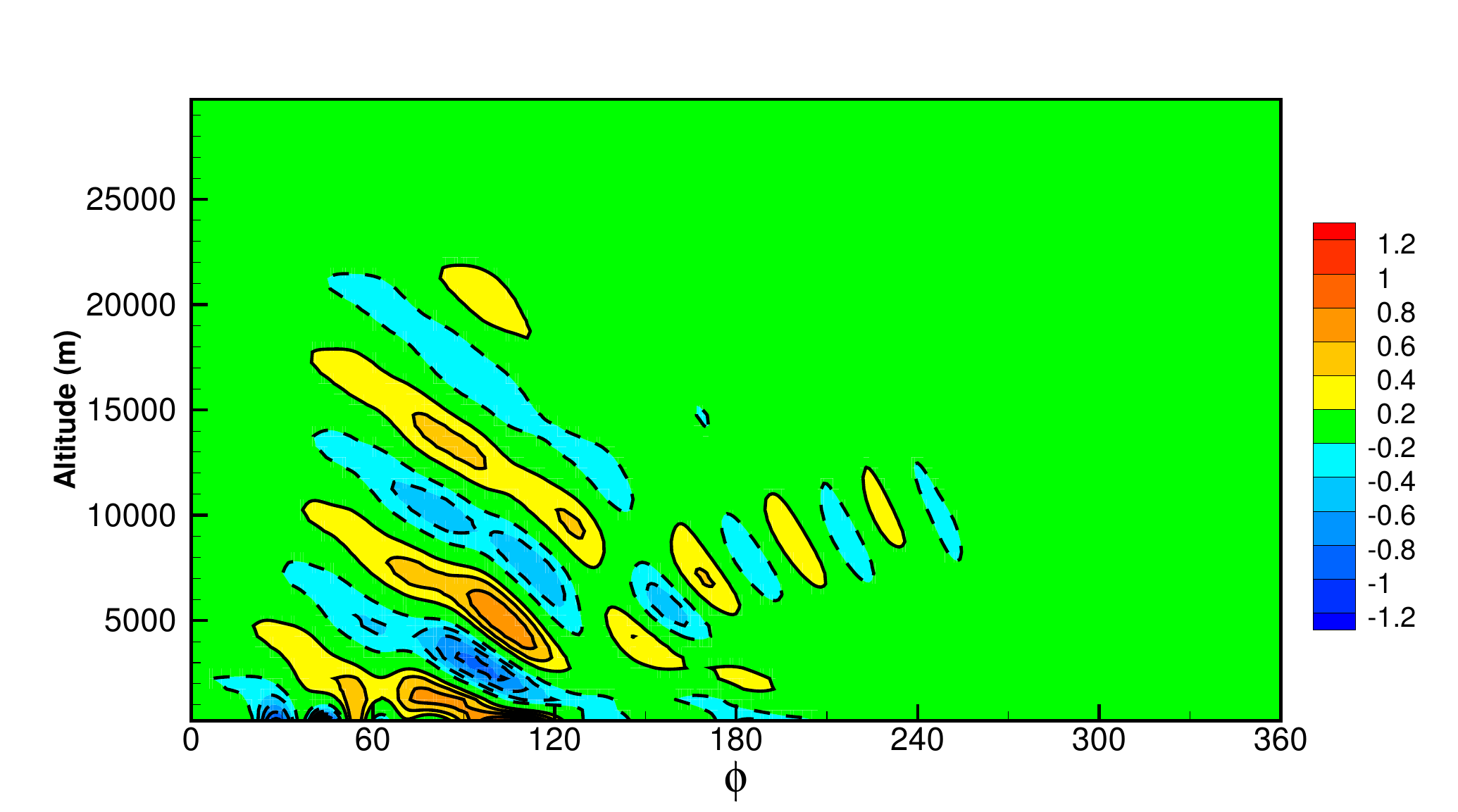}}
  \end{subfigure}
  \begin{subfigure}[Vertical wind at $t=2400$ s]
  { \includegraphics[width=0.48\textwidth]{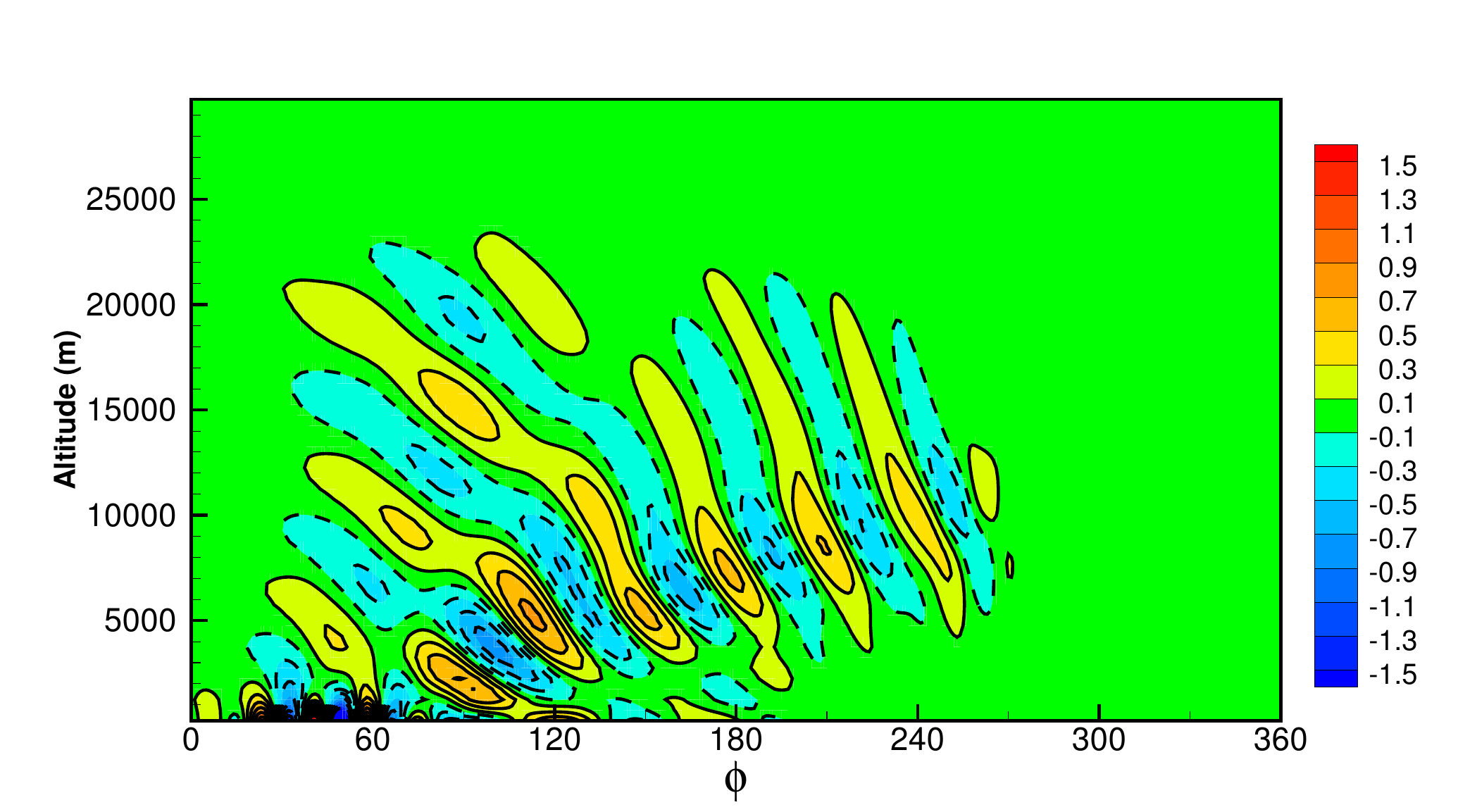}}
  \end{subfigure}
  \begin{subfigure}[Perturbation of temperature at $t=3600$ s]
  { \includegraphics[width=0.48\textwidth]{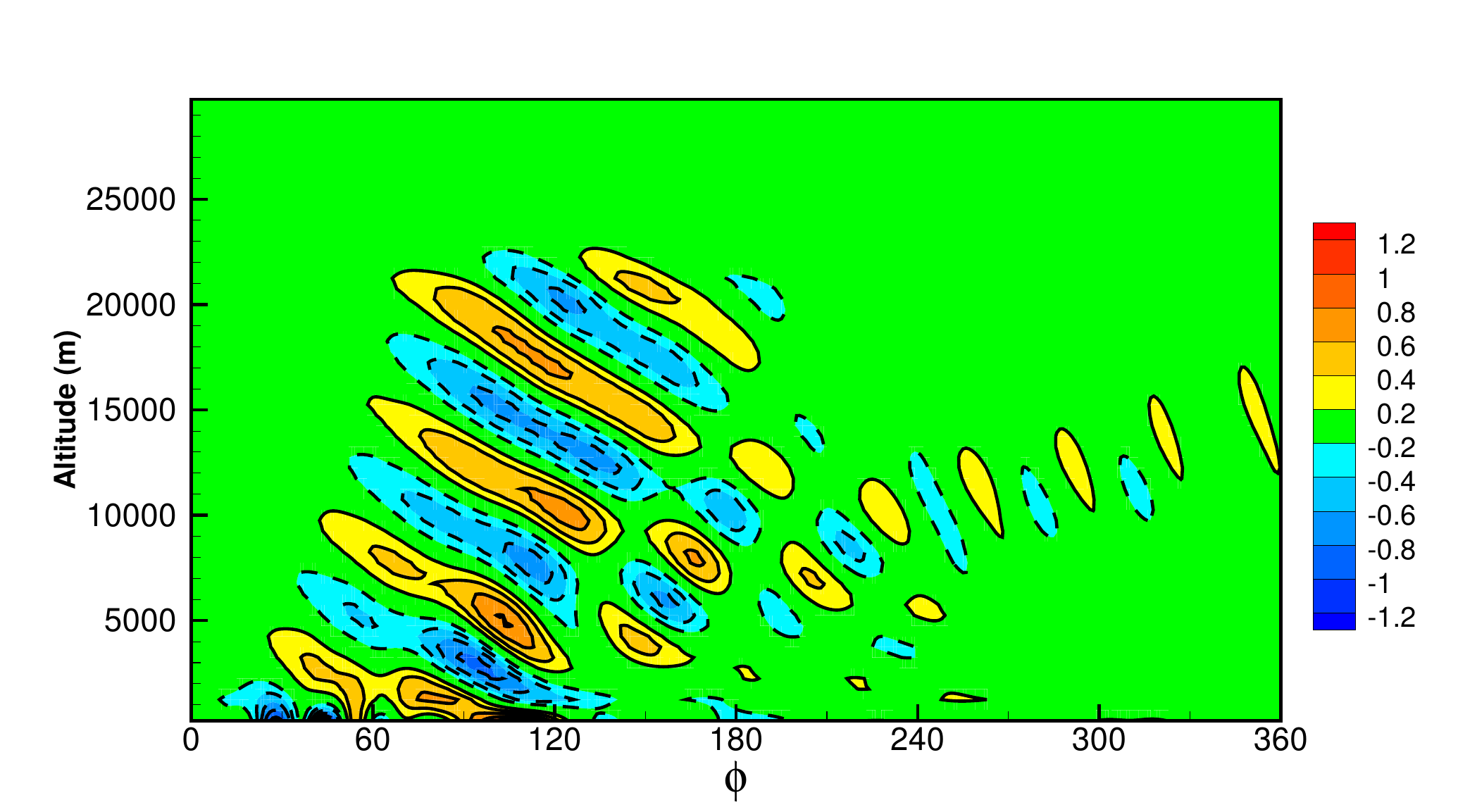}}
  \end{subfigure}
  \begin{subfigure}[Vertical wind at $t=3600$ s]
  { \includegraphics[width=0.48\textwidth]{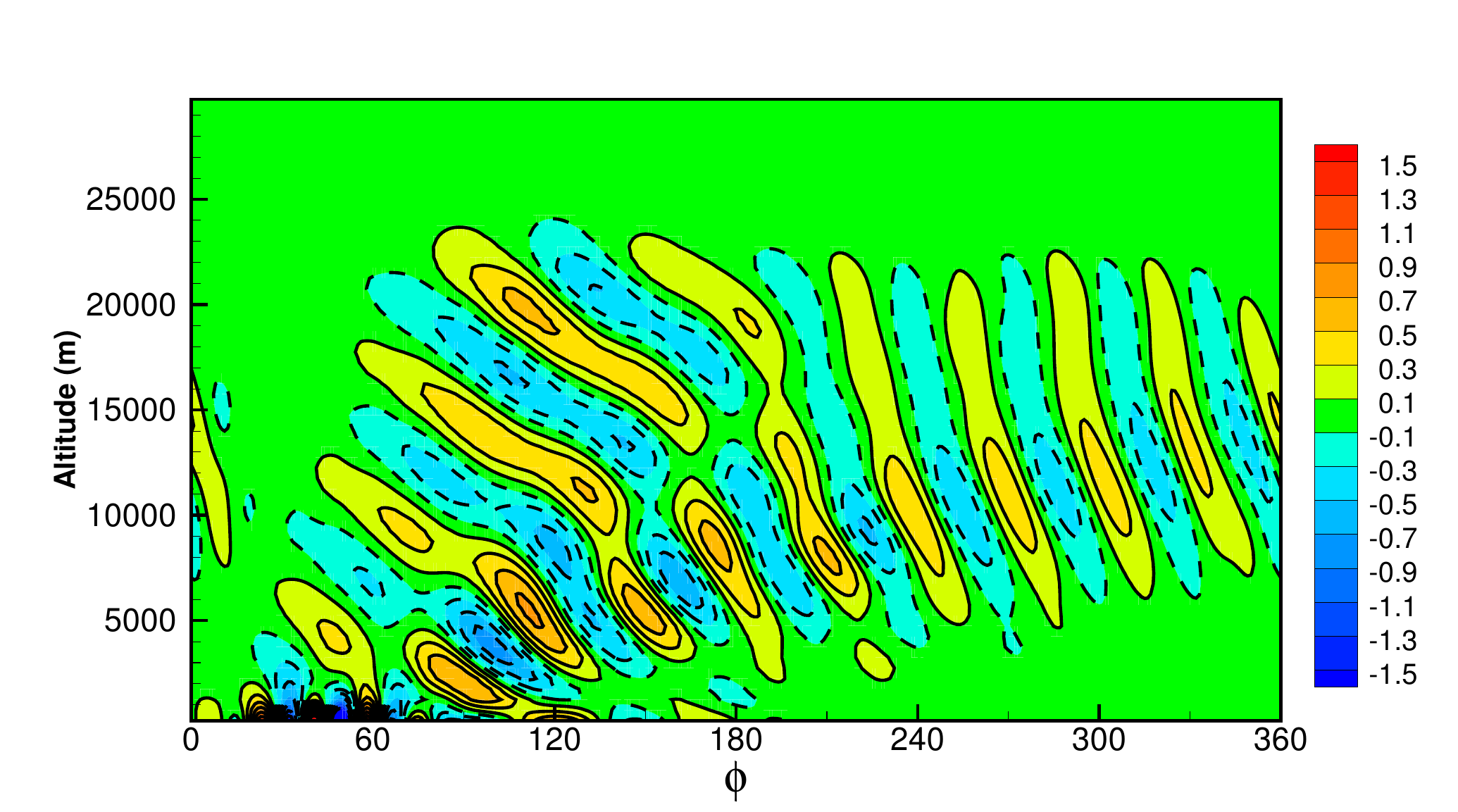}}
  \end{subfigure}
  \begin{subfigure}[Perturbation of temperature at $t=7200$ s]
  { \includegraphics[width=0.48\textwidth]{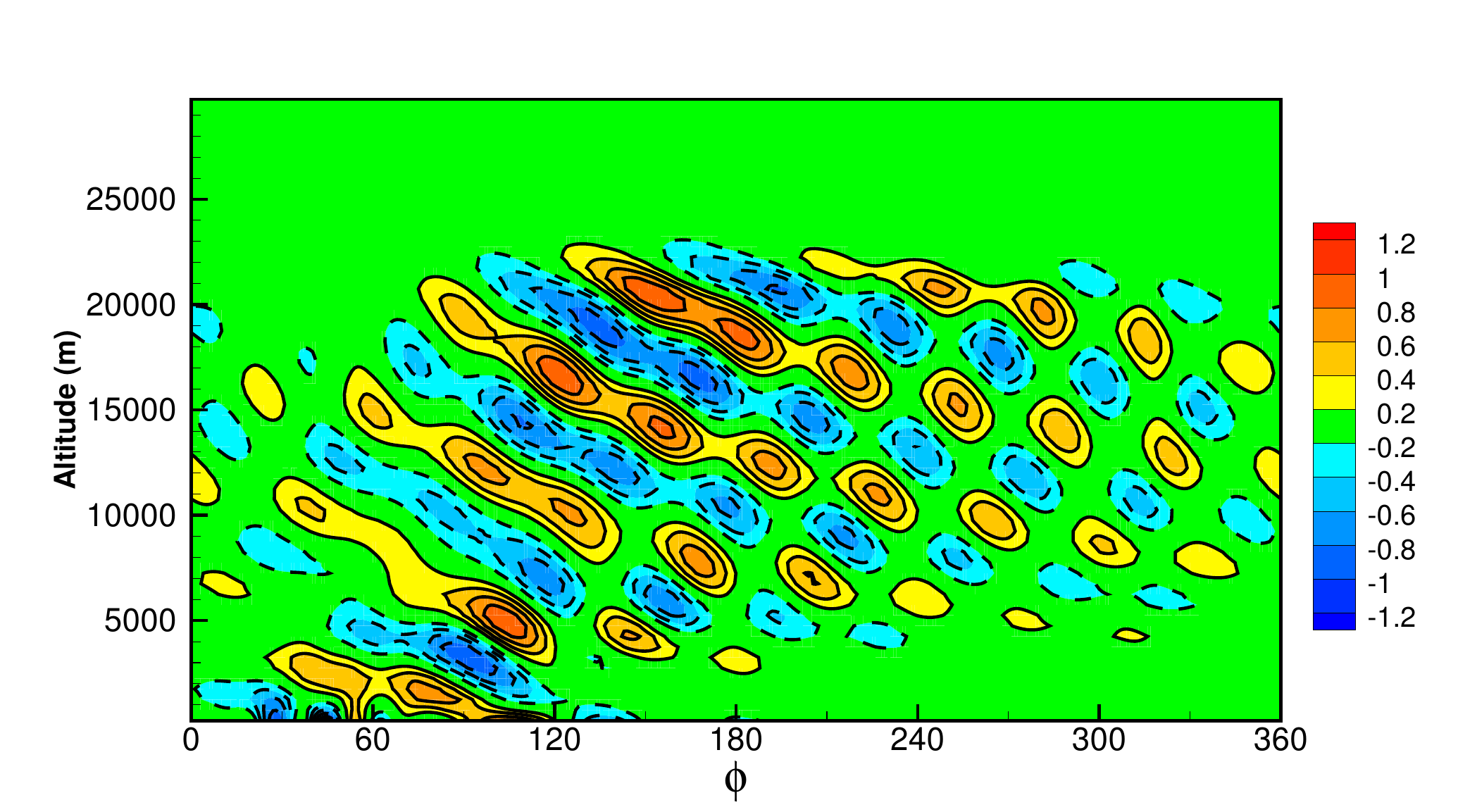}}
  \end{subfigure}
  \begin{subfigure}[Vertical wind at $t=7200$ s]
  { \includegraphics[width=0.48\textwidth]{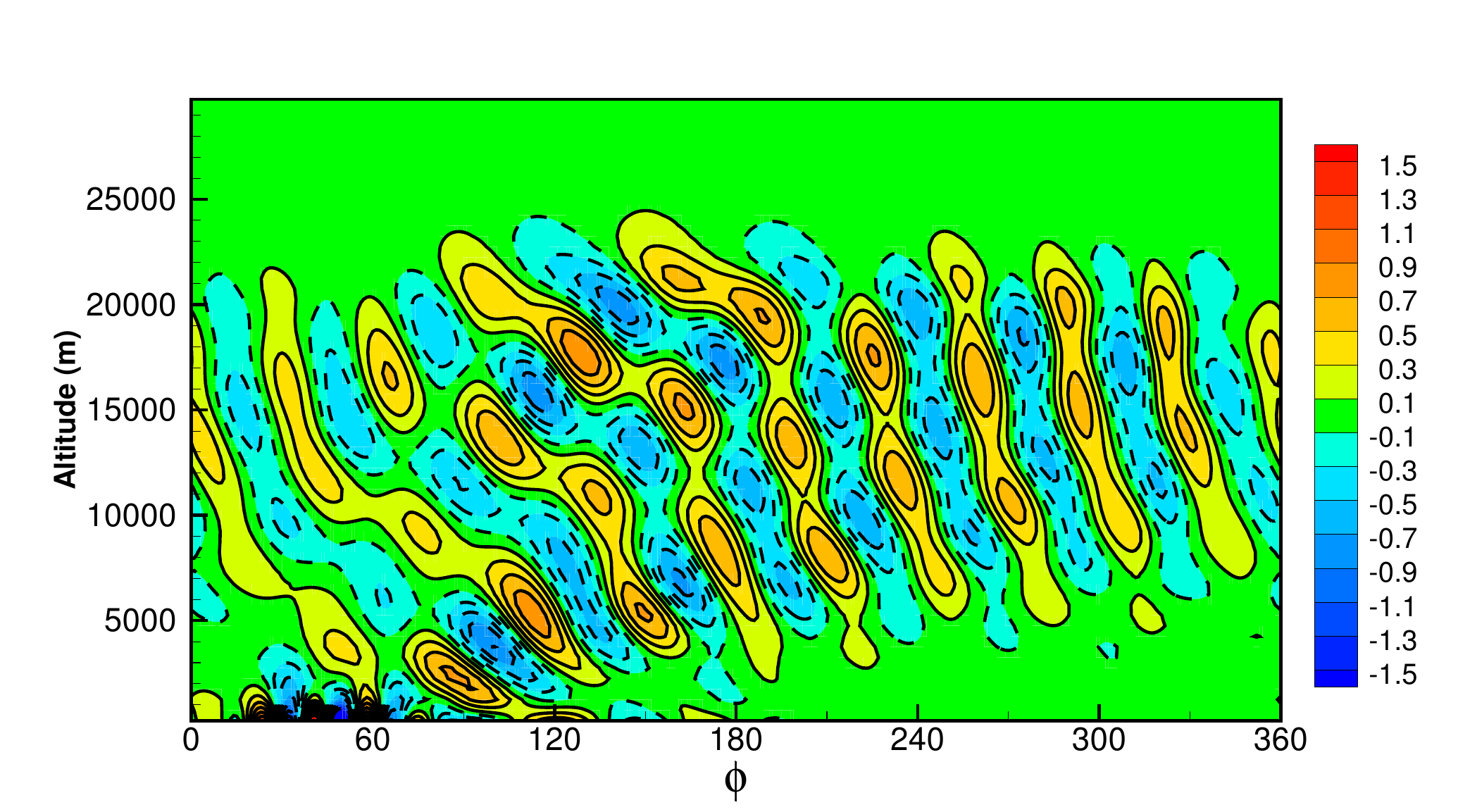}}
  \end{subfigure}
  \caption{Contour plots of numerical results of non-hydrostatic mountain waves (non-sheared case) at $t=2400$, $t=3600$ s at $t=7200$ s. Shown are perturbation of temperature (panels (a), (c) and (e)) and vertical wind (panels (b), (d) and (f)) along the Equator and the dashed lines denote the negative values.}\label{NHMountrain1}
\end{figure}

\clearpage

\begin{figure}[h]
 \centering
  \begin{subfigure}[Perturbation of temperature at $t=2400$ s]
  { \includegraphics[width=0.48\textwidth]{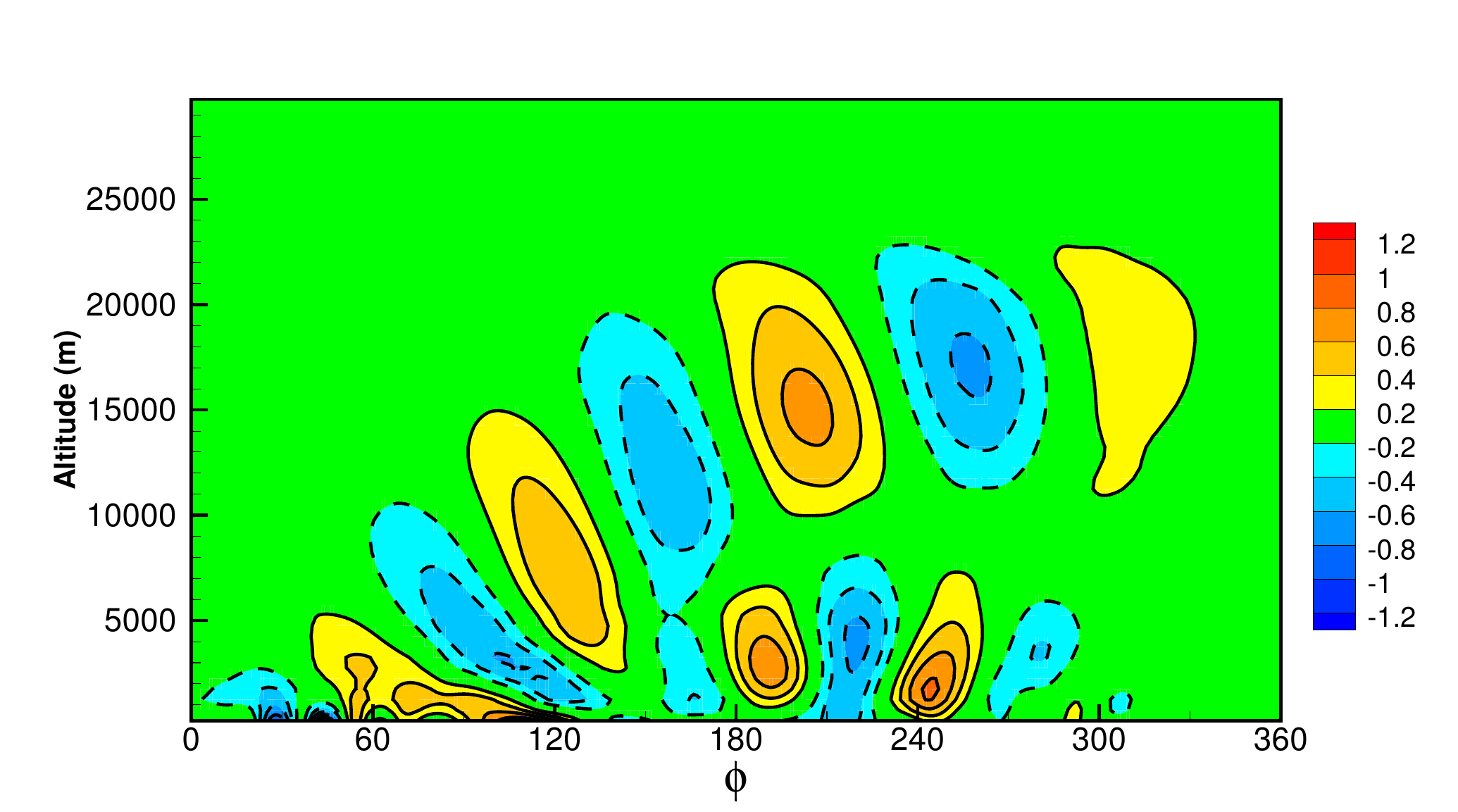}}
  \end{subfigure}
  \begin{subfigure}[Vertical wind at $t=2400$ s]
  { \includegraphics[width=0.48\textwidth]{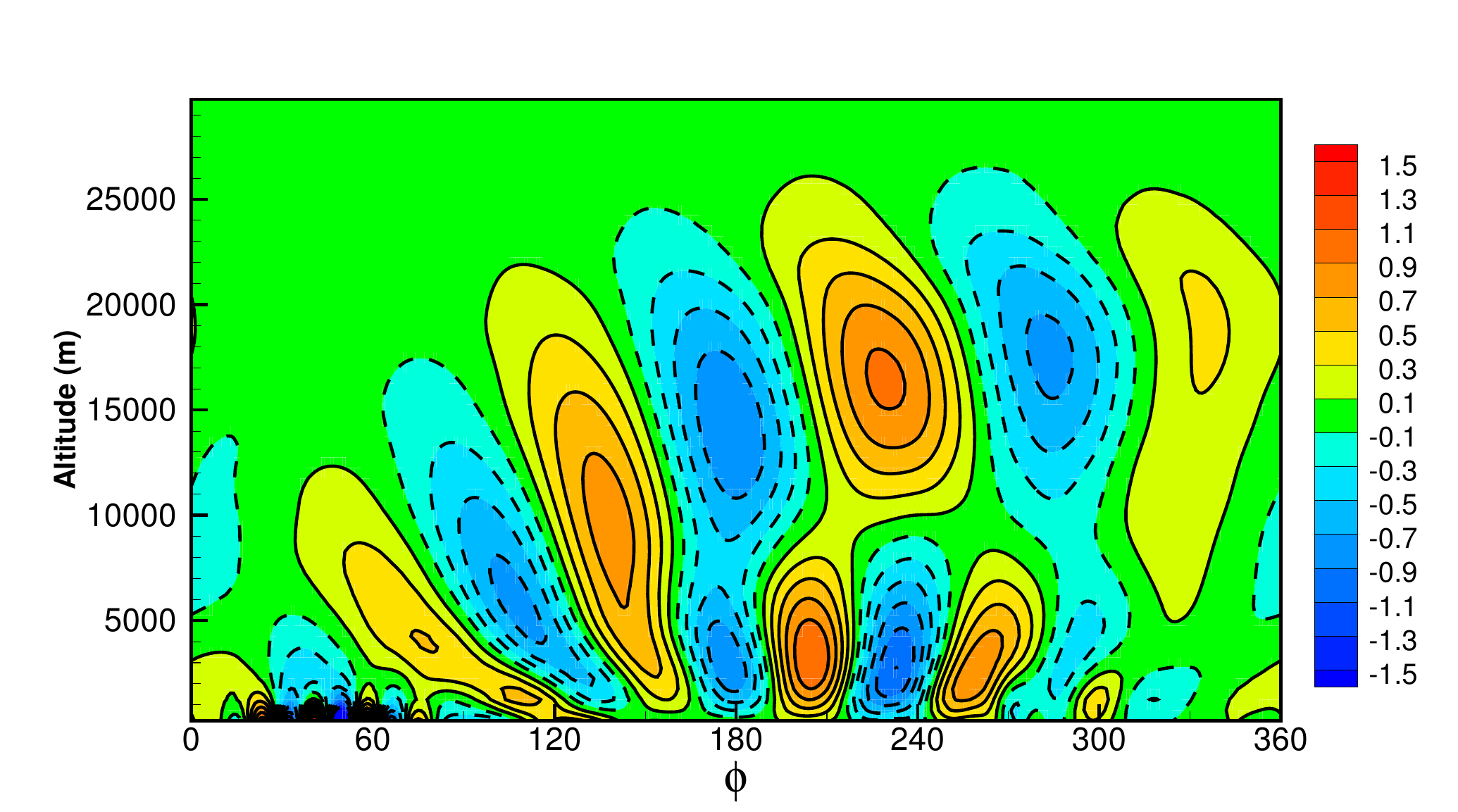}}
  \end{subfigure}
  \begin{subfigure}[Perturbation of temperature at $t=3600$ s]
  { \includegraphics[width=0.48\textwidth]{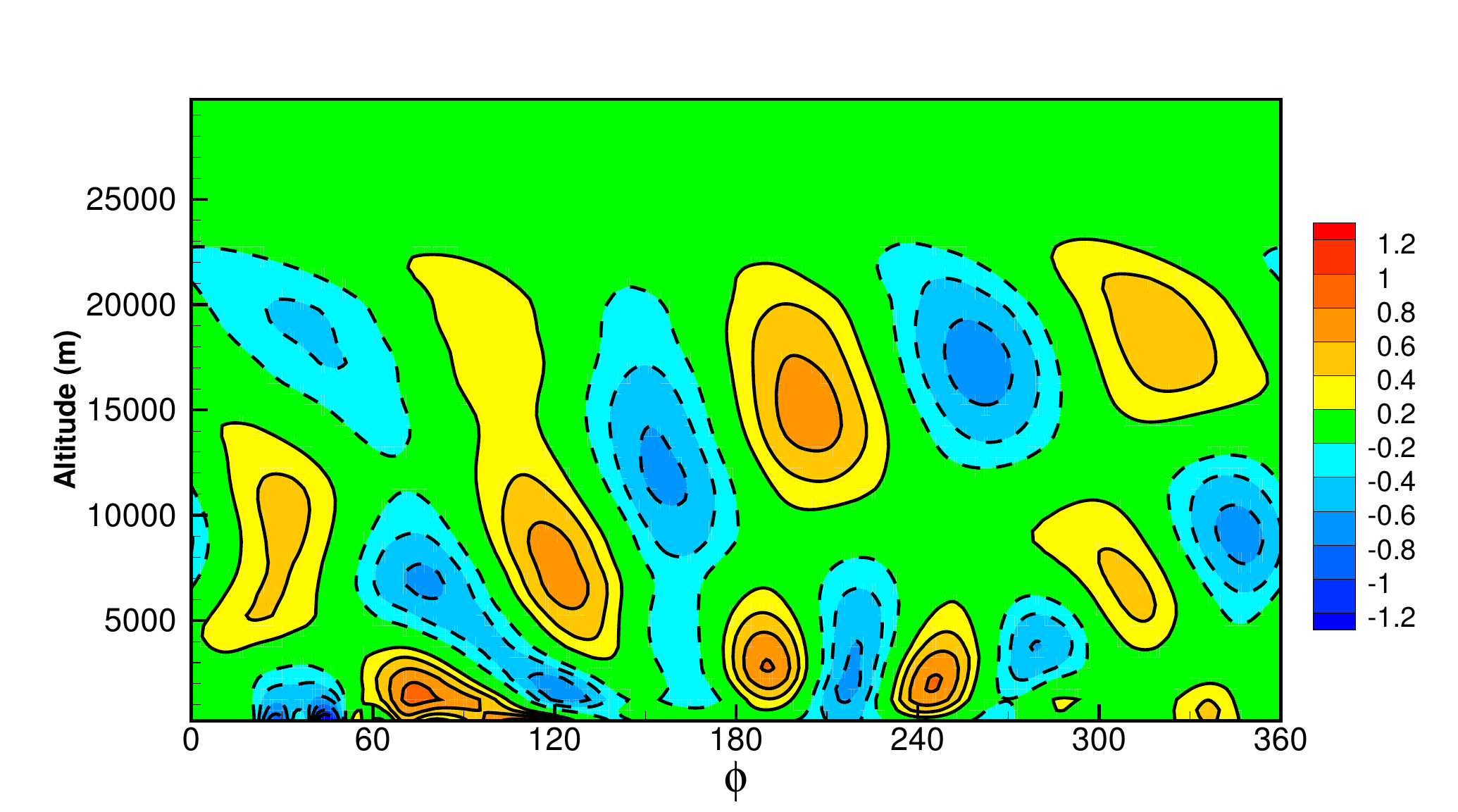}}
  \end{subfigure}
  \begin{subfigure}[Vertical wind at $t=3600$ s]
  { \includegraphics[width=0.48\textwidth]{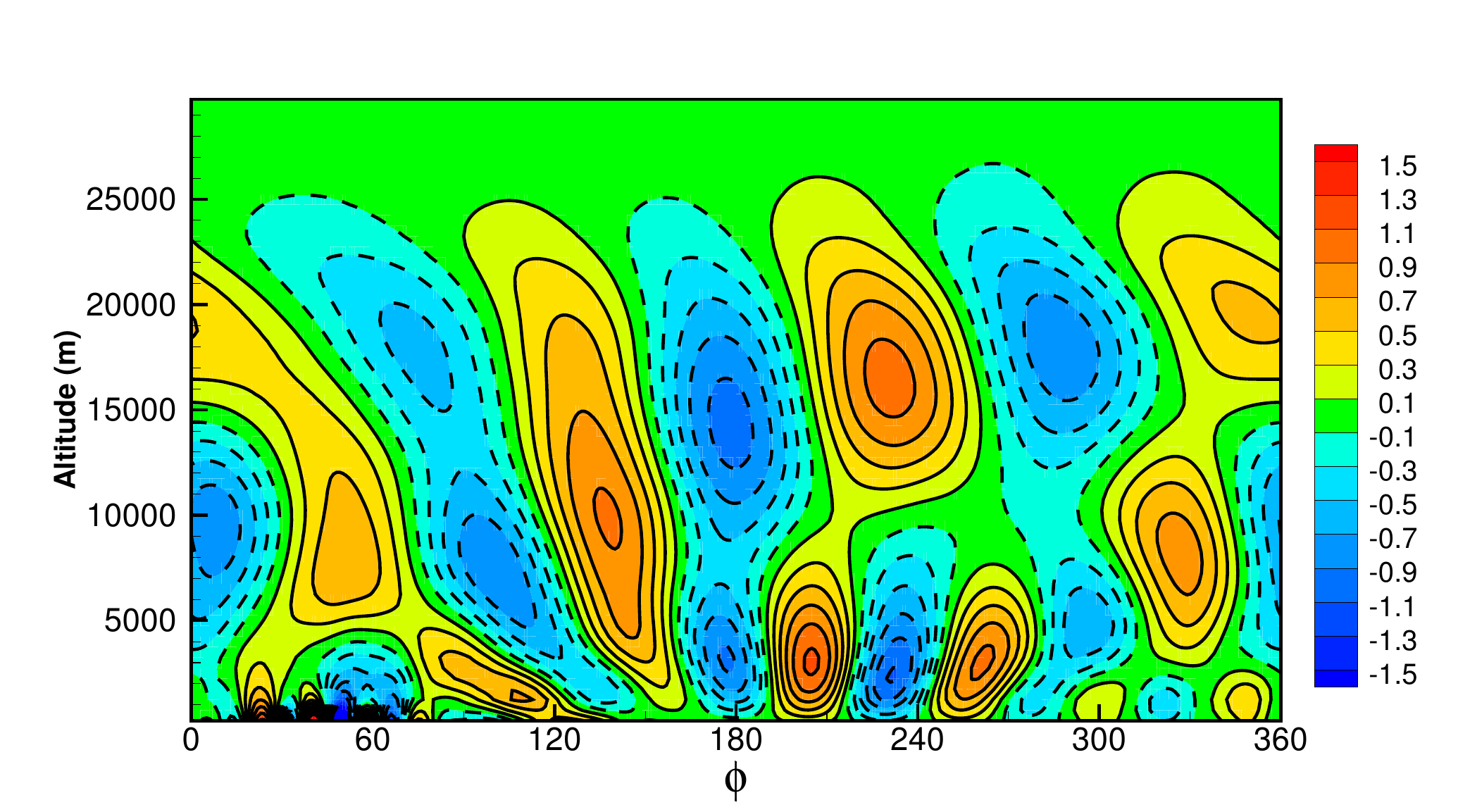}}
  \end{subfigure}
  \begin{subfigure}[Perturbation of temperature at $t=7200$ s]
  { \includegraphics[width=0.48\textwidth]{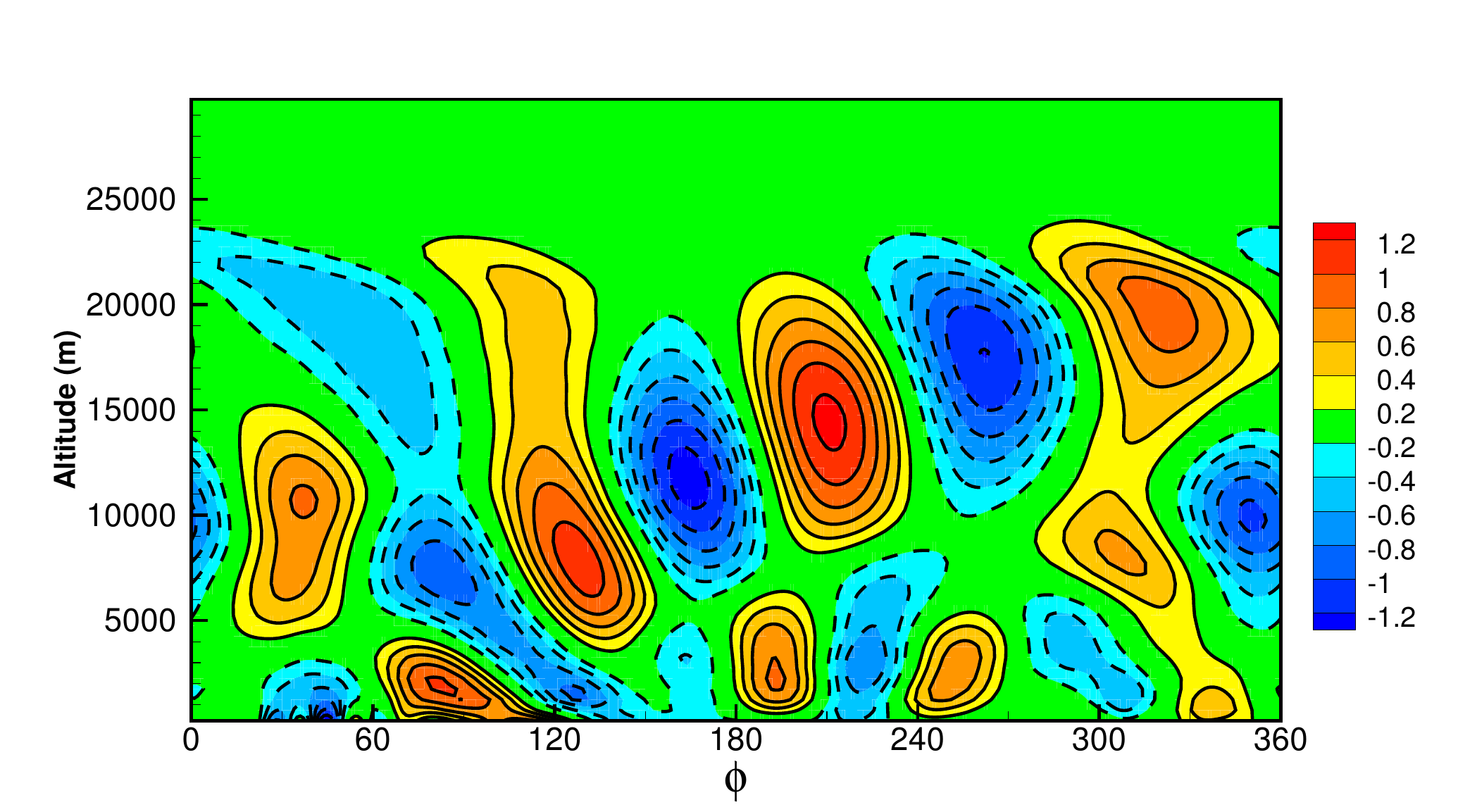}}
  \end{subfigure}
  \begin{subfigure}[Vertical wind at $t=7200$ s]
  { \includegraphics[width=0.48\textwidth]{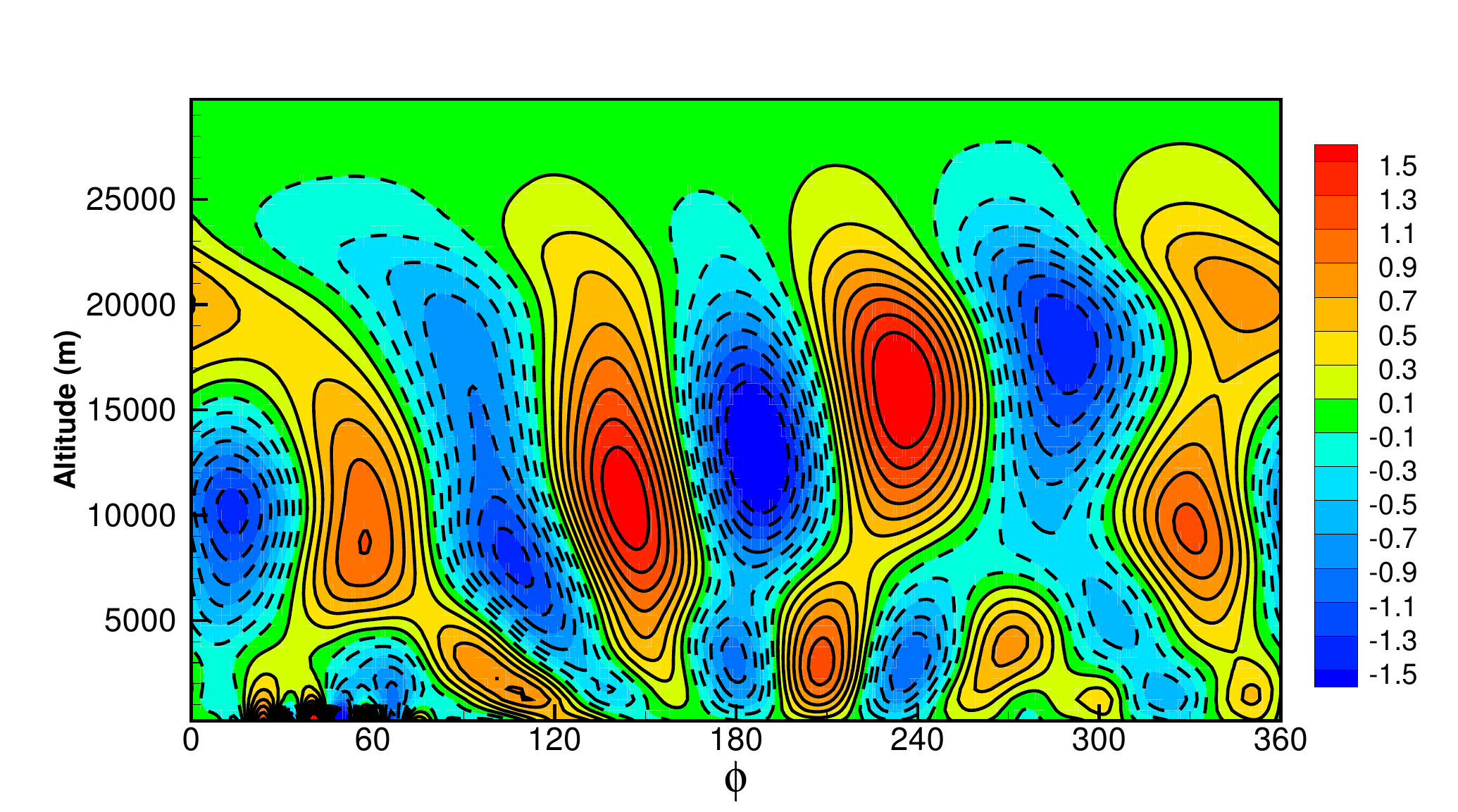}}
  \end{subfigure}
  \caption{Same as Fig. \ref{NHMountrain1}, but for the sheared case.}\label{NHMountrain2}
\end{figure}

\clearpage

\begin{figure}[h]
 \centering
  \begin{subfigure}[Temperature]
  { \includegraphics[width=0.48\textwidth]{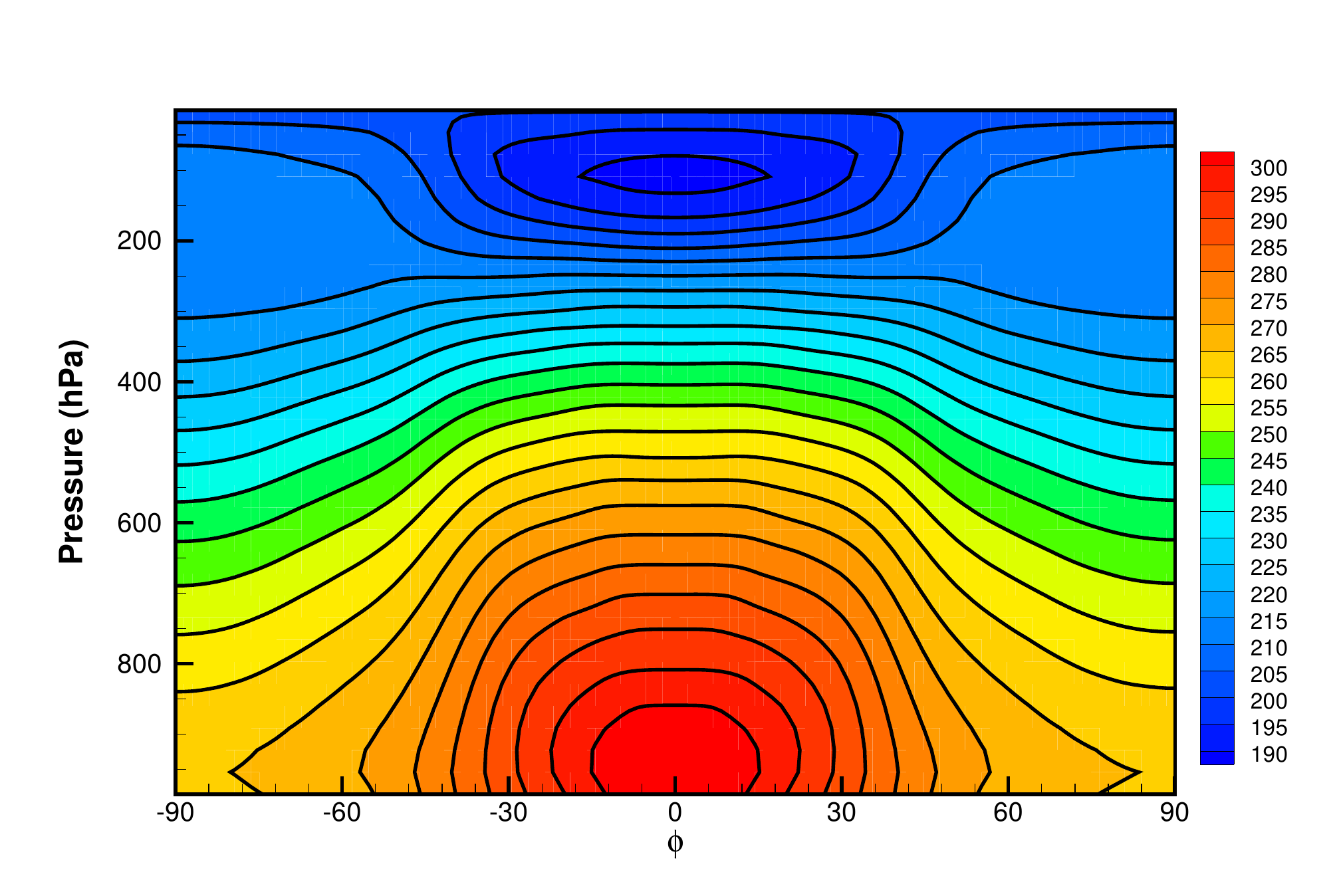}}
  \end{subfigure}
  \begin{subfigure}[Zonal wind]
  { \includegraphics[width=0.48\textwidth]{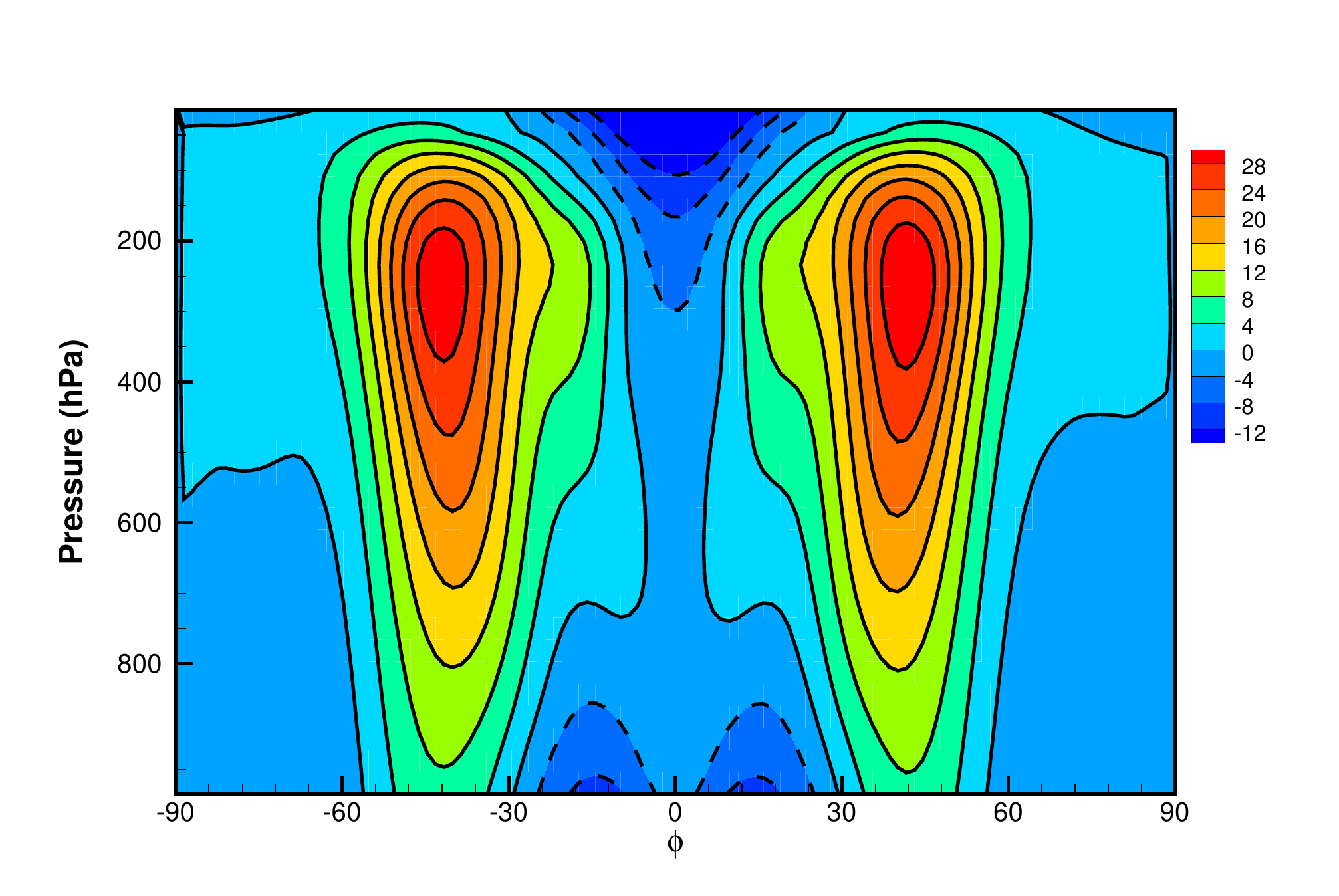}}
  \end{subfigure}
  \begin{subfigure}[Eddy momentum flux]
  { \includegraphics[width=0.48\textwidth]{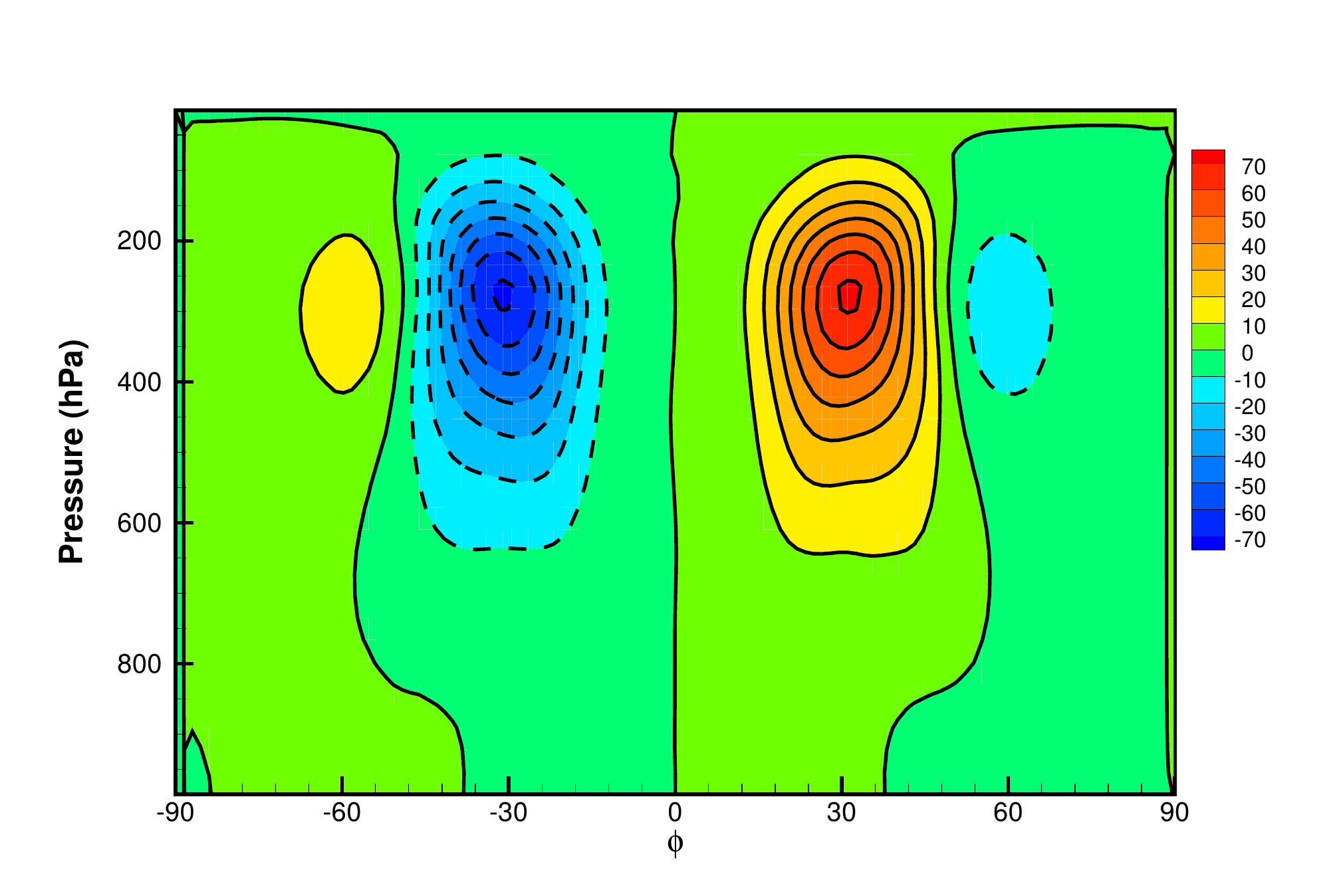}}
  \end{subfigure}
  \begin{subfigure}[Eddy kinetic energy]
  { \includegraphics[width=0.48\textwidth]{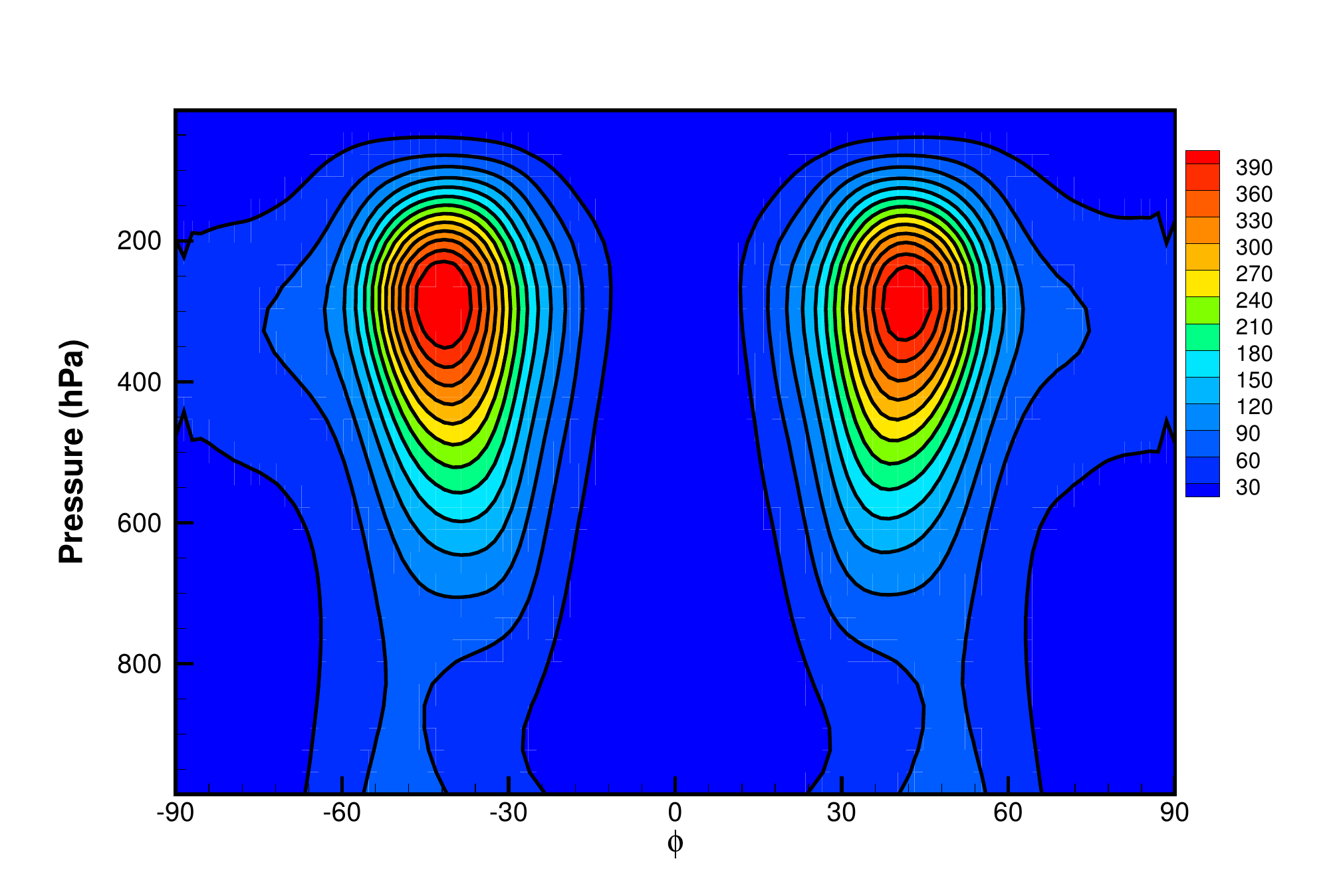}}
  \end{subfigure}
  \begin{subfigure}[Eddy heat flux]
  { \includegraphics[width=0.48\textwidth]{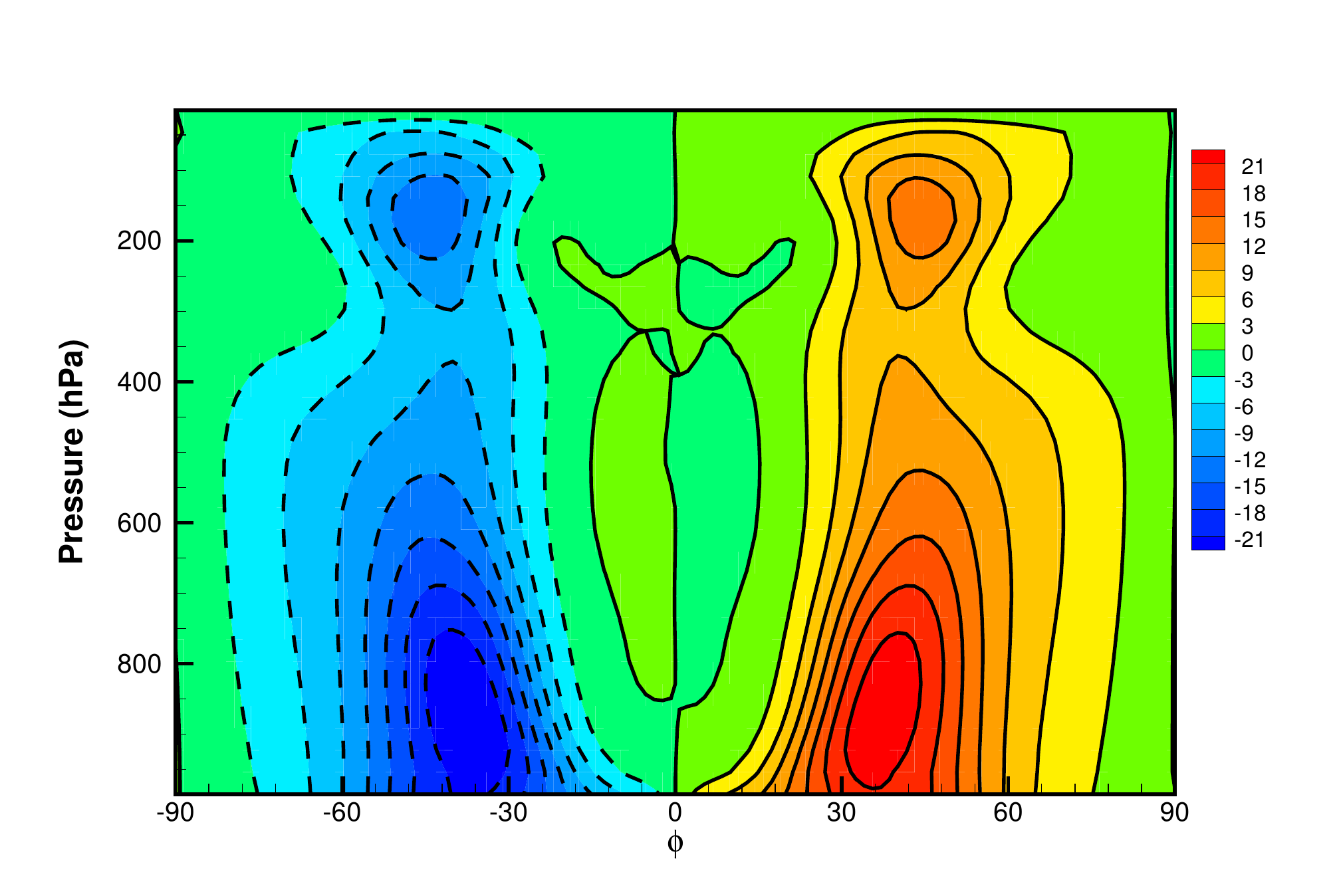}}
  \end{subfigure}
  \begin{subfigure}[Temperature variance]
  { \includegraphics[width=0.48\textwidth]{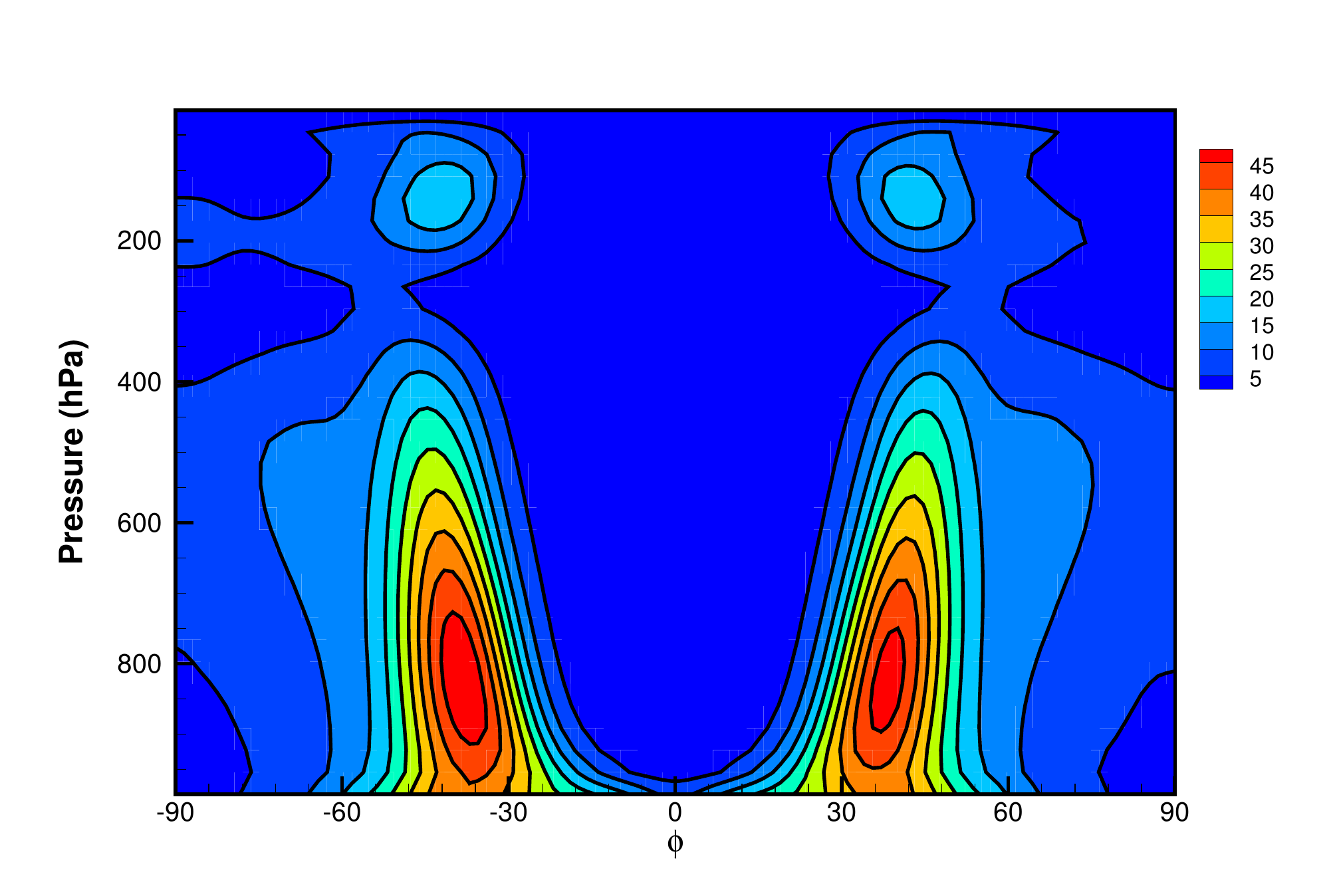}}
  \end{subfigure}
  \caption{Contour plots of numerical results of Held-Saurez test. Shown are 1000-day averages of zonal mean temperature (panel (a)), zonal velocity (panel (b)), eddy momentum flux (panel (c)), eddy kinetic energy (panel (d)), eddy heat flux (panel (e)) and temperature variance (panel (f)). The dashed lines denote the negative values.}\label{HS94}
\end{figure}

\end{document}